\documentclass[aps,pra,showpacs,floatfix,notitlepage]{revtex4-1}

\usepackage{graphicx,amsmath,amssymb,psfrag,empheq,hyperref,wasysym,siunitx}
\usepackage{bm}
\usepackage{subcaption}

\begin{document}

\title{Spectra of Neutron Wave Functions in Earth's Gravitational Field}

\author{Martin Suda$^{1,2}$}
\email{Correspondence: martin.suda@ait.ac.at}
\author{Manfried Faber$^2$}
\email{manfried.faber@tuwien.ac.at}
\author{Joachim Bosina$^{2,3}$}
\author{Tobias Jenke$^3$}
\author{Christian K\"ading$^2$}
\author{Jakob Micko$^{2,3}$}
\author{Mario Pitschmann$^2$}
%\author{Rene Sedmik$^2$}
\author{Hartmut Abele$^2$}
\email{hartmut.abele@tuwien.ac.at}
\affiliation{$^1$AIT, Austrian Institute of Technology, Giefinggasse 4, 1210 Vienna, Austria}
\affiliation{$^2$Technische Universit\"at Wien, Atominstitut, Stadionallee 2, 1020 Vienna, Austria}

\affiliation{$^3$Institut Laue-Langevin - 71 avenue des Martyrs
CS 20156, 38042 GRENOBLE Cedex 9 - France}

%\date{\today}

\begin{abstract}
The time evolution of a quantum wave packet in the linear gravity potential is known as Quantum Bouncing Ball. The qBounce collaboration recently observed such a system by dropping wave packets of ultracold neutrons by a height of roughly 30 microns. In this article, space and momentum spectra as well as Wigner functions of the neutron wave functions in the gravitational field of the Earth are analyzed. We investigate the quantum states in the "preparation region", into which they transition after exiting a narrow double-mirror system and where we would expect to observe free fall and bounces in classical physics. For this, we start from the stationary solutions and eigenvalues of the Schr\"odinger equation in terms of Airy functions and their zeros. Subsequently, we examine space and momentum distributions as well as Wigner functions in phase space for pure and mixed quantum states. The eventual influence of Yukawa-like forces for small distances of several micrometers from the mirror is included through first order perturbation calculations. Those allow us to study the resulting modifications of space and momentum distributions, and phase space functions. 
\end{abstract}


\maketitle

%\tableofcontents

%------------------------------------------------------------------------------
\section{Introduction}
%------------------------------------------------------------------------------
A quantum wave packet bouncing on a hard surface under the influence of gravity has drawn some attention in the literature  due to its departures from classical behaviour~\cite{Gea-Banacloche:1999,Gibbs:1975,Langhoff:1971a,Goodings:1991,Whineray:1992,Desko:1983,Dembinski:1996}. Other aspects of this quantum bouncer have also been studied to some extent. Of those, we would like to mention its chaotic behavior~\cite{Dembinski:1993}, the mathematical basis with orthonormal Airy eigenfunction solutions~\cite{Vallee:2010}, the Wigner phase space as an interface of gravity and quantum mechanics~\cite{Giese:2014}, quantum revivals in a periodically driven gravitational cavity~\cite{Saif:2000}, and inertial and gravitational mass in quantum mechanics~\cite{Kajari:2010}. The development of sufficient ultracold neutron sources at the Institut Laue Langevin (ILL) in Grenoble and techniques to manipulate neutrons with high precision have made the simple quantum bouncer experimentally realizable. Demonstrations of quantum states in the gravitational potential of the Earth can be found in~\cite{Nesvizhevsky:2002b,Nesvizhevsky:2003,Nesvizhevsky:2005a} and aspects from a more theoretical point of view in~\cite{Voronin:2006,Westphal:2007a}. From the beginning these experiments were used to constrain hypothetical gravity-like interactions~\cite{Abele:2003,Nesvizhevsky:2004,Baessler:2007}.

In this article, we will examine some details of the bounce of a neutron wave packet closely related to an experimental realization by the \textit{q}\textsc{Bounce} collaboration. More precisely, we will investigate the behavior of the momentum space wave packet solutions, the widths of the position and momentum space wave packets during the "bounce", and aspects of Yukawa-type interactions. Extensive use of the Wigner function formalism as a function of time is made as well.

The \textit{q}\textsc{Bounce} experiment has been performed at the UCN-beam position of the PF2 instrument at ILL, so far the 7th
strongest source for ultracold neutrons with high continuous fluence, which is ideal for quantum bouncer realizations. It tests gravity at small distances with quantum interference techniques. The experimental tool is a gravitationally interacting quantum system - an ultracold neutron in the gravitational potential of the Earth - and a reflecting mirror above which the neutron is bound in well-defined quantum states. The collaboration is continuously developing a gravity resonance spectroscopy (GRS)~\cite{Abele:2010a,Jenke:2011b,Jenke:2014a,Cronenberg:2018b} technique, which allows for a clear identification of the measured energy eigenstates states $|1\rangle$ $\longrightarrow$ $|2\rangle$, $|1\rangle$ $\longrightarrow$ $|3\rangle$, $|1\rangle$ $\longrightarrow$ $|4\rangle$, $|2\rangle$ $\longrightarrow$ $|3\rangle$, $|2\rangle$ $\longrightarrow$ $|4\rangle$, $|2\rangle$ $\longrightarrow$ $|5\rangle$, and most recently $|1\rangle$ $\longrightarrow$ $|6\rangle$. In this way, precisions are reached which enable us to search for hypothetical gravity-like interactions with relevance for cosmology. So far limits for axions~\cite{Jenke:2011b}, chameleon~\cite{Jenke:2014a} and symmetron fields~\cite{Cronenberg:2018b} have been placed.

For the purpose of this article, an important observable is the spatial density distribution of a free falling neutron above a reflecting mirror. A newly developed position-dependent neutron detector makes it possible to visualize the square of the Schr\"odinger wave function~\cite{Abele:2009a,Jenke:2009a}. Detailed descriptions of these processes can be found in \cite{A12}. We now have a high-precision gravitational neutron spectrometer with available spatial resolution of \SI{1.5}{\micro\meter} at our disposal. Neutrons are detected in CR-39 track detectors after neutron capture in a coated Boron-10 layer of \SI{100}{\nano\meter} thickness. An etching technique makes the tracks visible with a length of
about \SI{3}{\micro\meter} to \SI{6}{\micro\meter}~\cite{Jenke:2013a}.

Because of the Schr\"odinger equation, and therefore by means of the quantum mechanical description of particles in a gravitational field, a wave function is established exhibiting both, a local spreading and a momentum distribution. As is well-known, it is possible to describe this phenomenon using Airy functions. In doing so, it appears that, due to the reflection on the mirror surface, a ground state and excited states emerge. Moreover, the Wigner function allows for a combined view within the entire phase space. Further attention is especially put on marginal distribution functions of the Wigner distribution which correspond exactly to the space and momentum distributions. For all our numerical calculations we use the computer software \textit{Mathematica}. The space distribution had been measured using a track detector~\cite{Jenke:2013a}. Likewise, the momentum distribution should be determined experimentally using an appropriate detector. The main objective of our calculations is the comparison with these measurements. 
 
The article is organized as follows: In chapter II the Schr\"odinger equation including a gravitational potential is given and time-independent solutions are explored. Using an appropriate scaling, a differential equation is found, whose solutions can be expressed by Airy functions. The calculation of the Fourier transform of the ground state is presented and excited states are considered. Furthermore, the solutions using the Wigner function and the time dependence of the superposition of ground and first excited states are described. In chapter III we investigate the \textit{q}\textsc{Bounce}-system in which the neutron wave is enclosed between 2 mirrors. This chapter is subdivided into two sections, one dealing with the Fourier transformation of the wave function and the other one being concerned with the Wigner function. Chapter IV is dedicated to a wave function exiting the double mirror system and falling onto a subsequent mirror. One section describes the space distribution in this "free fall" region, another one is dedicated to the space distribution of mixtures, the third one deals with the calculation of the momentum distribution and the last section presents the related Wigner function. In chapter V we perform a first order perturbation calculation in order to describe a very small change in the potential near the mirror. At first, the mathematical background is presented. Afterwards, the Fourier transformations of the results are carried out, the momentum distribution including a Yukawa-like term described and the related space distribution evaluated. Finally, chapter VI gives a short summary.
%------------------------------------------------------------------------------

\section{Schr\"odinger equation for \textit{q}\textsc{Bounce}}
%------------------------------------------------------------------------------
The time-dependent Schr\"odinger equation for a neutron with mass $m_{\text{N}}$ in the gravitational field of the Earth with potential energy ($g$ is the gravitational acceleration, $z$ the distance above the mirror)
\begin{equation}\label{pot}
V(z)=m_{\text{N}} g z
\end{equation}
reads
\begin{equation}\label{SG}
\hat H\,\psi(z,t)=i\hbar\,\dot\psi(z,t)\,\,\,,
\end{equation}
where $\hat H$ is the Hamiltonian containing $V(z)$. The energy of the wave function $\psi(z,t)$ is quantized in the potential $V(z)$. Using the ansatz
\begin{equation}\label{Ans}
\psi_n(z,t)=\mathrm e^{- i\frac{E_n}{\hbar}t}\psi_n(z)
\end{equation}
for a stationary state of energy $E_n$ with $n=1,2,...$, we obtain the time-independent Schr\"odinger equation
\begin{equation}\label{zeitabh}
\left[-\frac{\hbar^2}{2m_{\text{N}}}\frac{\mathrm d^2}{\mathrm dz^2}+m_{\text{N}}gz\right]\psi_n(z)
=E_n\psi_n(z)\,\,\,.
\end{equation}
For negative values of $z$ we have $\psi(z)=0$ because the particles cannot enter the mirror surface. Therefore, the boundary condition for the solution of the differential equation is $\psi(0)=0$. For this reason, it is supposed that the surface of the mirror has an infinite Fermi potential and the quantum wave does not enter the surface. This is of course just an approximation.
 
At this point, it is appropriate to mention that the problem of two mirrors as well as the transition from an inertial frame $(z_0, t_0)$ to a non-inertial frame $(z,t)$ has already been described in~\cite{P18}.
 
Searching for solutions $\psi_n(z)$, we multiply Eq.~(\ref{zeitabh}) with the factor $\left(\frac{2}{\hbar^2m_{\text{N}}g^2}\right)^{1/3}$ and, using the substitutions~\cite{A12}
\begin{equation}\label{Einsetzung}
\zeta=z/z_0\,\,\,, \,\,\,z_0=\left(\frac{\hbar^2}{2m_{\text{N}}^2g}\right)^{1/3}\approx\SI{5.86796}{\micro\meter}\,\,\,,
\quad a_n=-E_n/E_0\,\,\,,\,\,\,E_0=\left(\frac{\hbar^2m_{\text{N}}g^2}{2}\right)^{1/3}\approx \SI{0.602}{\pico\electronvolt}\,\,\,,
\end{equation}
where $E_0$ is a characteristic gravitational energy scale,
we obtain the differential equation
\begin{equation}\label{DG}
\left(\frac{\mathrm d^2}{\mathrm d\zeta^2}-(\zeta+a_n)\right)\psi_n(\zeta)=0\,\,\,.
\end{equation}
\begin{figure}[h!]
\centering
\includegraphics[width=80mm]{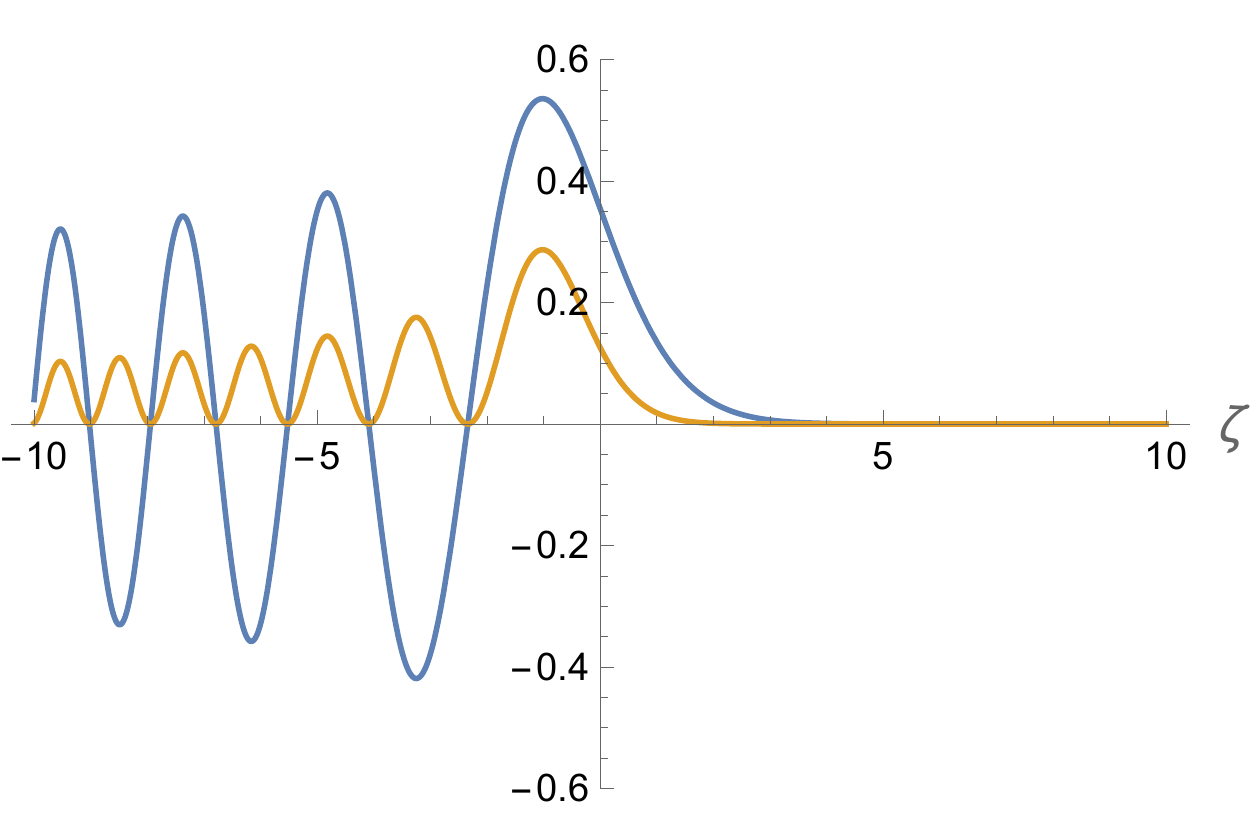}
\caption{Airy function $\mathrm{Ai}(\zeta)$ (blue) and $[\mathrm{Ai}(\zeta)]^2$ (orange)}\label{Fig1}
\end{figure}

Comparing this equation with the Airy equation
\begin{equation}\label{Ai}
  \left(\frac{\mathrm d^2}{\mathrm d\zeta^2}-\zeta\right)\mathrm{Ai}(\zeta)=0\,\,\,,
\end{equation}
we notice that the (non-normalized) eigenfunctions $\psi_n(\zeta)$ can be expressed through the Airy function $\mathrm{Ai}(\zeta)$ by moving the origin of coordinates to the $n^{th}$ zero point $a_n$:
\begin{equation}\label{EF}
\psi_n(\zeta)=\mathrm{Ai}(\zeta+a_n)\Theta(\zeta)\,\,\,.
\end{equation}
As described in~\cite{P18} the normalized wave function $\psi_n(z,t)$ is given by
\begin{equation}\label{EFnorm}
\psi_n(z,t)=\frac{1}{\sqrt{z_0}{\mathrm{Ai'}\left(-\frac{z_n}{z_0}\right)}}{\mathrm{Ai}}{\left(\frac{z-z_n}{z_0}\right)}\mathrm{e}^{-\frac{i}{\hbar}E_nt}=\psi_n(z)\,\mathrm{e}^{-\frac{i}{\hbar}E_nt}\,\,.
\end{equation}

The first zero point of the Airy function is located at $a_1\approx-2.3381$. This means, that $E_1=-a_1{}E_0\approx\SI{1.41}{\pico\electronvolt}$. Additional zero points are located along the negative axis, as can be seen in Fig.~\ref{Fig1}. They determine the energy eigenvalues $E_n$ according to Eq.~(\ref{Einsetzung}). We list some of them below ($z_n=-z_0a_n$):
\begin{eqnarray}\label{energies}
n&=&1\,\,\,,\,\,\,E_1\approx\SI{1.40672}{\pico\electronvolt}\,\,\,,\,\,\,z_1\approx\SI{13.71680}{\micro\meter}\,\,\,,\,\,\,a_1\approx-2.33810\,\,\,,\nonumber \\
n&=&2\,\,\,,\,\,\,E_2\approx\SI{2.45951}{\pico\electronvolt}\,\,\,,\,\,\,z_2\approx\SI{23.98246}{\micro\meter}\,\,\,,\,\,\,a_2\approx-4.08795\,\,\,,\nonumber \\
n&=&3\,\,\,,\,\,\,E_3\approx\SI{3.32144}{\pico\electronvolt}\,\,\,,\,\,\,z_3\approx\SI{32.38707}{\micro\meter}\,\,\,,\,\,\,a_3\approx-5.52056\,\,\,,\nonumber \\
n&=&4\,\,\,,\,\,\,E_4\approx\SI{4.08321}{\pico\electronvolt}\,\,\,,\,\,\,z_4\approx\SI{39.81502}{\micro\meter}\,\,\,,\,\,\,a_4\approx-6.78671\,\,\,,\nonumber \\
n&=&5\,\,\,,\,\,\,E_5\approx\SI{4.77958}{\pico\electronvolt}\,\,\,,\,\,\,z_5\approx\SI{46.60526}{\micro\meter}\,\,\,,\,\,\,a_5\approx-7.94412\,\,\,,\nonumber \\
n&=&6\,\,\,,\,\,\,E_6\approx\SI{5.42846}{\pico\electronvolt}\,\,\,,\,\,\,z_6\approx\SI{52.93243}{\micro\meter}\,\,\,,\,\,\,a_6\approx-9.02262\,\,\,.
\end{eqnarray}
The quantities $z_n$ are given, such that they can later be compared to  Eq.~(\ref{Wave2mpar}).
 
\begin{figure}[h!]
\centering
\begin{subfigure}{.5\textwidth}
  \centering
  \includegraphics[width=80mm]{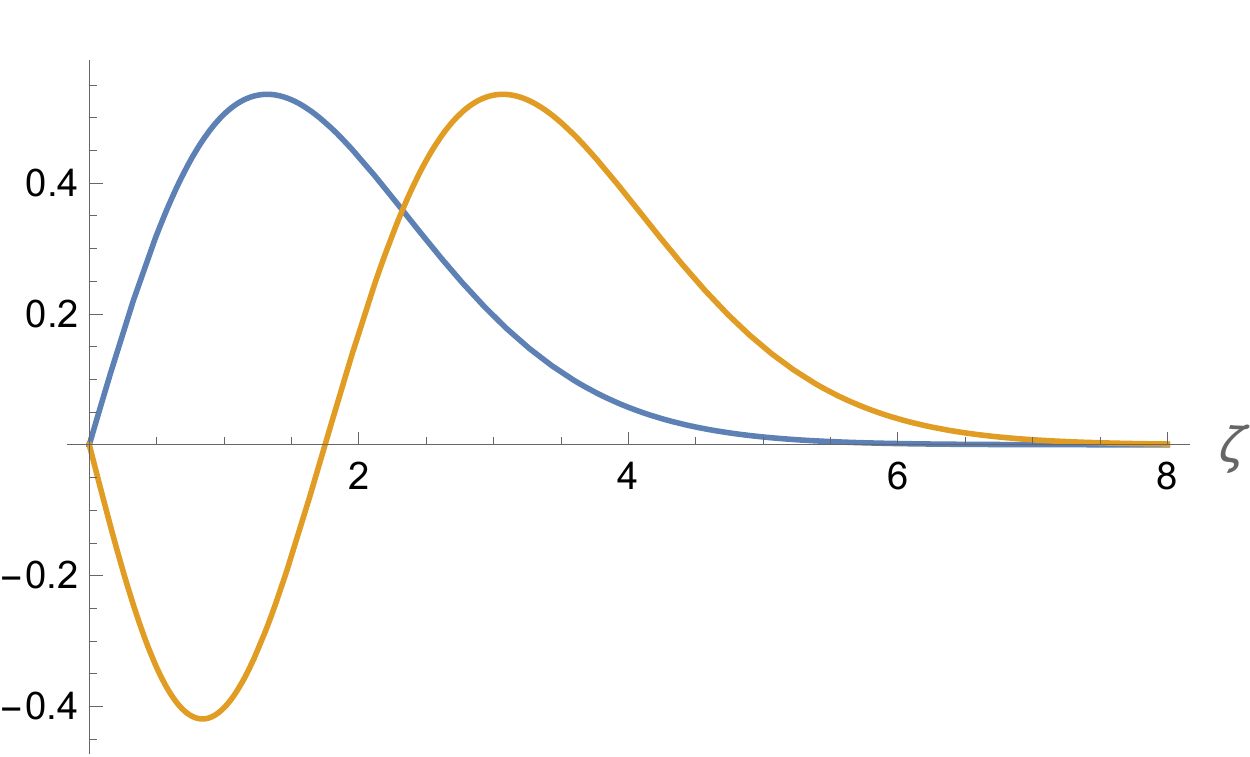}
\caption{ Wave functions $\psi_1(\zeta)=\mathrm{Ai}(\zeta+a_1)\Theta(\zeta)$ (blue) and $\psi_2(\zeta)=\mathrm{Ai}(\zeta+a_2)\Theta(\zeta)$ (orange)}\label{Fig_2_6}
\end{subfigure}%
\begin{subfigure}{.5\textwidth}
  \centering
  \includegraphics[width=80mm]{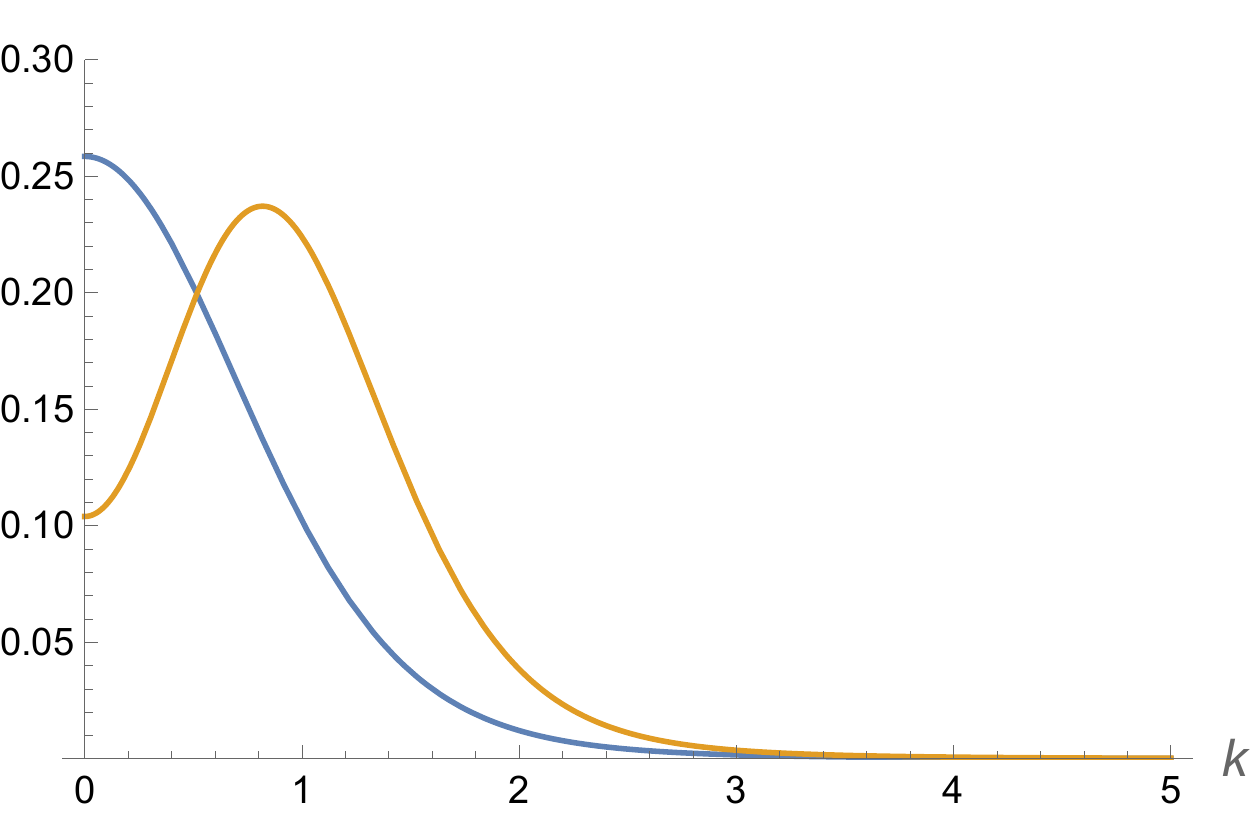}
\caption{Momentum distributions $|F(k,a_1)|^2$ (blue) and $|F(k,a_2)|^2$ (orange)} \label{Figs_5_7}
\end{subfigure}
\caption{Wave functions and momentum distributions of the ground state with $a_1\approx-2.3381$ and the first excited state with $a_2\approx-4.08795$}

\end{figure}

Concerning the calculation of the spectra and the Wigner function, the following formalism is developed using the example of the ground state. The wave function of the ground state $\psi_1(\zeta)$ can be written as $\psi_1(\zeta)=\mathrm{Ai}(\zeta+a_1)\Theta(\zeta)$, where $\Theta(\zeta)$ is Heaviside's step function, see Fig.~\ref{Fig_2_6}. This Heaviside step function is necessary in order to fulfill the boundary condition caused by the mirror whereupon the wave function has to be zero for negative $\zeta-$values.
 
The spatial distribution is given by $|\psi_{1}(\zeta)|^2$. This function can be taken from Fig.~\ref{Fig1}, orange curve, by imagining that the curve is shifted by $a_1$ to the positive $\zeta-$axis.

%------------------------------------------------------------------------------
\subsection{Calculation of Fourier transform of ground state}
\begin{figure}[h!]
\centering
\includegraphics[width=80mm]{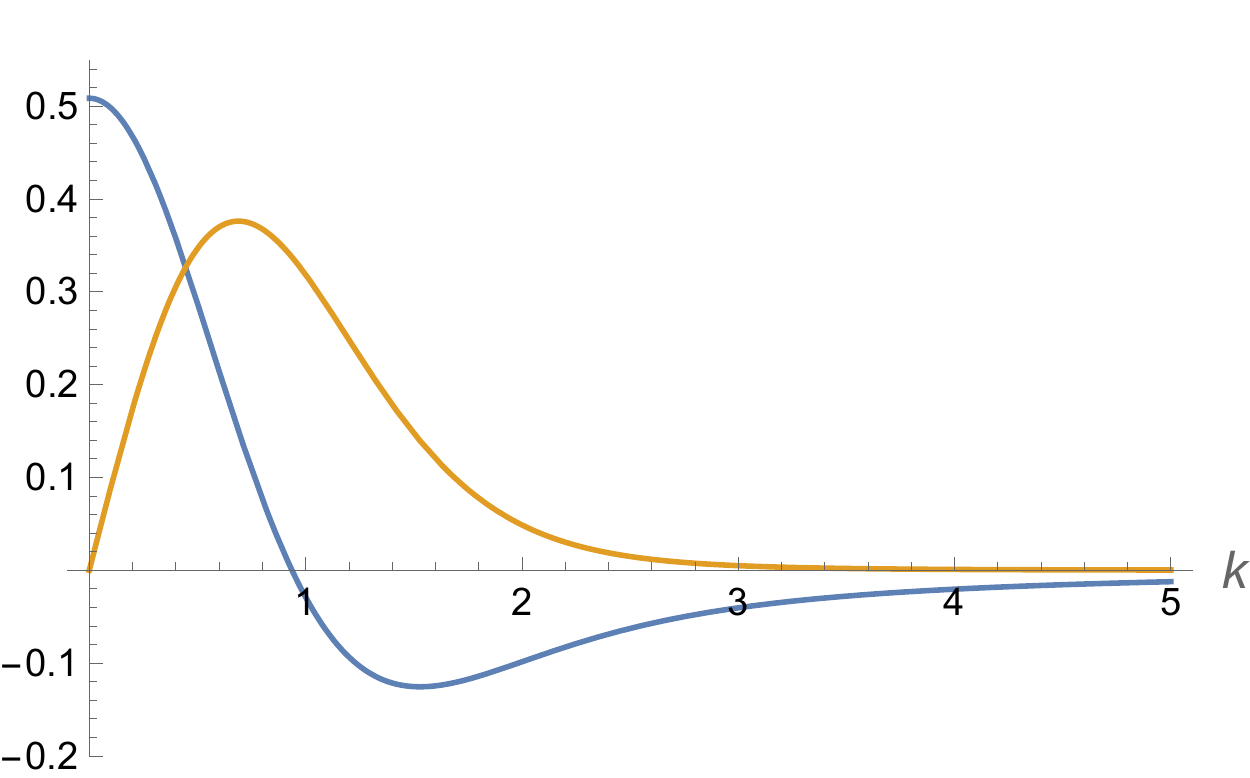}
\caption{$f_{c}(k,a_1)$ with $a_1\approx-2.3381$ (blue) and $f_{s}(k,a_1)$ with $a_1\approx-2.3381$ (orange)} \label{Figs_3_4}
\end{figure}

%------------------------------------------------------------------------------
In order to attain the momentum space (variable $k$), the wave function $\psi_1(\zeta)$ has to be Fourier transformed:
\begin{eqnarray}\label{Four}
F(k,a_1)&=&\frac{1}{\sqrt{2\pi}}\int_{-\infty}^{\infty} \mathrm{e}^{-i\zeta{}k}\mathrm{Ai}(\zeta+a_1)\Theta(\zeta)\,\mathrm{d}\zeta\nonumber\\
&=&\frac{1}{\sqrt{2\pi}}\int_{0}^{\infty}[\cos(\zeta{}k)-i\sin(\zeta{}k)]\mathrm{Ai}(\zeta+a_1)\,\mathrm{d}\zeta
\nonumber\\
&=:&
f_{c}(k,a_1)-if_{s}(k,a_1)\,\,\,,
\end{eqnarray}
where the two functions
\begin{eqnarray}\label{fcfs}
f_{c}(k,a_1)&:=&\frac{1}{\sqrt{2\pi}}\int_{0}^{\infty}\cos(\zeta{}k)\mathrm{Ai}(\zeta+a_1)\,\mathrm{d}\zeta\,\,\,,\nonumber\\
f_{s}(k,a_1)&:=&\frac{1}{\sqrt{2\pi}}\int_{0}^{\infty}\sin(\zeta{}k)\mathrm{Ai}(\zeta+a_1)\,\mathrm{d}\zeta
\end{eqnarray}
have been defined. They are displayed in Fig.~\ref{Figs_3_4}.  
 
The momentum spectrum is given by
\begin{eqnarray}\label{Imp}
|F(k,a_1)|^2=[f_{c}(k,a_1)]^2+[f_{s}(k,a_1)]^2\,\,\,,
\end{eqnarray}
see Fig.~\ref{Figs_5_7}. We have to stress that $k$ is actually a dimensionless variable and related to the physical momentum $k_p$ by 
\begin{eqnarray}
k &=& \frac{z_0}{\hbar}k_p\,\,\,.
\end{eqnarray}
 
%In order to rescale the $k-$axis of Fig.~\ref{Fig5}, two experimental parameters have to be included. As can be taken, e.g., from~\cite{Jenke:2014a},~\cite{Abele:2010a} or~\cite{Jenke:2011b}, the most probable horizontal velocity $v_0$ of the UCN neutrons is approximately \SI[per-mode=symbol]{6}{\meter\per\second}. This corresponds to a mean momentum $k_0=\SI{100}{\per\micro\meter}$. The maximum of $|F(k,a_1)|^2$ of Fig.~\ref{Fig5} appears at $k=0$, which therefore corresponds to $k_0$. The second parameter is $\delta{v}/v_0=\delta{k}/k_0\approx{1/10}$. This means, e.g., for $k=1$ we approximately obtain $k_0+\delta{k}/2\approx\SI{105}{\per\micro\meter}$. Thus, we have scheduled the $k-$scale. The $\zeta-$scale has already been fixed in Eq.~(\ref{Einsetzung}).
 
%The momentum spectrum  Fig.~\ref{Fig5} ought to be identified experimentally. 

%------------------------------------------------------------------------------
\subsection{Excited states}
%------------------------------------------------------------------------------
Here we look at the excited states by discussing their momentum spectra for a few selected example values of $n$. The first excited state is characterized by the second zero point $a_2\approx-4.08795$ of the  Airy function. Its wave function and momentum spectrum are depicted in Figs.~\ref{Fig_2_6} and ~\ref{Figs_5_7}, respectively.

\begin{figure}[h!]
\centering
\includegraphics[width=80mm]{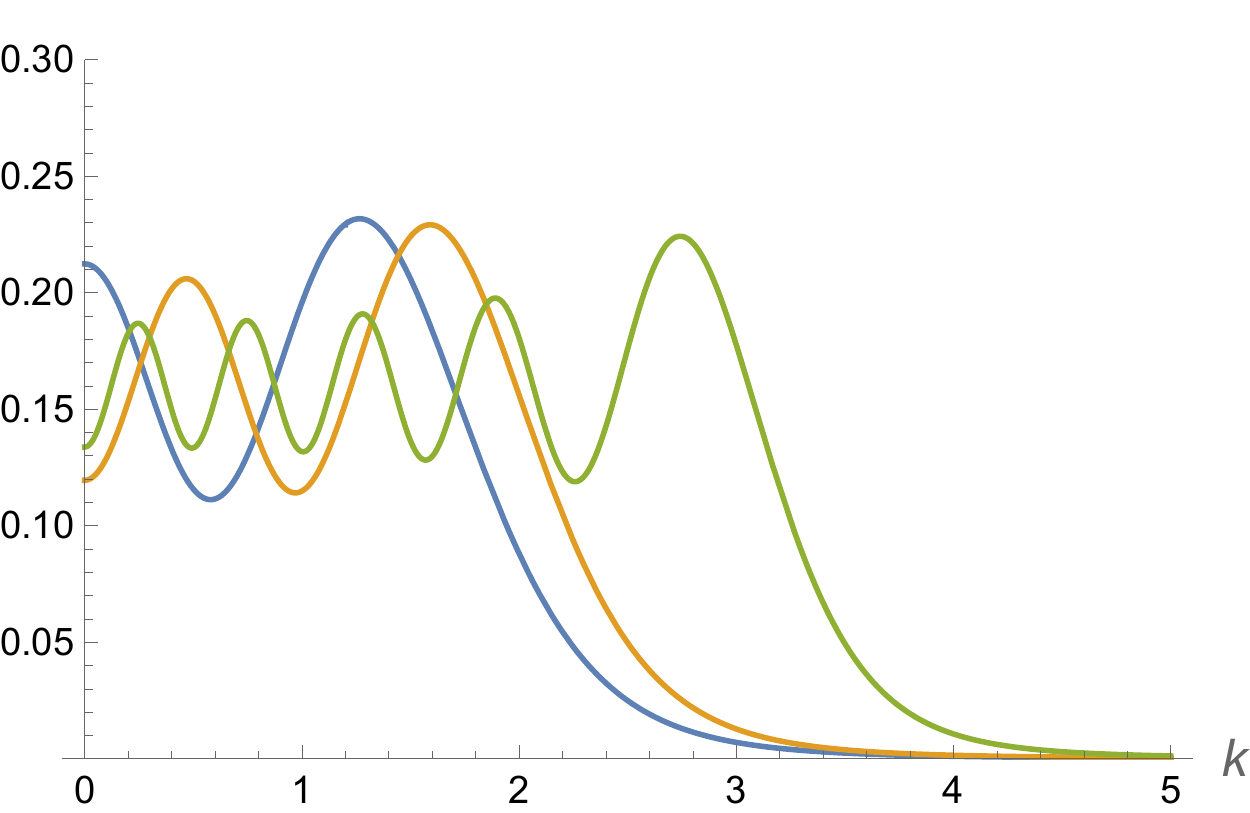}
\caption{Momentum distributions of the second excited state $|F(k,a_3)|^2$ with $a_3\approx-5.52056$ (blue), the third excited state $|F(k,a_4)|^2$ with $a_4\approx-6.78671$ (orange), and the ninth excited state $|F(k,a_{10})|^2$ with $a_{10}\approx-12.8288$ (green)} \label{Figs_8_9_10}
\end{figure}

The third zero point of the Airy function is located at $a_3\approx-5.52056$ ($2^{nd}$ excited state) and yields a momentum spectrum as given in Fig.~\ref{Figs_8_9_10}.
Besides, this figure also shows the results for the $3^{rd}$ and $9^{th}$ excited states with the corresponding fourth and tenth zero points $a_4\approx-6.78671$ and $a_{10}\approx-12.8288$ of the Airy function.  

In Fig.~\ref{Figs_8_9_10} we can see that the number of oscillations before the onset of the asymptotic behavior of the momentum spectra increases with $n$. In case of $n$ towards infinity, the amplitudes of the oscillations tend to zero and the momentum spectrum becomes a constant. %This has to be anticipated because the whole Airy function comes into operation.

%------------------------------------------------------------------------------
\subsection{Presentation using Wigner function}
%------------------------------------------------------------------------------
The 2-dimensional Wigner function is an important tool in quantum optics. It allows for a simultaneous view into space and momentum regions. The Wigner function is real but can be positive and negative as well. In this respect, it is not a classical 2-dim distribution function. Therefore, it is often called a quasi-distribution function. Most remarkably is the property that an integration of a Wigner function over momentum gives the spatial probability, while integration over the spatial coordinate gives the momentum probability. These two marginal distribution functions (spatial and momentum distributions) are, at least in principle, experimentally accessible. The Wigner function formalism has already been applied within the framework of investigations of the  gravitational potential of the Earth~\cite{Kajari:2010}. In addition, we would like to point to an article, in which the interface of gravity and quantum mechanics has been discussed with the Wigner phase space distribution function~\cite{G15}.
 
For our purposes, we will use the Wigner function in order to recover the momentum spectrum Eq.~(\ref{Imp}). The definition of the Wigner function is~\cite{S01}:
\begin{eqnarray}\label{WF}
W(\zeta,k):=\frac{1}{2\pi}\int_{-\infty}^{\infty}\mathrm{e}^{i\zeta'k}\,{\psi^*}{\left(\zeta+\frac{\zeta'}{2}\right)}\,{\psi}{\left(\zeta-\frac{\zeta'}{2}\right)}\,\mathrm{d}\zeta'\,\,\,.
\end{eqnarray}
Plugging Eq.~(\ref{EF}) into this definition, we find
\begin{eqnarray}\label{WFG}
W(\zeta,k,a_n)=\frac{1}{2\pi}\int_{-\infty}^{\infty}\mathrm{e}^{i\zeta'k}\,{\mathrm{Ai}}{\left(\zeta+\frac{\zeta'}{2}+a_n\right)}\,{\Theta}{\left(\zeta+\frac{\zeta'}{2}\right)}\,{\mathrm{Ai}}{\left(\zeta-\frac{\zeta'}{2}+a_n\right)}\,{\Theta}{\left(\zeta-\frac{\zeta'}{2}\right)}\,\mathrm{d}\zeta'\,\,\,,
\end{eqnarray}
which is obviously symmetric in $k$: $W(\zeta, k,a_n)=W(\zeta,-k,a_n)$.

Due to $\frac{1}{2\pi}\int_{-\infty}^{\infty}\mathrm{e}^{i\zeta' k}\,\mathrm{d}k=\delta(\zeta')$, we can easily see how integration over the momentum $k$ yields the space distribution:
\begin{eqnarray}\label{Ort}
\int_{-\infty}^{\infty}W(\zeta, k,a_n)\,\mathrm{d}k=[\mathrm{Ai}(\zeta+a_n)\,\Theta(\zeta)]^2=|\psi_n(\zeta)|^2\,\,\,.
\end{eqnarray}
%In principle, space distributions can be measured using track detectors~\cite{Jenke:2014a}.
The Wigner function in Eq.~(\ref{WFG}) can be rewritten as: \begin{eqnarray}\label{WFum}
W(\zeta, k,a_n)=\frac{1}{\pi}\int_{0}^{\infty}\cos(\zeta' k){\mathrm{Ai}}{\left(\zeta+\frac{\zeta'}{2}+a_n\right)}\,{\Theta}{\left(\zeta+\frac{\zeta'}{2}\right)}{\mathrm{Ai}}{\left(\zeta-\frac{\zeta'}{2}+a_n\right)}\,{\Theta}{\left(\zeta-\frac{\zeta'}{2}\right)}\,\mathrm{d}\zeta'\,\,\,.
\end{eqnarray}

Using Eq.~(\ref{WFum}), we can find the momentum distribution by integrating over $\zeta$:
\begin{eqnarray}\label{ImpWF}
|F( k,a_n)|^2&=&\int_{-\infty}^{\infty}W(\zeta, k,a_n)\,\mathrm{d}\zeta=\frac{1}{\pi}\int_{0}^{\infty}\cos(\zeta' k)f(\zeta',a_n)\,\mathrm{d}\zeta'\,\,\,,\nonumber\\
f(\zeta',a_n)&:=&\int_{\frac{\zeta'}{2}}^{\infty}{\mathrm{Ai}}{\left(\zeta+\frac{\zeta'}{2}+a_n\right)}{\mathrm{Ai}}{\left(\zeta-\frac{\zeta'}{2}+a_n\right)}\,\mathrm{d}\zeta\,\,\,,\,\,\zeta'\ge{0}\,\,\,.
\end{eqnarray}
Evaluating this expression numerically for the states considered in Figs.~\ref{Figs_5_7}, we obtain the same results as in these figures. Consequently, the Wigner function provides us with a second option to calculate $|F( k,a_n)|^2$. 
 
\begin{figure}[h!]
\begin{subfigure}{.49\linewidth}
\centering
\includegraphics[width=60mm]{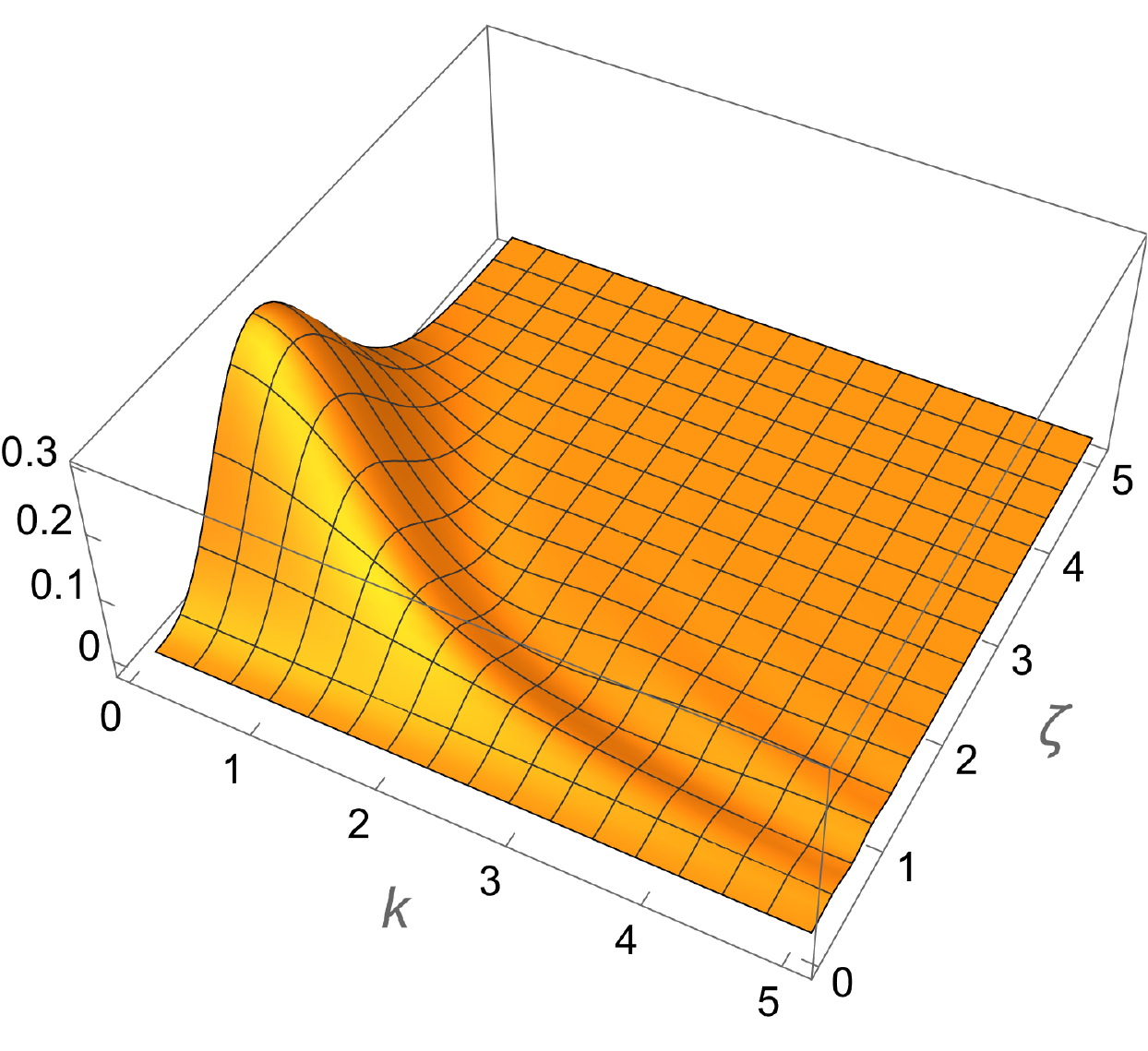}
\caption{Wigner function $W(\zeta, k,a_1)$ with $a_{1}\approx-2.3381$} \label{Fig11}
\end{subfigure}%

\begin{subfigure}{.49\linewidth}
\includegraphics[width=60mm]{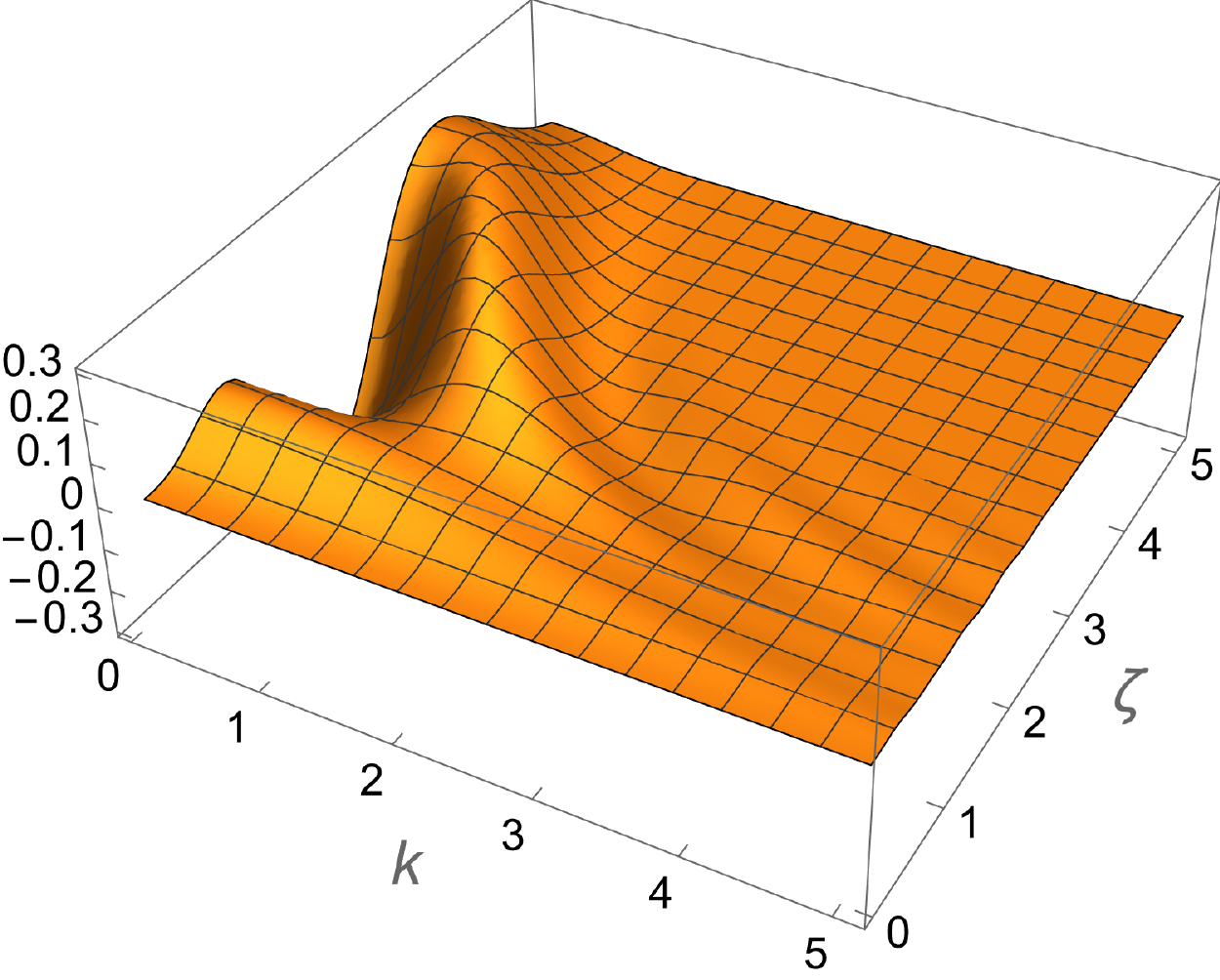}
\caption{Wigner function $W(\zeta, k,a_2)$ with $a_{2}\approx-4.08795$} \label{Fig12}
\end{subfigure}
\begin{subfigure}{.49\linewidth}
\includegraphics[width=60mm]{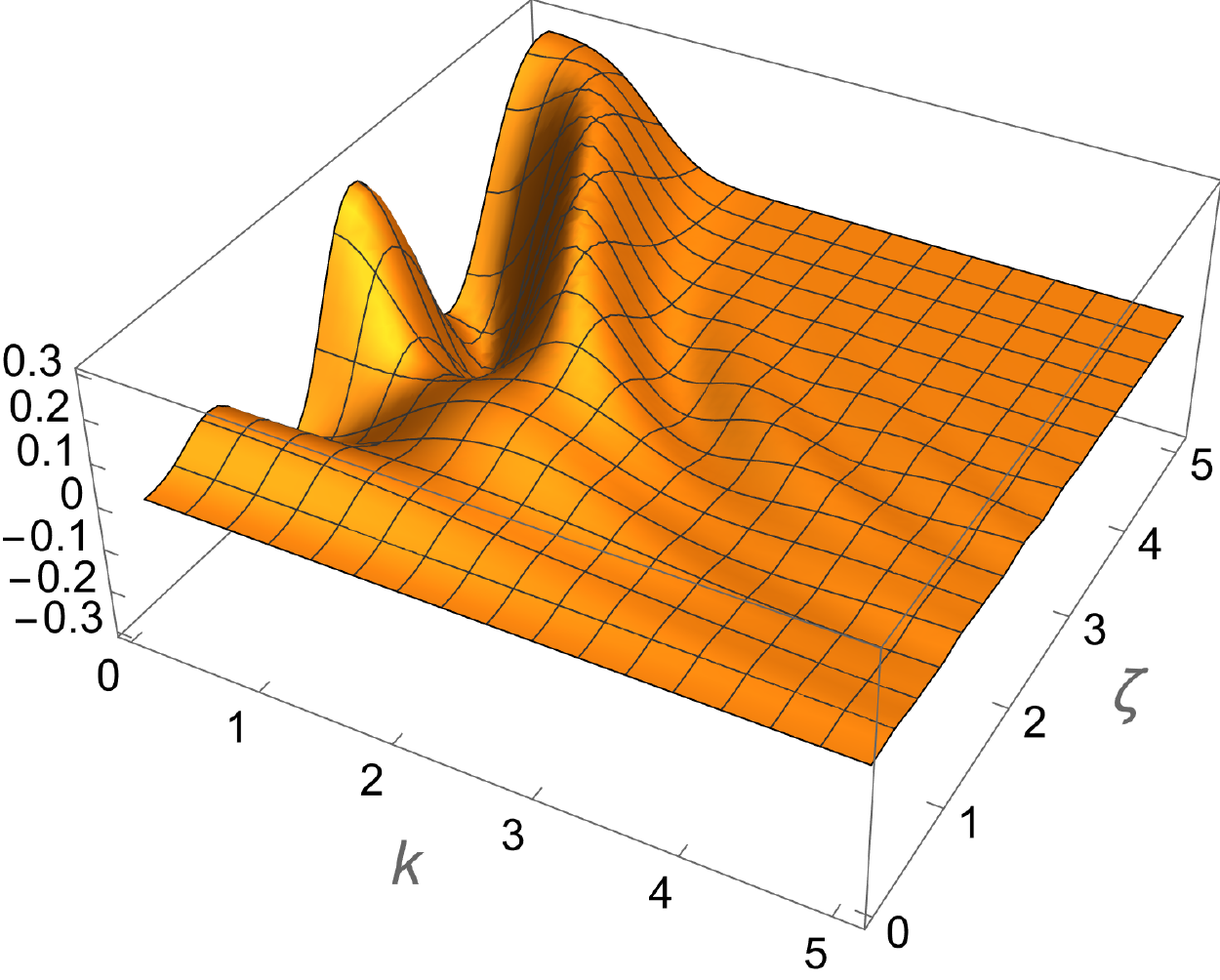}
\caption{Wigner function $W(\zeta, k,a_3)$ with $a_{3}\approx-5.52056$} \label{Fig13}
\end{subfigure}
\caption{Wigner functions of the ground state, and first and second excited state}
\end{figure}

The 2-dim Wigner function Eq.~(\ref{WFum}) of the ground state $W(\zeta, k,a_1)$ is plotted in Fig.~\ref{Fig11}. It is almost everywhere positive. There are only very small and hardly visible negative regions. 

The negative regions are much more visible for the Wigner functions of the first and second excited states, which are plotted in Figs.~\ref{Fig12} and \ref{Fig13}, respectively. This negativity distinguishes the Wigner distribution from the always strictly positive spatial and momentum distributions.
%-------------------------------------------------------------------------------
\subsection{Time-dependence of a mixture of ground state and first excited state}
\begin{figure}[h!]
\centering
\includegraphics[width=80mm]{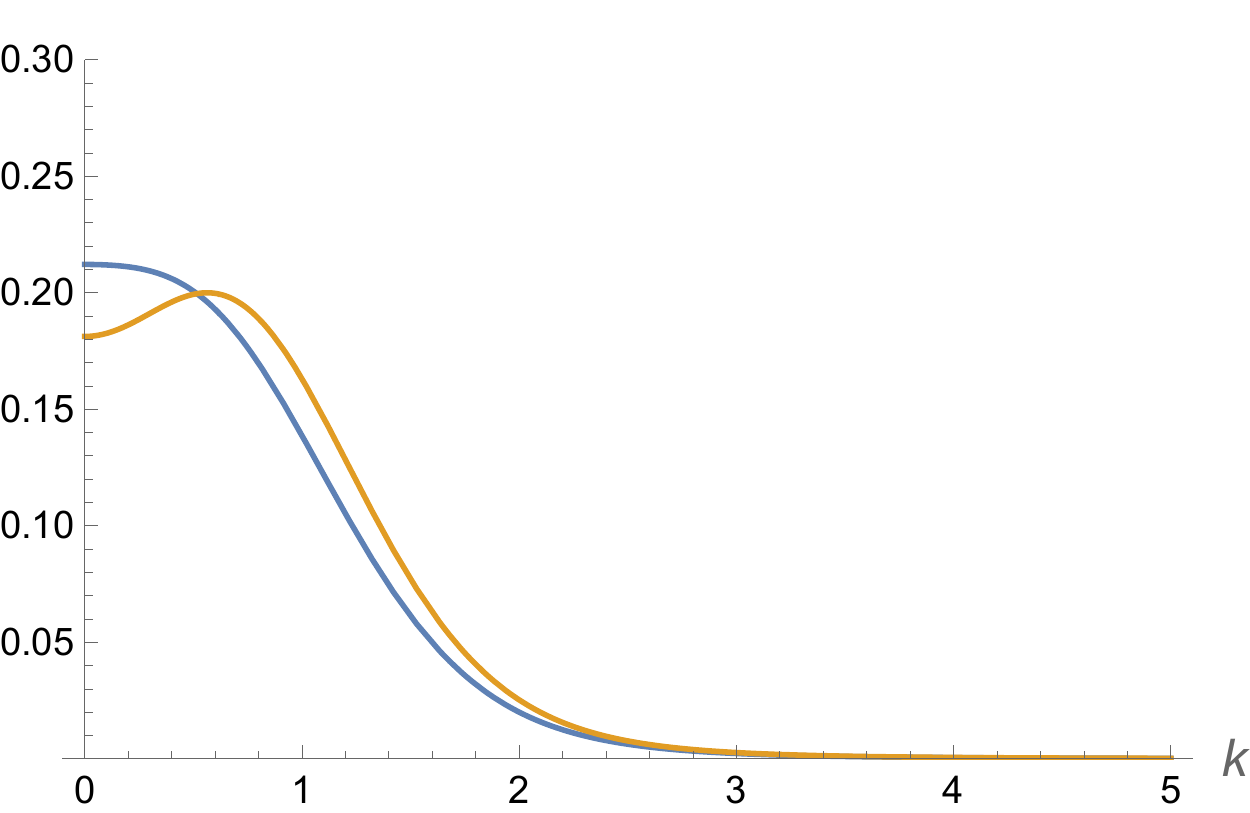}
\caption{Combined momentum distributions of the ground state and the first excited state $p_1\,|F( k,a_1)|^2+p_2\,|F( k,a_2)|^2$ with $p_1=0.7$ and $p_2=0.3$ (blue), and $p_1=p_2=0.5$ (orange)} \label{Figs_14_15}
\end{figure}

%-------------------------------------------------------------------------------
The spatial probability density has been experimentally verified for ultracold neutrons in~\cite{Jenke:2014a}. Here we suggest that the momentum probability distributions could be measured in a similar fashion. For example, if the ground state population amounts to $70\%$ ($p_1=0.7$), the first excited state amounts to $30\%$ ($p_2=0.3$), and no other excited states are populated, then extracting the total probability distributions from Fig.~\ref{Figs_5_7} is straightforward. This is because the relative contributions can be extracted from the figures: $\sum_{n}p_n|F( k,a_n)|^2$, see Fig.~\ref{Figs_14_15}. The orange line in Fig.~\ref{Figs_14_15} presents an example for $p_1=p_2=0.5$. In this case, the first excited state, represented by $|F( k,a_2)|^2$ and Fig.~\ref{Figs_5_7}, is clearly visible. It should be mentioned that this procedure corresponds to an incoherent superposition.
 
\begin{figure}[h!]
\begin{subfigure}{.49\linewidth}
\centering
\includegraphics[width=60mm]{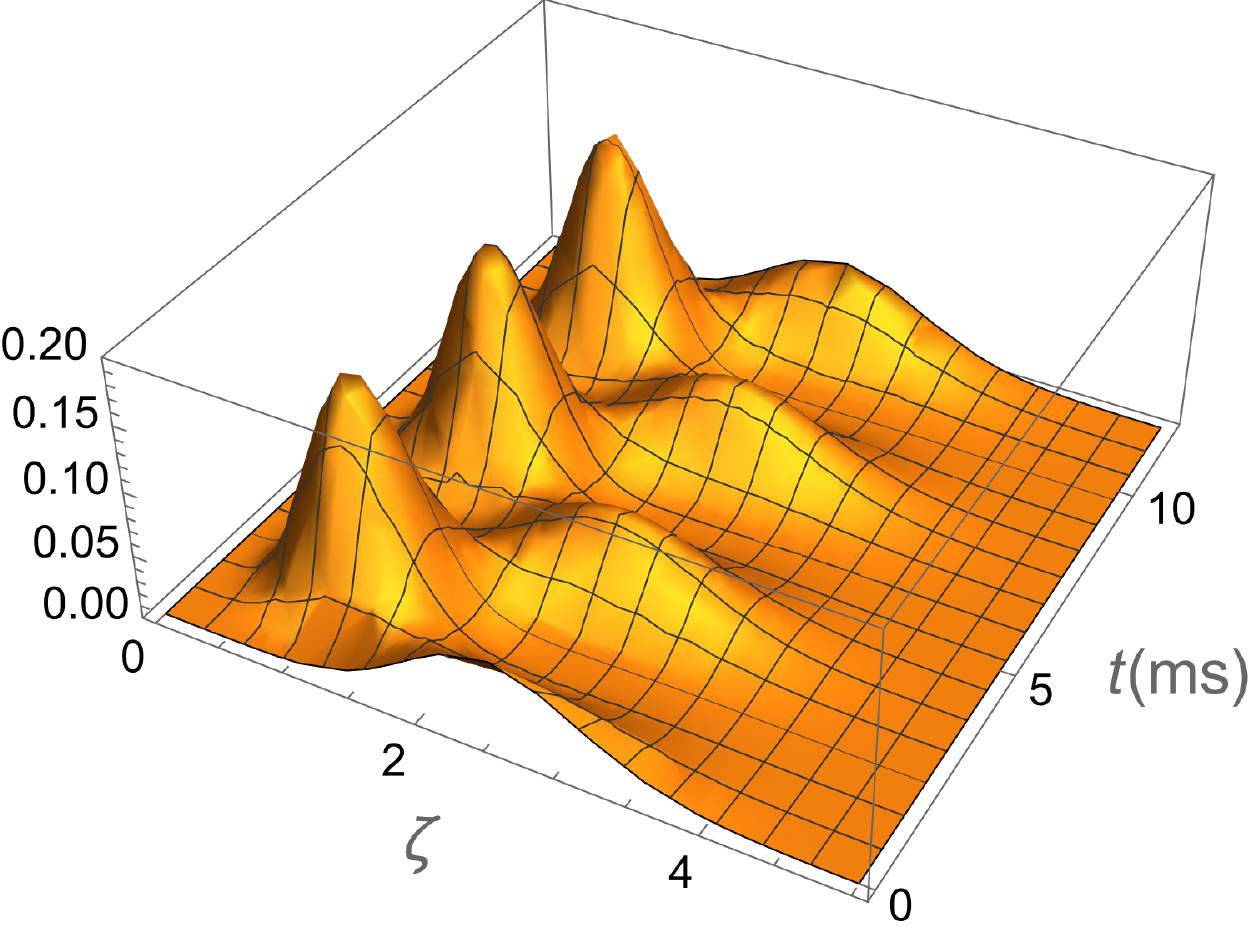}
\caption{$|\psi_s(\zeta,t)|^2$, see Eq.~(\ref{Possup})}
\label{Figosz12}
\end{subfigure}%
\begin{subfigure}{.49\linewidth}
\centering
\includegraphics[width=60mm]{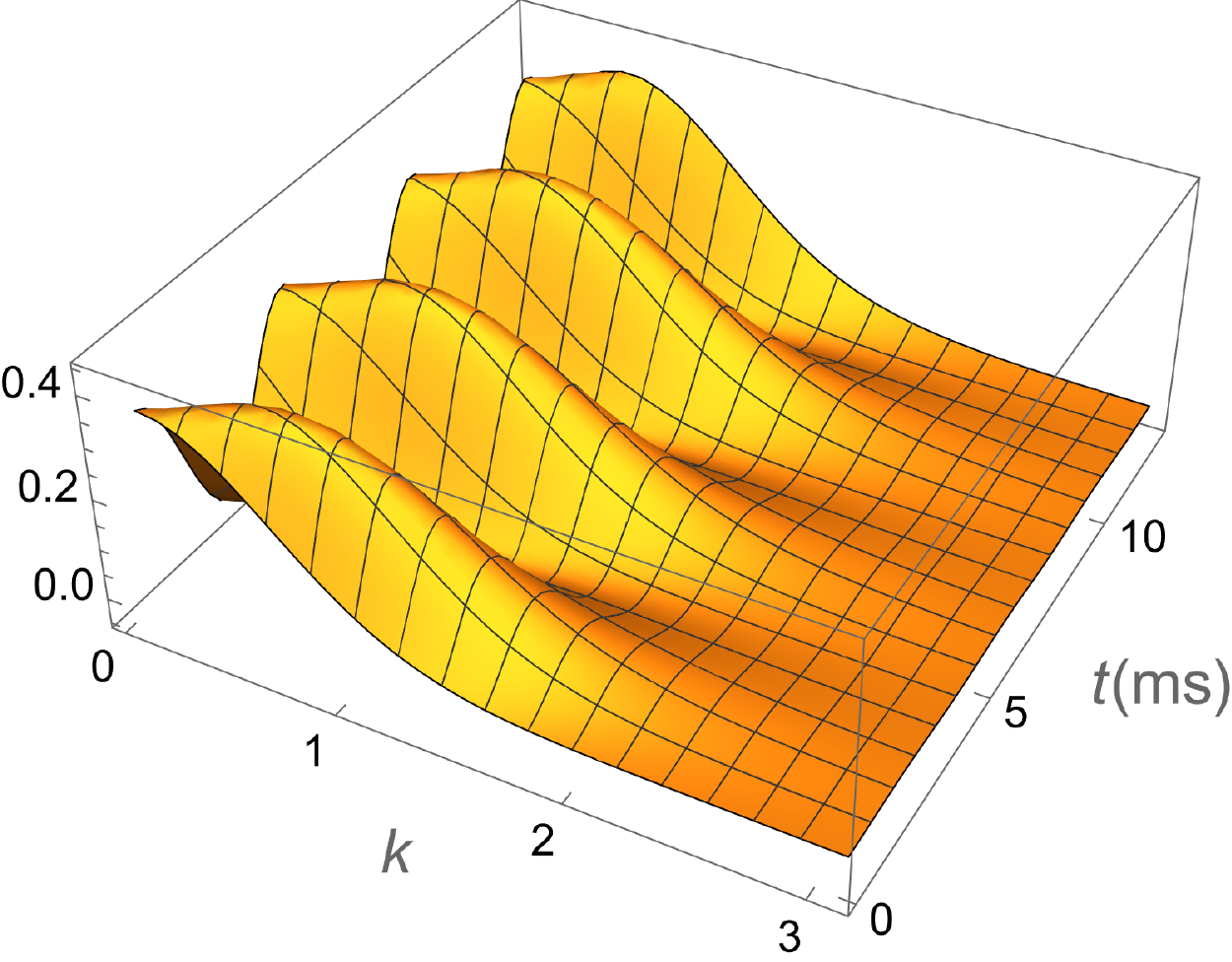}
\caption{$|F_s( k,t)|^2$, see Eq.~(\ref{Fs21})}
\label{Figoszimp73}
\end{subfigure}
\caption{Time-dependent position and momentum probabilities of superposition of the ground state and first excited state with $p_1=0.7$ and $p_2=0.3$}
\end{figure}

Therefore, the proposed procedure is not exact. We have to take the time dependence of the wave function, see Eq.~(\ref{Ans}), into account. Consequently, we will now go back to the time-dependent space distribution $|\psi(\zeta,t)|^2$ using the following ansatz of coherent superposition
\begin{eqnarray}\label{Sup}
\psi_s(\zeta,t)=\sqrt{p_1}\mathrm{e}^{-i\frac{E_1}{\hbar}t}\psi_1(\zeta)+\sqrt{p_2}\mathrm{e}^{-i\frac{E_2}{\hbar}t}\psi_2(\zeta)\,\,\,,
\end{eqnarray}
where $\psi_1(\zeta)=\mathrm{Ai}(\zeta+a_1)\,\Theta(\zeta)$ and $\psi_2(\zeta)=\mathrm{Ai}(\zeta+a_2)\,\Theta(\zeta)$ are real functions, and we ignore a potential phase between both terms for simplicity. The time-dependent position probability of this superposition state is therefore
\begin{eqnarray}\label{Possup}
|\psi_s(\zeta,t)|^2&=&p_1[\mathrm{Ai}(\zeta+a_1)\,\Theta(\zeta)]^2+p_2[\mathrm{Ai}(\zeta+a_2)\,\Theta(\zeta)]^2+\nonumber\\
&+&2\sqrt{p_1p_2}\,\mathrm{Ai}(\zeta+a_1)\,\mathrm{Ai}(\zeta+a_2)\,\Theta^2(\zeta)\,{\cos}{\left(\frac{E_1-E_2}{\hbar}t\right)}\,\,\,.
\end{eqnarray}
This function oscillates with time $t$. If $p_1=1$ and $p_2=0$, we recover $|\psi_1(\zeta)|^2$, which is the square of the function depicted in blue in Fig.~\ref{Fig_2_6}.
Furthermore, we have
\begin{eqnarray}\label{E1E2}
\frac{E_1-E_2}{\hbar}=\frac{E_0}{\hbar}(-a_1+a_2)\approx{\SI{1600.4}{\hertz}}\,\,\,.
\end{eqnarray}
This means, that $|\psi_s(\zeta,t)|^2$ oscillates in the time-range of milliseconds, as can be seen in the example shown in Fig.~\ref{Figosz12}. The first small peak at $t=0$ appears at $\zeta\approx2.5$ and not at $\zeta\approx1.3$, i.e., the maximum of $\psi_1(\zeta)$, see Fig.~\ref{Fig_2_6}, because the interference term in Eq.~(\ref{Possup}) contains $\mathrm{Ai}(\zeta+a_2)$,
which is negative in the region between $\zeta=0$ and $\zeta\approx{1.7}$, see Fig.~\ref{Fig_2_6}. However, the large maximum at $t\approx{\SI{0.002}{\second}}$ appears for $\zeta\approx{1.3}$ due to $\psi_1(\zeta)$ and $p_1=0.7$. We assume that such properties of time-dependent position probabilities for superpositions of ground and excited states can be measured.

\begin{figure}[h!]
\centering
\includegraphics[width=80mm]{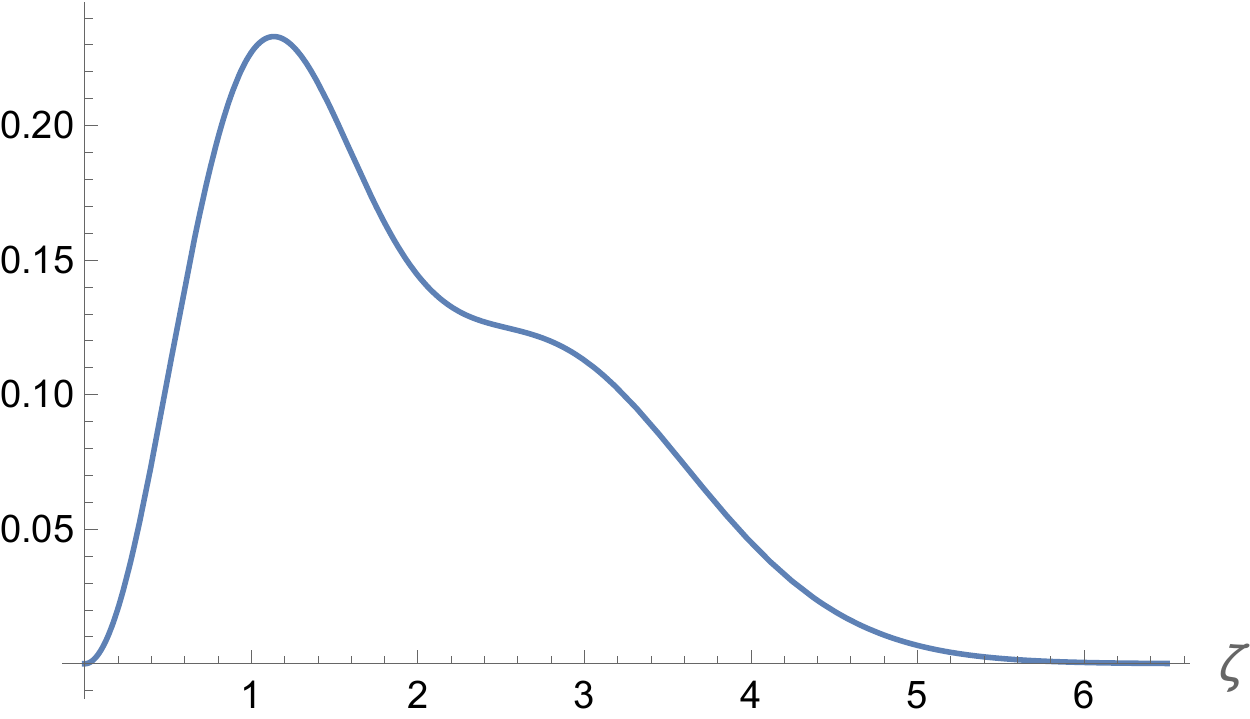}
\caption{Non-time-resolving position probability $|\psi|^2=p_1[\mathrm{Ai}(\zeta+a_1)\,\Theta(\zeta)]^2+p_2[\mathrm{Ai}(\zeta+a_2)\,\Theta(\zeta)]^2$ of  superposition of ground state and first excited state, see Eq.~(\ref{Possup}), with $p_1=0.7$ and $p_2=0.3$}
\label{pmix73}
\end{figure}

We have to point out that in case of a non-time-resolving measurement we have to integrate time $t$ in  Eq.~(\ref{Possup}) over one time period, e.g., from $0$ to $2\pi\hbar/(E_1-E_2)$. In this case, the interference term disappears and we obtain only the first two terms $|\psi|^2=p_1[\mathrm{Ai}(\zeta+a_1)\,\Theta(\zeta)]^2+p_2[\mathrm{Ai}(\zeta+a_2)\,\Theta(\zeta)]^2$. Fig.~\ref{pmix73} depicts an example. This spatial probability density is very similar to the results of an experiment using a track detector~\cite{Jenke:2014a}. %even though 2 mirrors at a distance of \SI{27}{\micro\meter} were used there.
 
The Fourier transform of $\psi_s(\zeta)$ in Eq.~(\ref{Sup}) reads
\begin{eqnarray}\label{SupFou}
F_s( k,t)=\sqrt{p_1}\mathrm{e}^{-i\frac{E_1}{\hbar}t}F( k,a_1)+\sqrt{p_2}\mathrm{e}^{-i\frac{E_2}{\hbar}t}F( k,a_2)\,\,\,.
\end{eqnarray}
Next, we want to calculate $|F_s( k,t)|^2$. We obtain
\begin{eqnarray}\label{Fs2}
|F_s( k,t)|^2&=&p_1|F( k,a_1)|^2+p_2|F( k,a_2)|^2+\sqrt{p_1p_2}(\alpha\beta^*+\alpha^*\beta)\,\,\,,\nonumber\\
\alpha&=&\mathrm{e}^{i\frac{(E_1-E_2)}{\hbar}t}\,\,\,,\,\,\,\beta^*=F^*( k,a_1)F( k,a_2)\,\,\,.
\end{eqnarray}
Using Eq.~(\ref{Four}) in order to decompose $\beta^*$, we find
\begin{eqnarray}\label{b*}
\beta^*=f_c( k,a_1)f_c( k,a_2)+f_s( k,a_1)f_s( k,a_2)+i\,[f_s( k,a_1)f_c( k,a_2)-f_c( k,a_1)f_s( k,a_2)]\,\,\,.
\end{eqnarray}
Since $\alpha\beta^*+\alpha^*\beta=2[\mathrm{Re}(\alpha)\mathrm{Re}(\beta)+\mathrm{Im}(\alpha)\mathrm{Im}(\beta)]$, the final expression for $|F_s( k,t)|^2$ is
\begin{eqnarray}\label{Fs21}
|F_s( k,t)|^2&=&p_1|F( k,a_1)|^2+p_2|F( k,a_2)|^2
+2\sqrt{p_1p_2}\left\{{\cos}{\left[\frac{E_1-E_2}{\hbar}t\right]}[f_c( k,a_1)f_c( k,a_2)+f_s( k,a_1)f_s( k,a_2)]\right.\phantom{xxxx}\nonumber\\
&\phantom{=}&\phantom{p_1|F( k,a_1)|^2+p_2|F( k,a_2)|^2
+2\sqrt{p_1p_2}}
\left.-{\sin}{\left[\frac{E_1-E_2}{\hbar}t\right]}[f_s( k,a_1)f_c( k,a_2)-f_c( k,a_1)f_s( k,a_2)]\right\}\,\,\,.
\end{eqnarray}
The time dependence of this momentum spectrum could also be measured experimentally.
 
Fig.~\ref{Figoszimp73} gives an example of Eq.~(\ref{Fs21}) with $p_1=0.7$ and $p_2=0.3$. If we chose $p_1=1$ and $p_2=0$, or $p_1=0$ and $p_2=1$, we would instead recover Fig.~\ref{Figs_5_7}. In the case of a non-time-resolving measurement, we have to integrate time $t$ in  Eq.~(\ref{Fs21}) over one time period. Hence, the interference term disappears and we obtain only the first two terms $p_1|F( k,a_1)|^2+p_2|F( k,a_2)|^2$, which, using $p_1=0.7$ and $p_2=0.3$, is the function in blue in Fig.~\ref{Figs_14_15}.
 
%The Wigner function of coherent superposition can be written as
%\begin{eqnarray}\label{WFs}
%W_s(\zeta, k,t)=\frac{1}{2\pi}\int_{-\infty}^{\infty}\mathrm{e}^{\mathrm{i}\zeta' k}{\psi^*_s}{\left(\zeta+\frac{\zeta'}{2},t\right)}\,{\psi_s}{\left(\zeta-\frac{\zeta'}{2},t\right)}\,\mathrm{d}\zeta'\,\,\,.
%\end{eqnarray}
Using Eqs.~(\ref{WFG}) and (\ref{Sup}), the Wigner function of the coherent superposition can be found to be
\begin{eqnarray}\label{WFss}
W_s(\zeta, k,t)&=&p_1W(\zeta, k,a_1)+p_2W(\zeta, k,a_2)\phantom{xx}\nonumber\\
&\phantom{=}&+\sqrt{p_1p_2}\frac{1}{\pi}\int_{-\infty}^{\infty}{\mathrm{Ai}}{\left(\zeta-\frac{\zeta'}{2}+a_2)\right)}\,{\mathrm{Ai}}{\left(\zeta+\frac{\zeta'}{2}+a_1\right)}\,{\Theta}{\left(\zeta-\frac{\zeta'}{2}\right)}\,{\Theta}{\left(\zeta+\frac{\zeta'}{2}\right)}\,{\cos}{\left(\frac{E_1-E_2}{\hbar}t+\zeta' k\right)}\,\mathrm{d}\zeta'\,\,\,.\nonumber\\
\,
\end{eqnarray}
The single Wigner functions $W(\zeta, k,a_1)$ and $W(\zeta, k,a_2)$ have already been depicted in Figs.~\ref{Fig11} and \ref{Fig12}, respectively, while the integral on the right-hand side of Eq.~(\ref{WFss}), the interference term, can be evaluated similarly to Eq.~(\ref{ImpWF}).
 
\begin{figure}[h!]
\centering
\includegraphics[width=60mm]{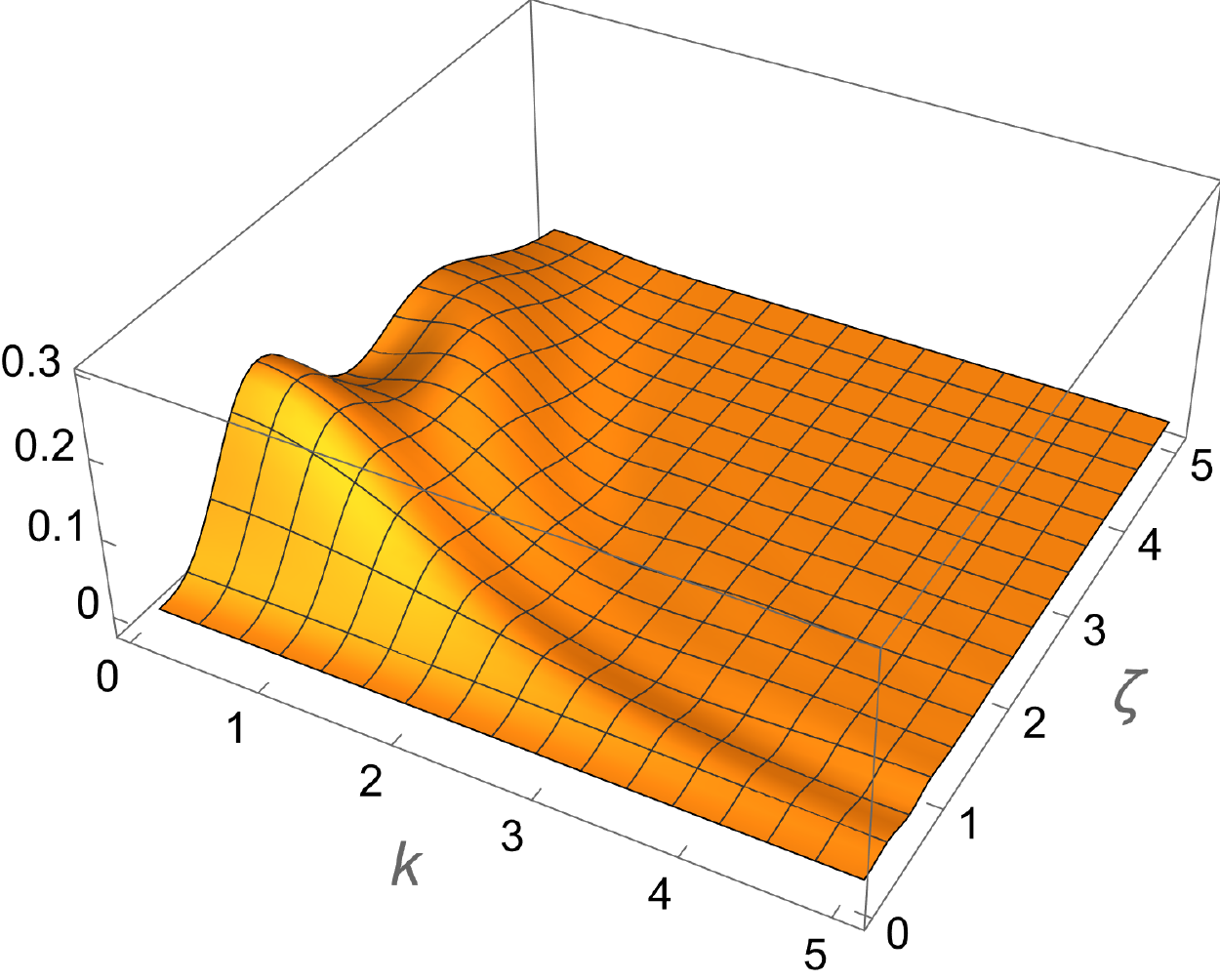}
\caption{Wigner function of superposition of ground state and first excited state for non-time-resolving measurement $p_1W(\zeta, k,a_1)+p_2W(\zeta, k,a_2)$, see Eq.~(\ref{WFss}), with $p_1=0.7$ and $p_2=0.3$ }
\label{Wigner73}
\end{figure}

If the time $t$ cannot be resolved experimentally, we have to integrate over one time period, which causes the interference term to disappear. The resulting Wigner function for an example population is given in Fig.~\ref{Wigner73}.
 
\section{Wave function in a double mirror system (region I)}
\begin{figure}[h!]
\centering
\includegraphics[scale=0.6]{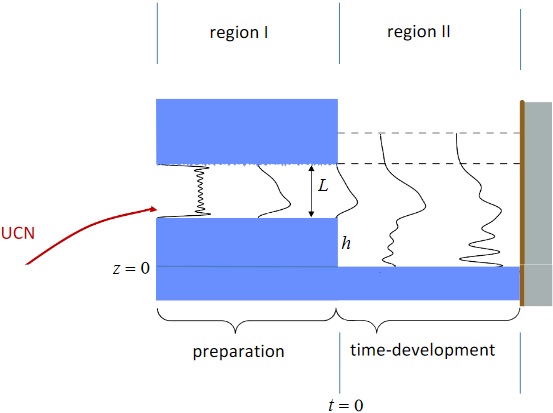}
\caption{Sketch of the two regions I and II, in which the horizontal direction (abscissa) is the time $t$-axis and the vertical direction (ordinate) is the $z-$axis. UCN (ultracold neutrons) enter a double mirror system of distance $L$, called the preparation region (or region I). The lower mirror is smooth, the upper one rough. Due to the roughness of the upper mirror, at the end of region I, only the ground state of the quantum wave is left ($m=1$). In region II, the quantum wave falls down a step of height $h$ onto a smooth mirror located at $z=0$. The transition point from region I to II is where we choose to set $t=0$. At the end of region II, a track detector measures the vertical space distribution $|\psi_{m,\text{II}}(z,t)|^2$.} \label{expaufbau}
\end{figure}

In this chapter we are considering the experimental setting depicted in Fig.~\ref{expaufbau}, and focus on the states in region I.
We can take the normalized wave function of the \textit{q}\textsc{Bounce} problem in the case of two mirrors with fixed separation $L$ from article~\cite{Pitschmann:2019boa} equation (10):
\begin{eqnarray}\label{Wave2m}
\psi_m^{(0)}(z,t)&=&\frac{1}{\sqrt{z_0}}\frac{1}{N_m}\mathrm{e}^{-\frac{i}{\hbar}\bar{E}_mt}\left[\,b_m{\mathrm{Ai}}{\left(\frac{z-\bar{z}_m}{z_0}\right)}-a_m{\mathrm{Bi}}{\left(\frac{z-\bar{z}_m}{z_0}\right)}\right]\,\,\,,\nonumber \\
N_m^2&=&\left[\,b_m{\mathrm{Ai'}}{\left(\frac{-\bar{z}_m}{z_0}\right)}-a_m{\mathrm{Bi'}}{\left(\frac{-\bar{z}_m}{z_0}\right)}\right]^2-\left[\,b_m{\mathrm{Ai'}}{\left(\frac{L-\bar{z}_m}{z_0}\right)}-a_m{\mathrm{Bi'}}{\left(\frac{L-\bar{z}_m}{z_0}\right)}\right]^2\,\,\,,\nonumber\\
a_m&=&{\mathrm{Ai}}{\left(\frac{-\bar{z}_m}{z_0}\right)}\,\,\,,\,\,\,b_m={\mathrm{Bi}}{\left(\frac{-\bar{z}_m}{z_0}\right)}\,\,\,.
\end{eqnarray}
Here the prime denotes derivatives with respect to the argument, i.e., $z_0\,\mathrm{d}/\mathrm{d}z$, and $\mathrm{Ai}(x)$ and $\mathrm{Bi}(x)$ are the two independent solutions of Airy's equation. 
Due to the experimental setting, we assume that the wave function in region I has support only on $[0,L] \times (-\infty,0]$, but we keep this assumption implicit for notational convenience.
Below we present a selection of possible numerical values for the parameters used in Eq.~(\ref{Wave2m}):
\begin{eqnarray}\label{Wave2mpar}
L&=&\SI{28}{\micro\meter}\,\,\,,\nonumber \\
m&=&1\,\,\,,\,\,\,\bar{E}_1\approx\SI{1.40821}{\pico\electronvolt}\,\,\,,\,\,\,\bar{z}_1\approx\SI{13.73133}{\micro\meter}\,\,\,,\,\,\,\nonumber \\
m&=&2\,\,\,,\,\,\,\bar{E}_2\approx\SI{2.53045}{\pico\electronvolt}\,\,\,,\,\,\,\bar{z}_2\approx\SI{24.67419}{\micro\meter}\,\,\,,\,\,\,\nonumber \\
m&=&3\,\,\,,\,\,\,\bar{E}_3\approx\SI{3.84125}{\pico\electronvolt}\,\,\,,\,\,\,\bar{z}_3\approx\SI{37.45569}{\micro\meter}\,\,\,,\,\,\,\nonumber \\
m&=&4\,\,\,,\,\,\,\bar{E}_4\approx\SI{5.64658}{\pico\electronvolt}\,\,\,,\,\,\,\bar{z}_4\approx\SI{55.05930}{\micro\meter}\,\,\,,\,\,\,\nonumber \\
m&=&5\,\,\,,\,\,\,\bar{E}_5\approx\SI{7.98191}{\pico\electronvolt}\,\,\,,\,\,\,\bar{z}_5\approx\SI{77.83089}{\micro\meter}\,\,\,,\,\,\,\nonumber \\
m&=&6\,\,\,,\,\,\,\bar{E}_6\approx\SI{10.8441}{\pico\electronvolt}\,\,\,,\,\,\,\,\bar{z}_6\approx\SI{105.7399}{\micro\meter}\,\,\,.\,\,\,
\end{eqnarray}
Here we have $\bar{E}_m=\bar{z}_m m_{\text{N}}  g$. The energy spectrum $\bar{E}_m$ is obtained by the conditions that the wave functions vanish at the lower as well as upper mirror surface, i.e., $\psi_m^{(0)}(0)=\psi_m^{(0)}(L)=0$.
\subsection{Fourier transformation of the wave function}
\begin{figure}[h!]
\begin{subfigure}{.49\linewidth}
\centering
\includegraphics[width=80mm]{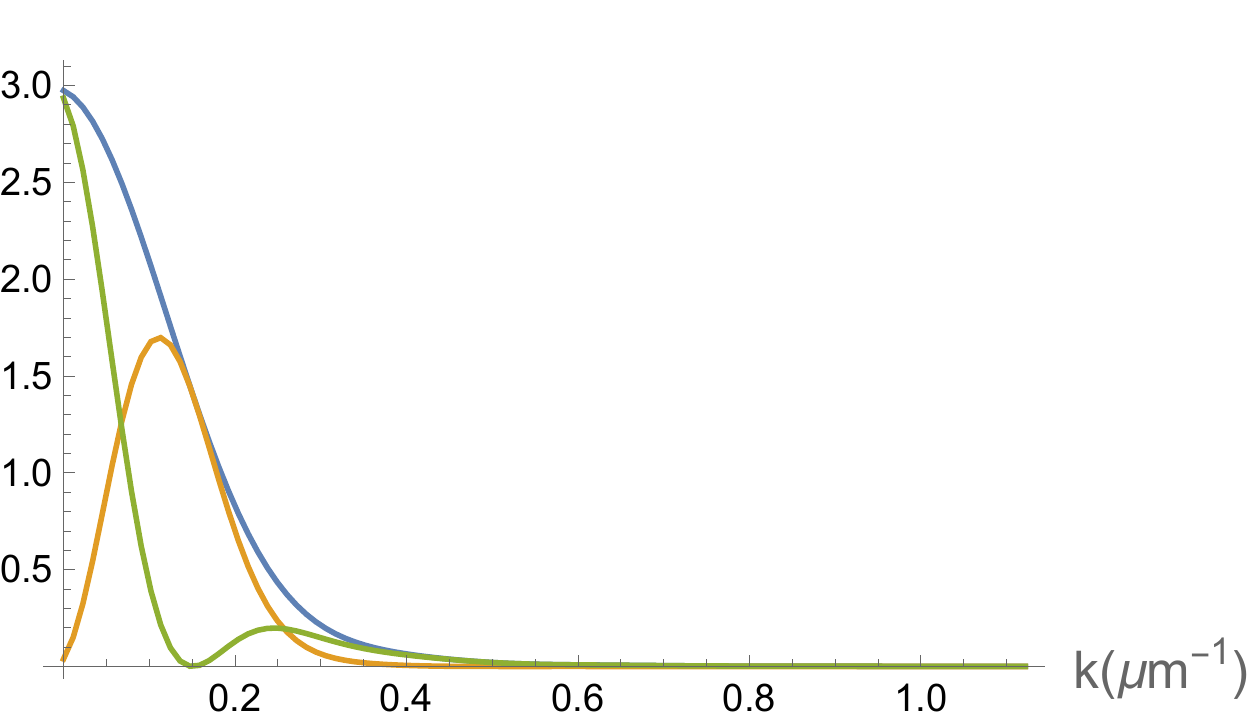}
\caption{$|F_1^{(0)}(k)|^2$ (blue); $|C_1|^2\alpha_s^2(k,1)$ (orange); $|C_1|^2\alpha_c^2(k,1)$ (green) }
\label{Fig20}
\end{subfigure}%

\begin{subfigure}{.49\linewidth}
\includegraphics[width=80mm]{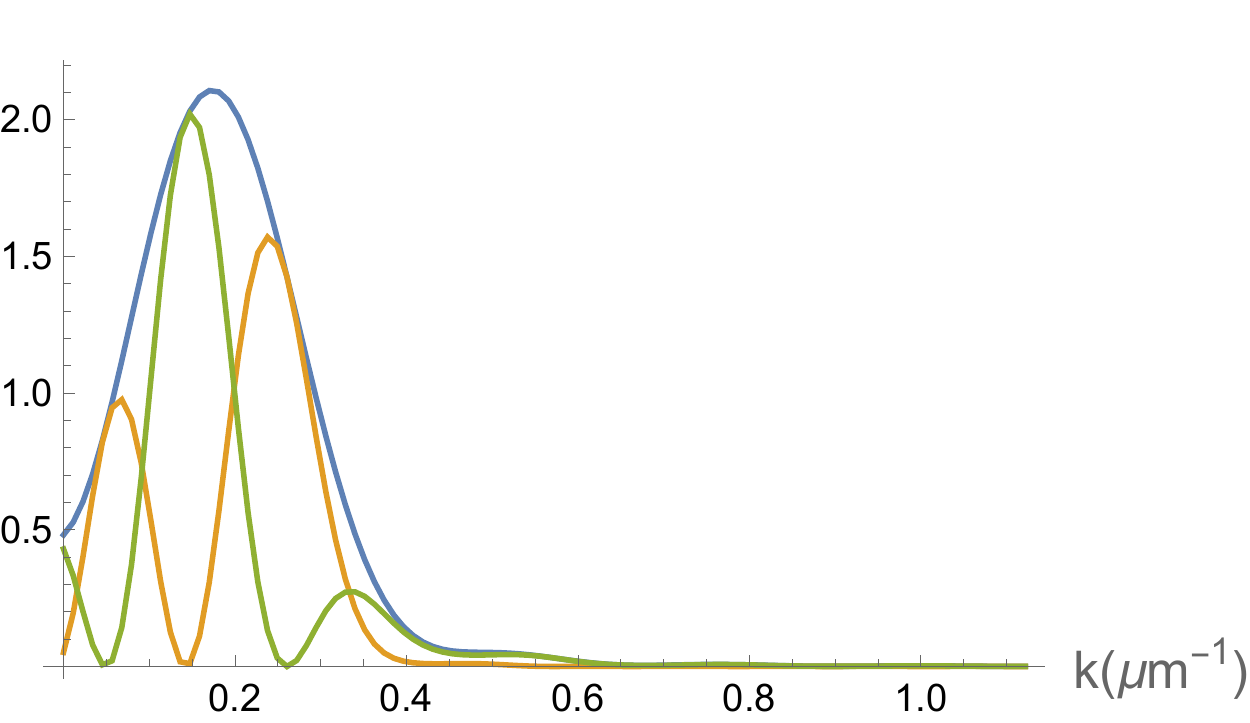}
\caption{$|F_2^{(0)}(k)|^2$ (blue); $|C_2|^2\alpha_s^2(k,2)$ (orange); $|C_2|^2\alpha_c^2(k,2)$ (green)}
\label{Fig21}
\end{subfigure}
\begin{subfigure}{.49\linewidth}
\includegraphics[width=80mm]{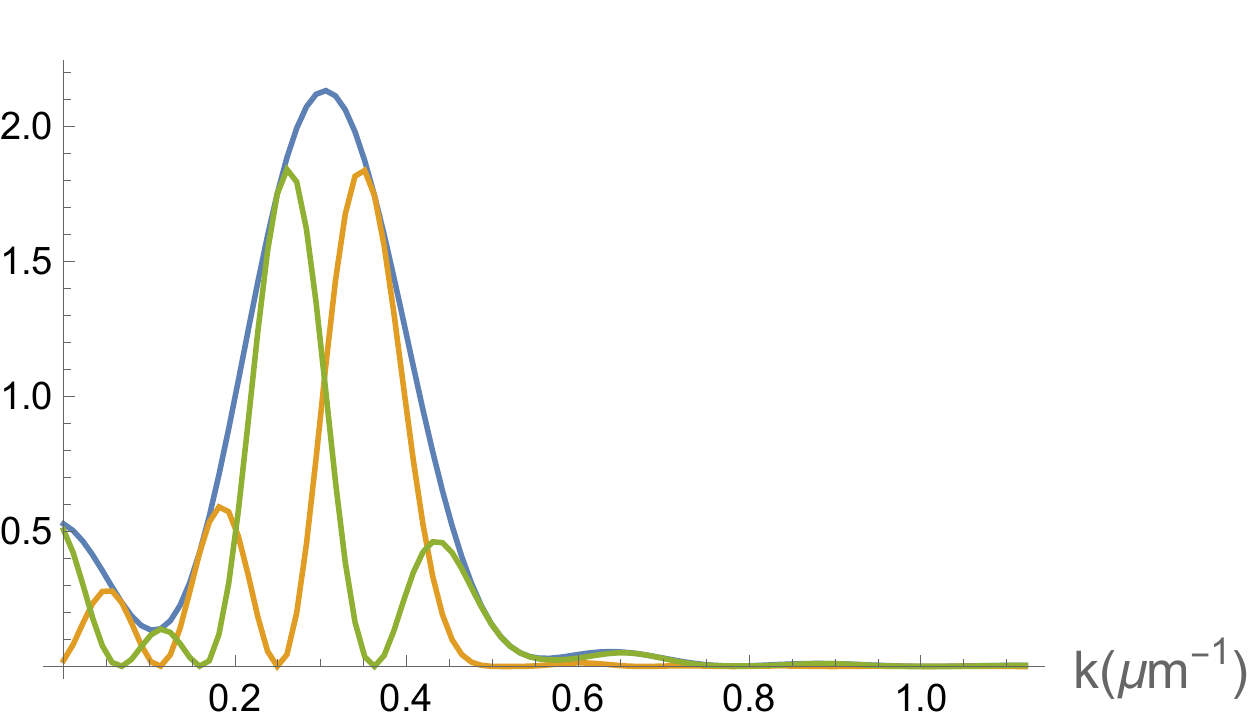}
\caption{$|F_3^{(0)}(k)|^2$ (blue); $|C_3|^2\alpha_s^2(k,3)$ (orange); $|C_3|^2\alpha_c^2(k,3)$ (green)}
\label{Fig22}
\end{subfigure}
\caption{Spectral functions of the ground state, and first and second excited state, see Eq.~(\ref{Spectrum2m})}
\end{figure}

The Fourier transformation of the wave function in Eq.~(\ref{Wave2mpar}) is given by
\begin{eqnarray}\label{FourWave2m}
F_m^{(0)}(k,t)=\frac{1}{\sqrt{2\pi}}\int_{-\infty}^{\infty}\mathrm{e}^{-ikz}\psi_m^{(0)}(z,t)\,\mathrm{d}z
=C_m(t)\int_0^L\mathrm{e}^{-ikz}\left[\,b_m{\mathrm{Ai}}{\left(\frac{z-\bar{z}_m}{z_0}\right)}-a_m{\mathrm{Bi}}{\left(\frac{z-\bar{z}_m}{z_0}\right)}\right]\,\mathrm{d}z\,\,\,,
\end{eqnarray}
where $C_m(t)=\frac{1}{\sqrt{2\pi}\sqrt{z_0}N_m}\mathrm{e}^{-\frac{i}{\hbar}\bar{E}_mt}$ and we made use of the restrictions on the support of the wave function that we mentioned earlier. Since $z$ has the dimension of a length, here the variable $k$  must have the dimension of an inverse length and is related to the physical momentum $k_p$ by 
\begin{eqnarray}
k &=& \frac{k_p}{\hbar}\,\,\,.
\end{eqnarray}
We define the following stationary quantities:
%\begin{eqnarray}\label{fcsAiBi}
%f_c^{\mathrm{Ai}}(k,m)&=&\int_0^L\cos(kz){\mathrm{Ai}}{\left(\frac{z-\overline{z}_m}{z_0}\right)}\,\mathrm{d}z\nonumber\\
%f_c^{\mathrm{Bi}}(k,m)&=&\int_0^L\cos(kz){\mathrm{Bi}}{\left(\frac{z-\overline{z}_m}{z_0}\right)}\,\mathrm{d}z\nonumber\\
%f_s^{\mathrm{Ai}}(k,m)&=&\int_0^L\sin(kz){\mathrm{Ai}}{\left(\frac{z-\overline{z}_m}{z_0}\right)}\,\mathrm{d}z\nonumber\\
%f_s^{\mathrm{Bi}}(k,m)&=&\int_0^L\sin(kz){\mathrm{Bi}}{\left(\frac{z-\overline{z}_m}{z_0}\right)}\,\mathrm{d}z\nonumber\\
%\alpha_c(k,m)&=&b_mf_c^{\mathrm{Ai}}(k,m)-a_mf_c^{\mathrm{Bi}}(k,m)\nonumber\\
%\alpha_s(k,m)&=&b_mf_s^{\mathrm{Ai}}(k,m)-a_mf_s^{\mathrm{Bi}}(k,m)\,\,\,,\nonumber\\
%F_m^{(0)}(k,t)&=&C_m(t)\,[\,\alpha_c(k,m)-i\alpha_s(k,m)\,]\,\,\,.
%\end{eqnarray}
\begin{eqnarray}\label{fcsAiBi}
\alpha_c(k,m) &:=& \frac{1}{C_m(t)}\mathrm{Re}[F_m^{(0)}(k,t)]\,\,\,,\nonumber\\
\alpha_s(k,m) &:=& -\frac{1}{C_m(t)}\mathrm{Im}[F_m^{(0)}(k,t)]\,\,\,,
\end{eqnarray}
such that 
\begin{eqnarray}
F_m^{(0)}(k,t)&=&C_m(t)\,[\,\alpha_c(k,m)-i\alpha_s(k,m)\,]\,\,\,,
\end{eqnarray}
and the spectral function is given through
\begin{eqnarray}\label{Spectrum2m}
|F_m^{(0)}(k)|^2=|C_m|^2[\,\alpha_c^2(k,m)+\alpha_s^2(k,m)\,]
\end{eqnarray}
with $|C_m|^2=\frac{1}{2\pi{}z_0{}N_m^2}$. Notice that this momentum distribution is stationary.
It is depicted for the cases $m=1$, $m=2$, and $m=3$ in Figs.~\ref{Fig20}, \ref{Fig21}, and \ref{Fig22}, respectively. These figures also explicitly show the respective $|C_m|^2\alpha_s^2(k,m)$ and $|C_m|^2\alpha_c^2(k,m)$. %One can therefore realize the contributions of these quantities for $|F_m^{(0)}(k)|^2$ in Eq.~(\ref{Spectrum2m}).

The wave function for two mirrors $\psi_m^{(0)}(z,t)$ in Eq.~(\ref{Wave2m}) is normalized as
\begin{eqnarray}\label{normalizationWF}
\int_{-\infty}^{\infty}|\psi_m^{(0)}(z,t)|^2\,\mathrm{d}z=\int_{0}^{L}|\psi_m^{(0)}(z,t)|^2\,\mathrm{d}z=1\,\,\,.
\end{eqnarray} In a similar way, the spectral function $|F_m^{(0)}(k)|^2$ has to be normalized, such that
\begin{eqnarray}\label{normalization}
\int_{-\infty}^{\infty}|F_m^{(0)}(k)|^2\,\mathrm{d}k=1\,\,\,.
\end{eqnarray}
We can easily show that this normalization is indeed valid here by using the first equality in Eq.~(\ref{FourWave2m}):
\begin{eqnarray}
\int^\infty_{-\infty}|F_m^{(0)}(k,t)|^2dk &=& \frac{1}{2\pi}\int^\infty_{-\infty}\mathrm{e}^{ikz'}\psi_m^{(0)\ast}(z',t)\mathrm{e}^{-ikz}\psi_m^{(0)}(z,t)\,\mathrm{d}z'\mathrm{d}z\mathrm{d}k
\nonumber
\\
&=& \int^\infty_{-\infty}\delta(z-z')\psi_m^{(0)\ast}(z',t)\psi_m^{(0)}(z,t)\,\mathrm{d}z'\mathrm{d}z
\nonumber
\\
&=& \int^\infty_{-\infty}|\psi_m^{(0)}(z,t)|^2\,\mathrm{d}z
\nonumber
\\
&=&1\,\,\,.
\end{eqnarray}

\subsection{Wigner function}
Using Eq.~(\ref{Wave2m}) with Eq.~(\ref{WF}) gives the Wigner function
\begin{eqnarray}\label{Wig2m}
W_m^{(0)}(z,k)&=&\frac{1}{2\pi}\int_{-\infty}^{\infty}\mathrm{e}^{iz'k}{\psi_m^{(0)\ast}}{\left(z+\frac{z'}{2},t\right)}{\psi_m^{(0)}}{\left(z-\frac{z'}{2},t\right)}\,\mathrm{d}z'\nonumber\\
&=&\frac{1}{2\pi{}z_0N_m^2}\int_{A(z)}^{B(z)}\mathrm{e}^{iz'k}D(z,z')\,\mathrm{d}z'
\end{eqnarray}
with
\begin{eqnarray}
D(z,z')&:=&\left[\,b_m{\mathrm{Ai}}{\left(\frac{z+\frac{z'}{2}-\bar{z}_m}{z_0}\right)}-a_m{\mathrm{Bi}}{\left(\frac{z+\frac{z'}{2}-\bar{z}_m}{z_0}\right)}\right]
\left[\,b_m{\mathrm{Ai}}{\left(\frac{z-\frac{z'}{2}-\bar{z}_m}{z_0}\right)}-a_m{\mathrm{Bi}}{\left(\frac{z-\frac{z'}{2}-\bar{z}_m}{z_0}\right)}\right]\,\,\,,
\nonumber
\\
\end{eqnarray}
and $A(z)$ and $B(z)$ are the limits of integration that appear due to the restrictions on the support of $\psi_m^{(0)}$. It can directly be seen that $D(z,z')=D(z,-z')$.

\begin{figure}[h!]
\begin{subfigure}{.49\linewidth}
\centering
\includegraphics[width=60mm]{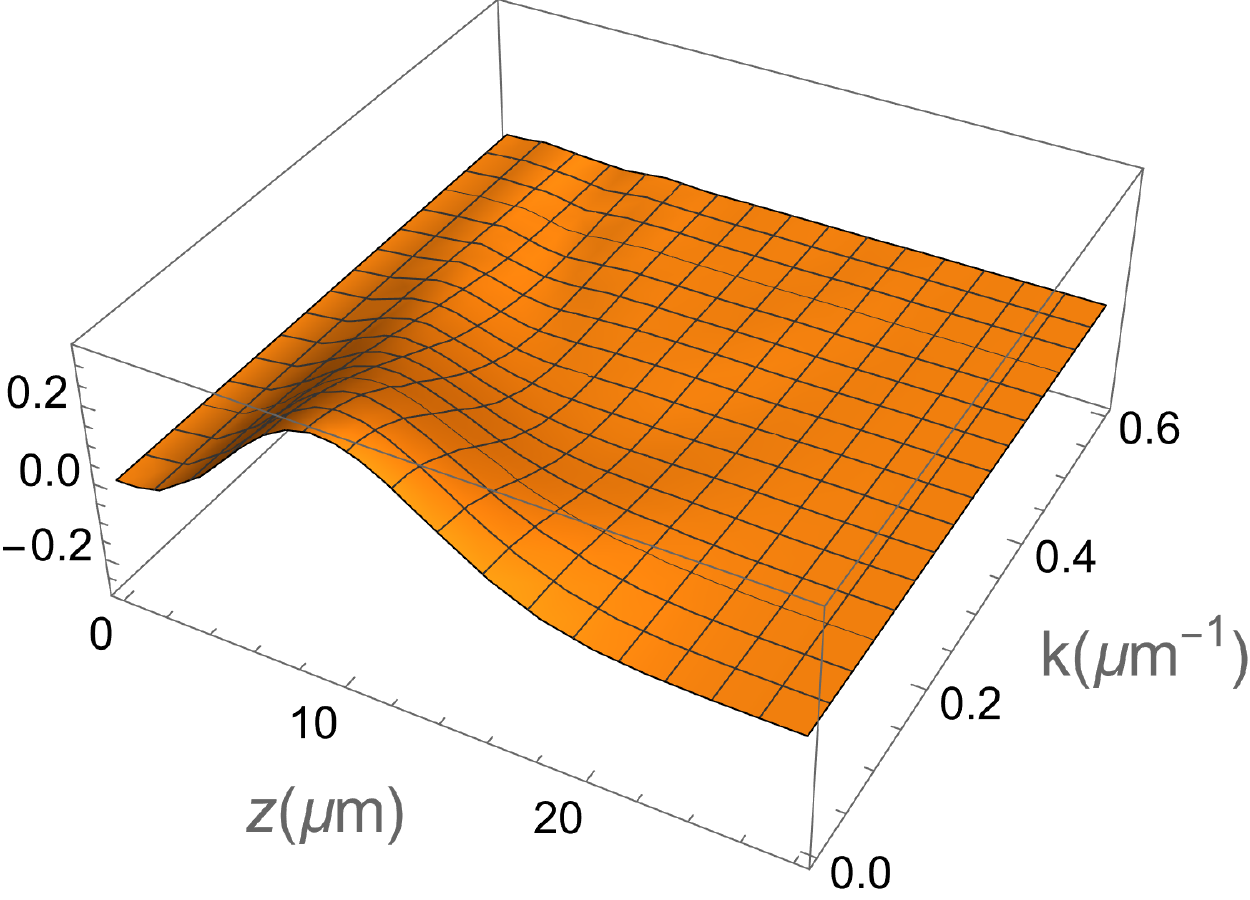}
\caption{Wigner function $W_1^{(0)}(z,k)$} \label{W1zweiSp}
\end{subfigure}%

\begin{subfigure}{.49\linewidth}
\includegraphics[width=60mm]{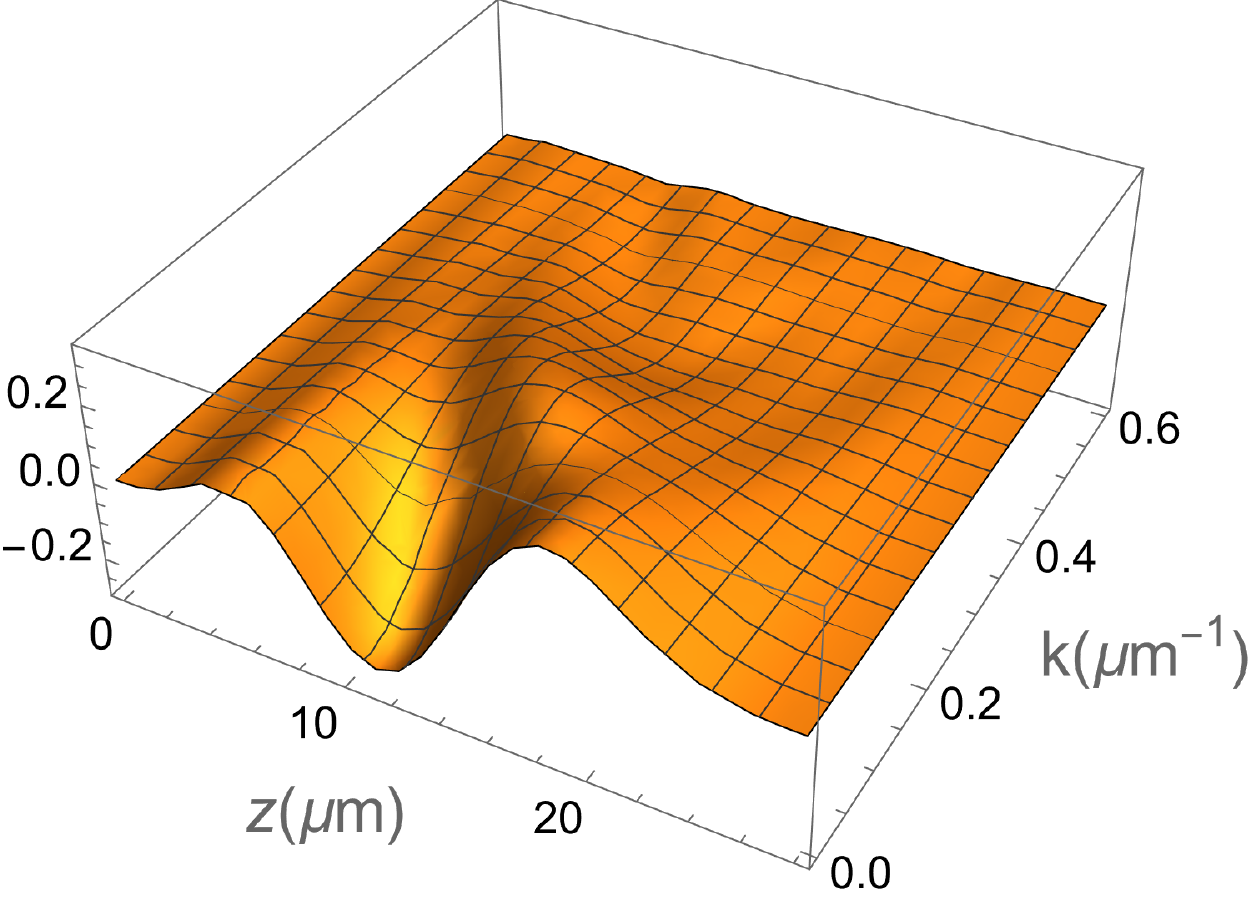}
\caption{Wigner function $W_2^{(0)}(z,k)$} \label{W2zweiSp}
\end{subfigure}
\begin{subfigure}{.49\linewidth}
\includegraphics[width=60mm]{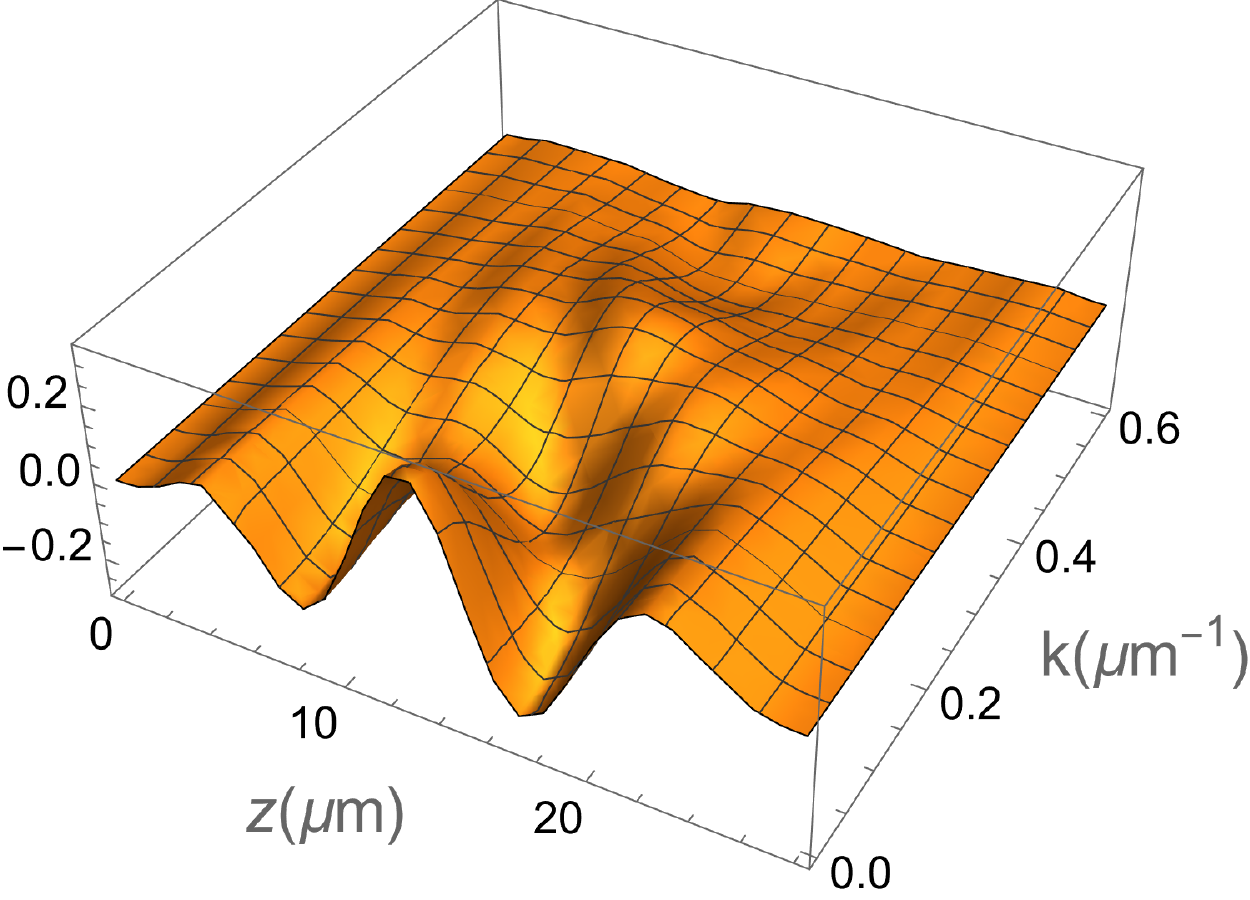}
\caption{Wigner function $W_3^{(0)}(z,k)$} \label{W3zweiSp}
\end{subfigure}
\caption{Wigner functions of the ground state, and first and second excited state, see Eq.~(\ref{WFfin})}
\end{figure}

We will now determine the limits of integration. For this, we consider that $z$ is limited between $0$ and $L$, leading us to the following 4 equations
\begin{eqnarray}\label{boundAB}
&z&+\frac{z'}{2}=0\,\,\,\Rightarrow\,\,\,z'=-2z\,\,\,;\,\,\,\,\,\,\,\,\,\,\,\,\,\,\,z-\frac{z'}{2}=0\,\,\,\Rightarrow\,\,\,z'=2z\,\,\,;\nonumber\\
&z&+\frac{z'}{2}=L\,\,\,\Rightarrow\,\,\,z'=2(L-z)\,\,\,;\,\,\,z-\frac{z'}{2}=L\,\,\,\Rightarrow\,\,\,z'=2(z-L)\,\,\,;
\end{eqnarray}
which represent 4 straight lines in the $(z,z')$-diagram and generate a rhombus. The values inside of this rhombus are the allowed values for integration. Therfore, we conclude
\begin{eqnarray}\label{AB}\,\,\,\,\,\,
0\leq z\leq\frac{L}{2}\,\,\,&\Rightarrow&\,\,\,A(z)=-2z\leq z'\leq 2z=B(z)\,\,\,;\nonumber\\
\frac{L}{2}\leq z \leq L\,\,\,&\Rightarrow&\,\,\,A(z)=-2(L-z)\leq z'\leq 2(L-z)=B(z)\,\,\,.
\end{eqnarray}
Since $A(z)=-B(z)$, the Wigner function can finally be written as
\begin{eqnarray}\label{WFfin}
W_m^{(0)}(z,k)&=&\frac{1}{\pi{}z_0N_m^2}\int_{0}^{B(z)}\cos(z'k)D(z,z')\,\mathrm{d}z'
\nonumber
\\
&=&\frac{1}{\pi{}z_0N_m^2}
    \begin{cases}
    \int_{0}^{2z}\cos(z'k)D(z,z')\,\mathrm{d}z'\,\,\,,\,\,\, \text{for}\,\,\,0<z<\frac{L}{2}
    \\\\
    \int_{0}^{2(L-z)}\cos(z'k)D(z,z')\,\mathrm{d}z'\,\,\,,\,\,\, \text{for}\,\,\,\frac{L}{2}<z<L
    \end{cases}\,\,\,.
\end{eqnarray}
It is depicted in  Figs.~\ref{W1zweiSp}, \ref{W2zweiSp}, and \ref{W3zweiSp} for $m=1$, $m=2$, and $m=3$, respectively. 
%The following boundary distributions and normalization have to be applied and are valid:
%\begin{eqnarray}\label{BCN}
%\int_{-\infty}^{\infty}W_m^{(0)}(z,k)\,\mathrm{d}k&=&|\psi_m^{(0)}(z)|^2\,\,\,,\nonumber\\
%\int_{-\infty}^{\infty}W_m^{(0)}(z,k)\,\mathrm{d}z&=&|F_m^{(0)}(k)|^2\,\,\,,\nonumber\\
%\int_{-\infty}^{\infty}\int_{-\infty}^{\infty}W_m^{(0)}(z,k)\,\mathrm{d}k\,\mathrm{d}z&=&1\,\,\,.
%\end{eqnarray}
As can be seen from Fig.~\ref{W1zweiSp}, the Wigner function $W_1^{(0)}(z,k)$ of the ground state ($m=1$) is positive everywhere. This is not the case for the Wigner function $W_2^{(0)}(z,k)$ of the first excited  state ($m=2$), see Fig.~\ref{W2zweiSp}, and the Wigner function $W_3^{(0)}(z,k)$ of the second excited state ($m=3$), see Fig.~\ref{W3zweiSp}. For these states we can clearly observe negative regions of the Wigner functions. This is a very characteristic property of excited states in quantum mechanics. At $z=0$ (bottom mirror) and at $z=\SI{28}{\micro\meter}$ (top mirror) the Wigner functions vanish exactly because of the boundary conditions. %In case of $m=2$ or $m=3$ the zero points of the wave functions for excited states are reproduced.
 
In summary, in this chapter, we considered the eigenstates of the neutron wave function in a double mirror system, calculated the spectral functions and evaluated the corresponding Wigner functions. The latter enabled us to look into the complete phase space in order to study momentum and position simultaneously.

%------------------------------------------------------------------------------
\section{"Free fall" of wave function after double mirror (region II)}
%------------------------------------------------------------------------------
 
%
In this chapter we consider the "free fall" of a wave function which exits a double mirror system (we denote this region by I, see Fig.~\ref{expaufbau}). The wave function reaches a second region II, where it falls down a height $h = \SI{27}{\micro\meter}$ on a subsequent static mirror located below the double mirror system. This case has been investigated theoretically in~\cite{P18}. The wave function in region I has been given in Eq.~(\ref{Wave2m}). Since we are now also going to consider region II, and for convenience, we apply the coordinate shift $z \to z - h$ to the result from Eq.~(\ref{Wave2m}), such that we have
\begin{eqnarray}\label{psiI}
\psi_{m,\text{I}}(z,t)
&:=&\bar{C}_m\mathrm{e}^{-\frac{i}{\hbar}\bar{E}_mt}\left[\,b_m{\mathrm{Ai}}{\left(\frac{z-h-\bar{z}_m}{z_0}\right)}-a_m{\mathrm{Bi}}{\left(\frac{z-h-\bar{z}_m}{z_0}\right)}\right]
\,\,\,,
\end{eqnarray}
where we introduced the notation $\bar{C}_m:=\frac{1}{\sqrt{z_0}N_m}$, and the wave function has support only on $[h,L+h]\times(-\infty,0]$.
 
In region II the wave function takes on the following form
\begin{eqnarray}\label{psiII}
\psi_{m,\text{II}}(z,t)=\bar{C}_m\sum_{n=1}^{\infty}D_{n,m}{\mathrm{Ai}}{\left(\frac{z-z_n}{z_0}\right)}\mathrm{e}^{-\frac{i}{\hbar}E_nt}
\end{eqnarray}
and includes coefficients
\begin{eqnarray}\label{Dnm1}
D_{n,m}&:=&\left\{\,\left[b_m{\mathrm{Ai'}}{\left(\frac{L-\bar{z}_m}{z_0}\right)}-a_m{\mathrm{Bi'}}{\left(\frac{L-\bar{z}_m}{z_0}\right)}\right]\left[{\mathrm{Ai}}{\left(\frac{L+h-z_n}{z_0}\right)}-{\mathrm{Ai}}{\left(\frac{L-\bar{z}_m}{z_0}\right)}\right]\right. \nonumber\\
&\phantom{=}&\left.-\left[b_m{\mathrm{Ai'}}{\left(-\frac{\bar{z}_m}{z_0}\right)}-a_m{\mathrm{Bi'}}{\left(-\frac{\bar{z}_m}{z_0}\right)}\right]\left[{\mathrm{Ai}}{\left(\frac{h-z_n}{z_0}\right)}-a_m\right]\,\right\}\,\left(\frac{z_0}{z_n-\bar{z}_m-h}\right)\left[{\mathrm{Ai'}}{\left(-\frac{z_n}{z_0}\right)}\right]^{-2}\,\,\,.
\end{eqnarray}
Formula Eq.~(\ref{psiII}) together with Eq.~(\ref{Dnm1}) corresponds exactly to equation $(70)$ in article~\cite{P18}.
 
The wave function in region II, given in Eq.~(\ref{psiII}), should have support only on $[0,\infty)\times(0,\infty)\cup [h,L+h] \times \{0\}$. This results from the requirement of a continuous transition between regions I and II expressed by
\begin{eqnarray}\label{psiII1}
\psi_{m,\text{I}}(z,t=0)=\psi_{m,\text{II}}(z,t=0)\,\,\,,
\end{eqnarray}
which can only be fulfilled if both $\psi_{m,\text{I}}(z,t)$ and $\psi_{m,\text{II}}(z,t)$ have the same support at $t=0$.

\begin{figure}[h!]
\centering
\includegraphics[width=80mm]{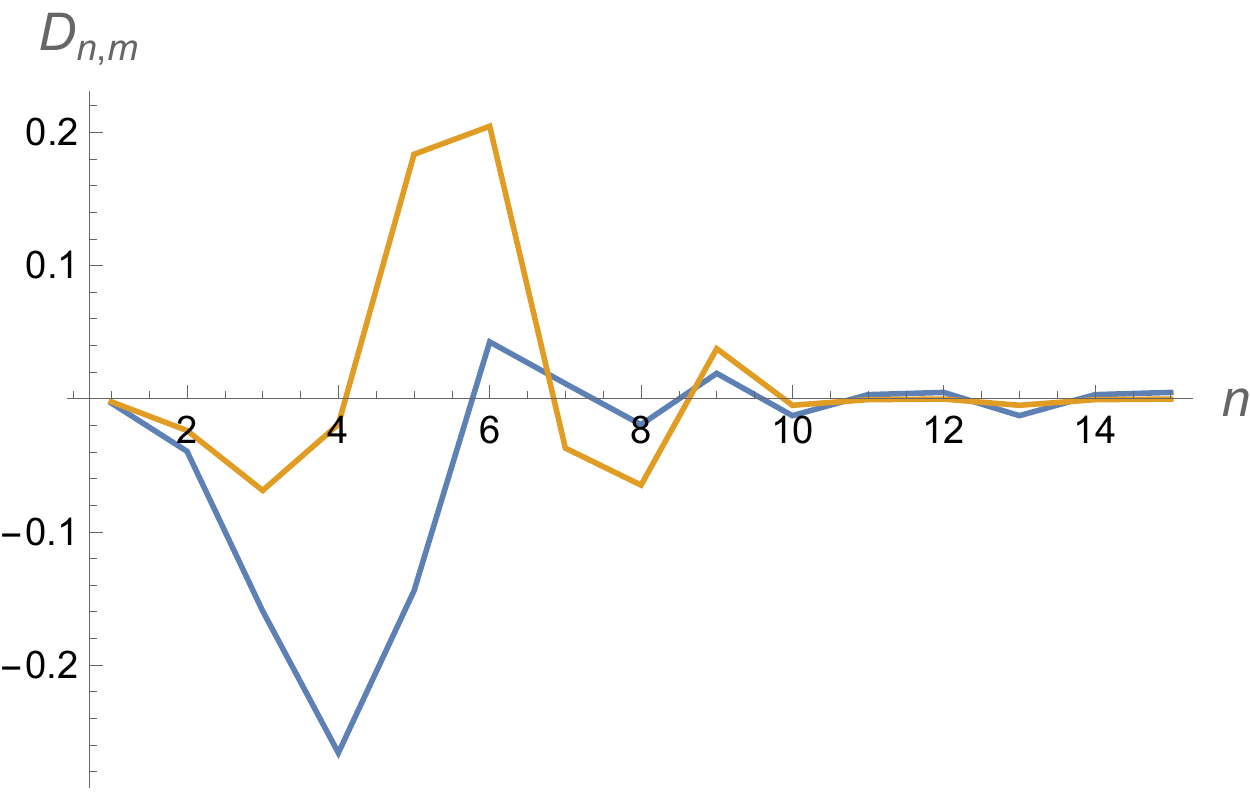}
\caption{Coefficients $D_{n,m}$ from Eq.~(\ref{Dnm1}) for ground state with $m=1$ (blue) and first excited state with $m=2$ (orange); For these cases, the coefficients are very small for $n>12$.} \label{Dnm15}
\end{figure}

Next, we examine the coefficients $D_{n,m}$ for $m=1,2$ and $n=1,2,...\,\,$. For this, in what follows, we present the numerical values of a few selected parameters relevant for Eq.~(\ref{Dnm1}):
\begin{eqnarray}\label{coeffDnm}
&&z_1=\SI{13.71680}{\micro\meter}\,\,,\,\,\,\,z_2=\SI{23.98246}{\micro\meter}\,\,,\,\,\,\,z_3=\SI{32.38707}{\micro\meter}\,\,,\,\,\,\,z_4=\SI{39.81502}{\micro\meter}\,\,,
\nonumber
\\
&&
z_5=\SI{46.60526}{\micro\meter}\,\,,\,\,\,\,z_6=\SI{52.93243}{\micro\meter}\,\,,
\,\,\,\,z_7=\SI{58.90210}{\micro\meter}\,\,,\,\,\,\,z_8=\SI{64.58300}{\micro\meter}\,\,,\nonumber
\\
&&
z_9=\SI{70.02430}{\micro\meter}\,\,,\,\,\,\,z_{10}=\SI{75.26180}{\micro\meter}\,\,,\,\,\,z_{11}=\SI{80.23200}{\micro\meter}\,\,,\,\,\,z_{12}=\SI{85.22950}{\micro\meter}\,\,\,.
\end{eqnarray}
Fig.~\ref{Dnm15} shows some of the first coefficients $D_{n,m}$ for the cases $m=1$ and $m=2$. From there it can be seen that for $n>12$ the coefficients $D_{n,m}$ become very small and can therefore be neglected.
%
%\begin{figure}[h!]
%\centering
%\includegraphics[scale=1.0]{bilder/Dnm12}
%\caption{The coefficients $D_{n,m}$ from Eq.~(\ref{Dnm1}) for $m=1$ (ground state) and for $m=2$ (first excited state). The index $n$ is defined in Eq.~(\ref{psiII}). For $m=1$ or $m=2$ the coefficients $D_{n,m}$ are very small if $n>12$.} \label{Dnm12}
%\end{figure}
%
%
%Because of their importance we compile the values $D_{n,m}$ displayed in %Fig.~\ref{Dnm} numerically:
%\begin{eqnarray}\label{numerically}
%&\,&D_{1,1}=-0.003254\,,\,D_{2,1}=-0.051828\,,\,D_{3,1}=-0.242690\,,\,D_{4,1}=-0.44864%5\,,\,\nonumber\\
%&\,&D_{5,1}=-0.263219\,,\,D_{6,1}=+0.082965\,,\,D_{7,1}=+0.023248\,,\,D_{8,1}=-0.04180%7\,,\,\nonumber\\
%&\,&D_{9,1}=+0.042597\,,\,D_{10,1}=-0.029548\,,\,D_{11,1}=+0.007320\,,\,D_{12,1}=+0.01%1655\,,\,\nonumber\\
%&\,&\nonumber\\
%&\,&D_{1,2}=-0.001870\,,\,D_{2,2}=-0.025309\,,\,D_{3,2}=-0.084678\,,\,D_{4,2}=-0.02621%0\,,\,\nonumber\\
%&\,&D_{5,2}=+0.270434\,,\,D_{6,2}=+0.321231\,,\,D_{7,2}=-0.061443\,,\,D_{8,2}=-0.11252%7\,,\,\nonumber\\
%&\,&D_{9,2}=+0.067833\,,\,D_{10,2}=-0.009242\,,\,D_{11,2}=-0.001380\,,\,D_{12,2}=-0.00&0566\,,\,\nonumber
%\end{eqnarray}
%
 
\begin{figure}[h!]
\begin{subfigure}[t]{.49\linewidth}
\centering
\includegraphics[width=80mm]{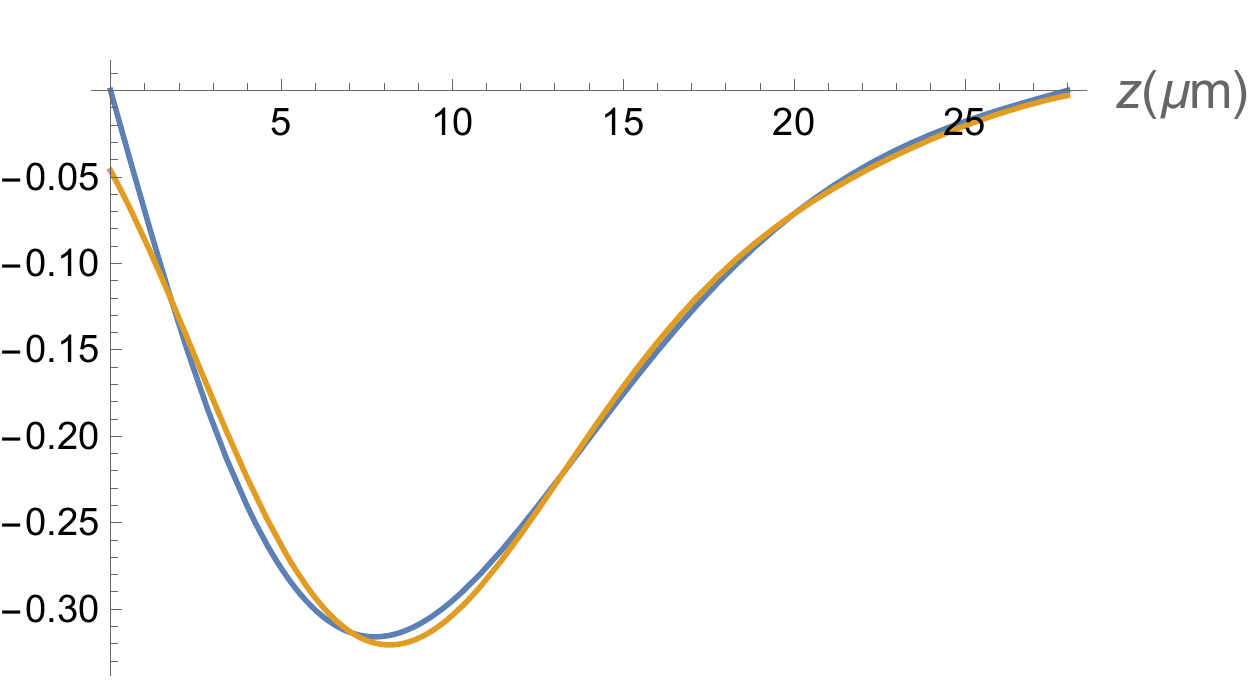}
\caption{Using only coefficients $D_{n,1}$ for $n=1$ to $n=12$, the agreement is already considerable.} \label{psig12}
\end{subfigure}%
\begin{subfigure}[t]{.49\linewidth}
\centering
\includegraphics[width=80mm]{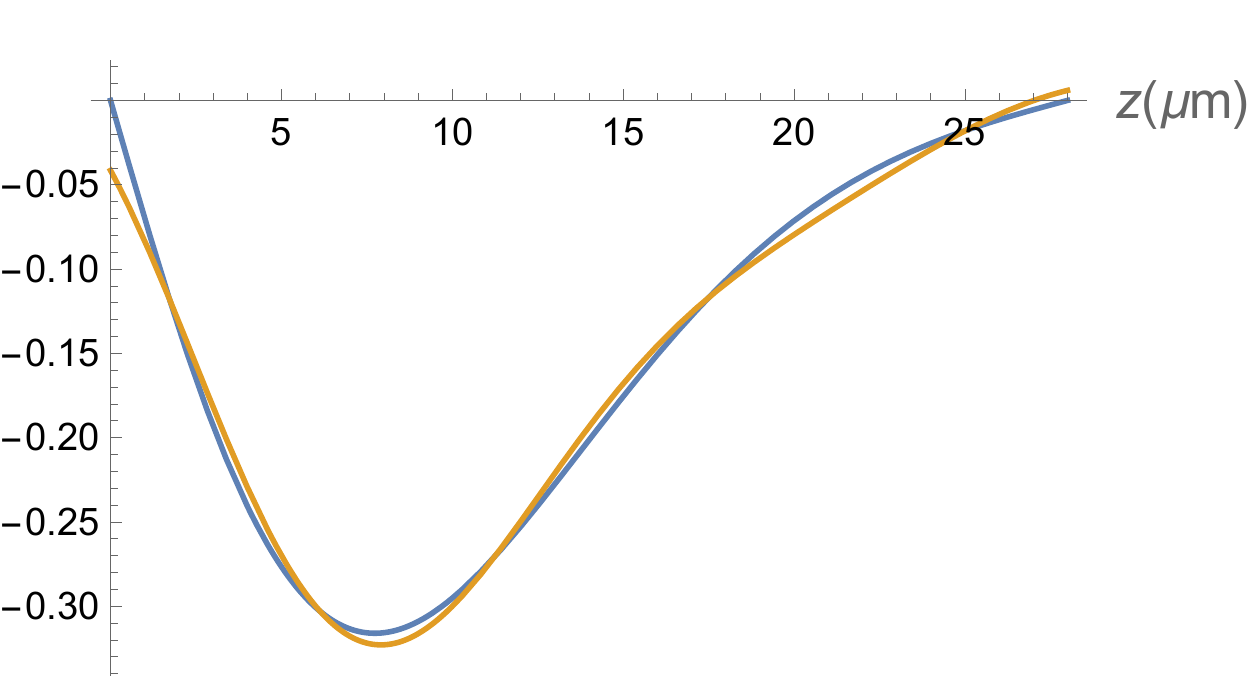}
\caption{Using only coefficients $D_{n,1}$ for $n=1$ to $n=15$} \label{psig15}
\end{subfigure}
\caption{Comparison of $\psi_{1,\text{I}}(z,t=0)$  (blue) with $\psi_{1,\text{II}}(z,t=0)$ (orange) for $m=1$, see  Eq.~(\ref{psiII1}); The coordinate $z=h$ has been shifted to the point of origin.}
\end{figure}

Now we can check whether $\psi_{1,\text{I}}(z,t=0)$ and $\psi_{1,\text{II}}(z,t=0)$ fulfill the condition in  Eq.~(\ref{psiII1}). We do this for the ground state $m=1$. For this, we consider only a finite number of coefficients $D_{n,1}$. The results including $n$ up to $12$ and $15$ are presented in  Figs.~\ref{psig12} and \ref{psig15}, respectively. There the abscissa represents the $z-$coordinate in $\si{\micro\meter}$ shifted to the point of origin of the coordinate system. Therefore, in region I there are values of $z$ from $0$ to $L$. The agreement between $\psi_{1,\text{I}}(z,t=0)$ and the plotted approximation of $\psi_{1,\text{II}}(z,t=0)$ is very good except near certain regions, e.g., around $z=0$. These small differences supposedly are due to using only a finite number of $D_{n,1}$-coefficients. We expect the differences to become smaller when more $D_{n,1}$-coefficients are considered.
\subsection{Spatial distribution (SD) in "free-fall" region }
%--------------------------------------------------------
The spatial distribution in region II follows from Eq.~(\ref{psiII}):
\begin{eqnarray}\label{sdII}
|\psi_{m,\text{II}}(z,t)|^2=\left|\bar{C}_m\sum_{n=1}^{\infty}D_{n,m}{\mathrm{Ai}}{\left(\frac{z-z_n}{z_0}\right)}\mathrm{e}^{-\frac{i}{\hbar}E_nt}\right|^2\,\,\,.
\end{eqnarray}
We define
\begin{eqnarray}\label{sdII1}
G_m^c(z,t)&:=&\bar{C}_m\sum_{n=1}^{\infty}D_{n,m}{\mathrm{Ai}}{\left(\frac{z-z_n}{z_0}\right)}{\cos}{\left(\frac{E_n}{\hbar}t\right)}\,\,\,,\nonumber\\
G_m^s(z,t)&:=&\bar{C}_m\sum_{n=1}^{\infty}D_{n,m}{\mathrm{Ai}}{\left(\frac{z-z_n}{z_0}\right)}\,{\sin}{\left(\frac{E_n}{\hbar}t\right)}\,\,\,,
\end{eqnarray}
such that the spatial distribution in region II reads
\begin{eqnarray}\label{sdII2}
|\psi_{m,\text{II}}(z,t)|^2=\,[G_m^c(z,t)]^2+[G_m^s(z,t)]^2\,\,\,.
\end{eqnarray}
Whenever performing numerical calculations we will only consider the sums in Eq.~(\ref{sdII1}) from $n=1$ to $15$ due to the smallness of later coefficients $D_{n,m}$.

\begin{figure}[h!]
\begin{subfigure}[t]{.49\linewidth}
\centering
\includegraphics[width=60mm]{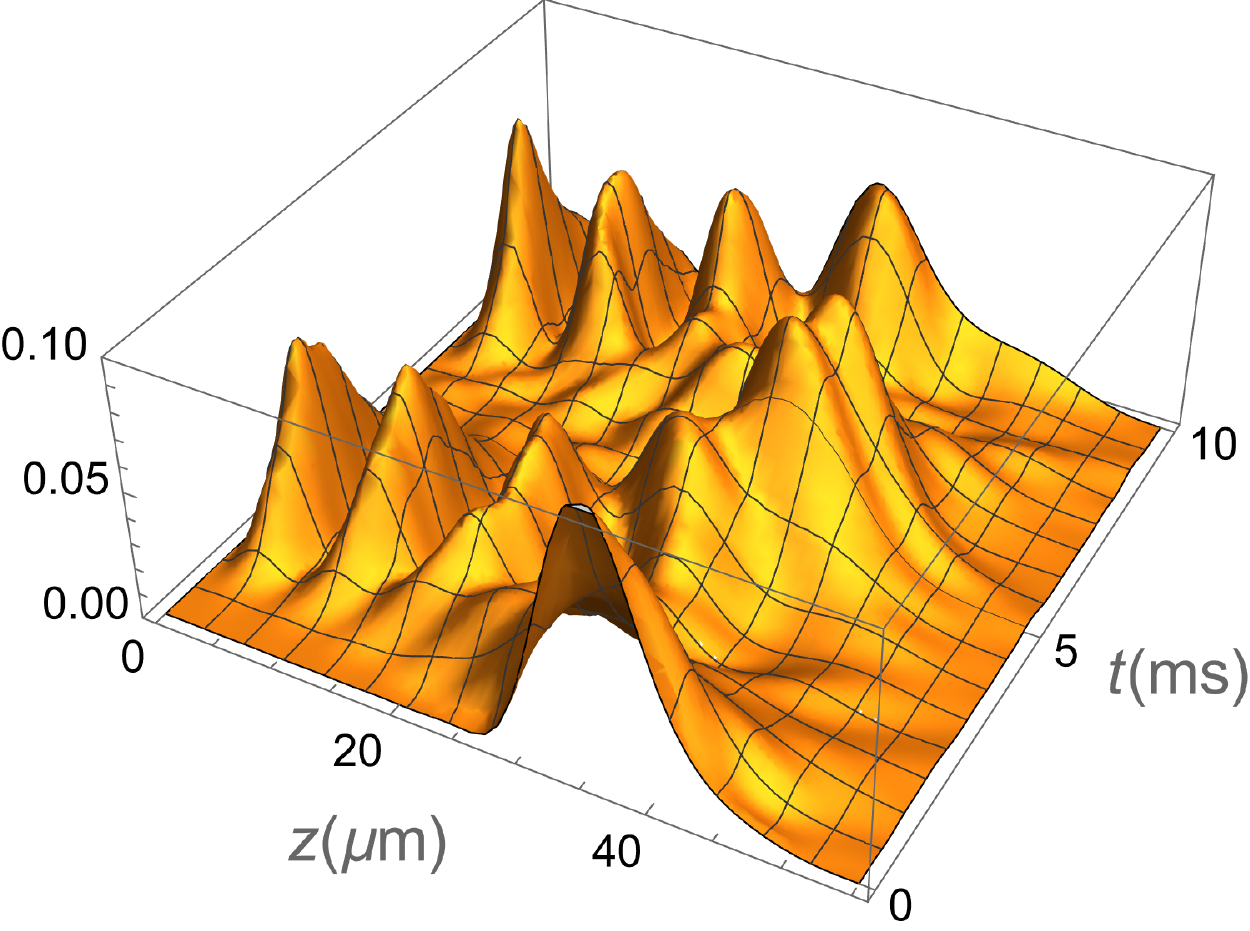}
\caption{Spatial distribution $|\psi_{1,\text{II}}(z,t)|^2$; The ground state ($m=1$) appears at $t=0$. The density distribution peaks nicely along the classical parabola.} \label{ortsvert1}
\end{subfigure}%
\begin{subfigure}[t]{.49\linewidth}
\centering
\includegraphics[width=60mm]{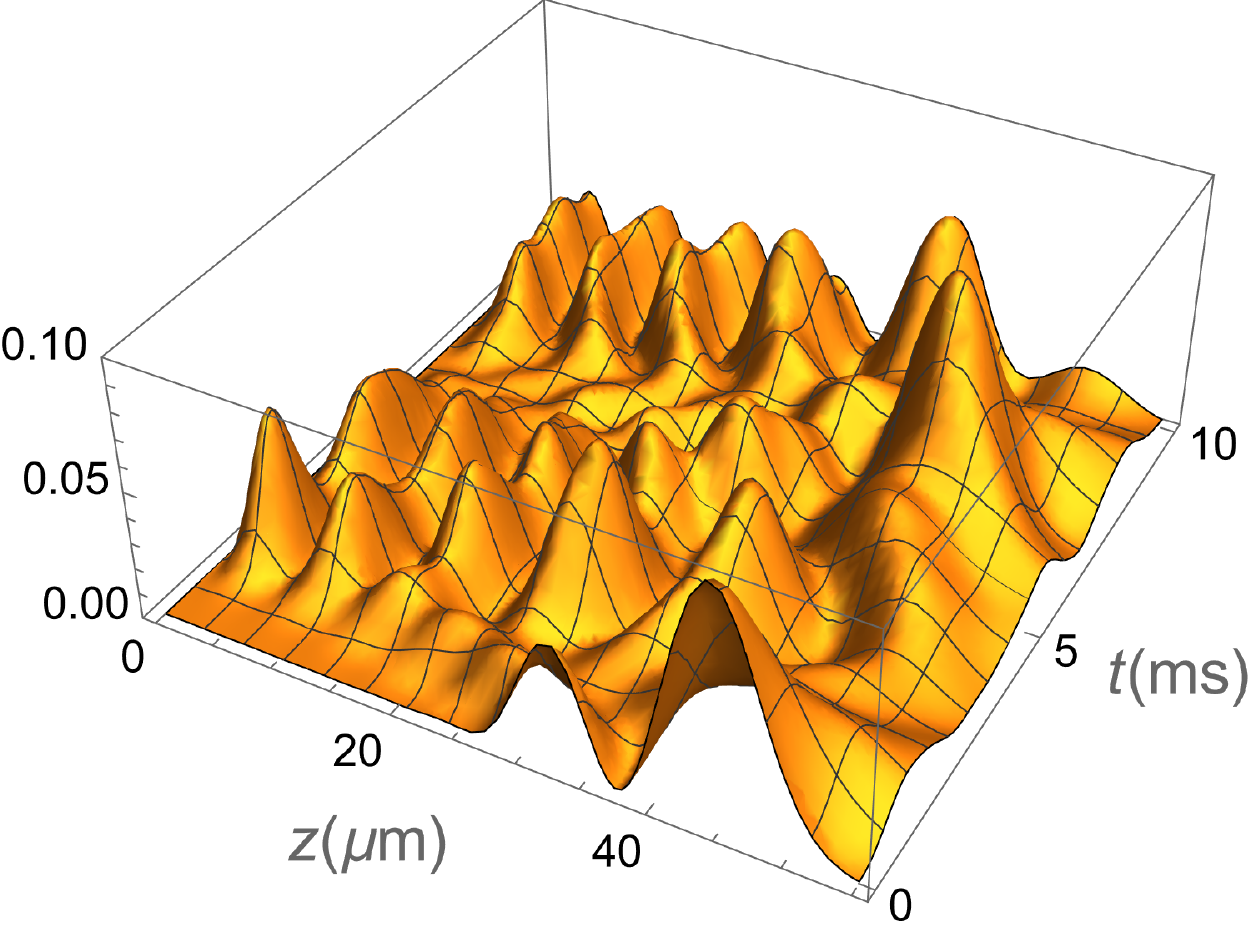}
\caption{Spatial distribution $|\psi_{2,\text{II}}(z,t)|^2$; The first excited state ($m=2$) appears at $t=0$.} \label{ortsvert2}
\end{subfigure}
\caption{Spatial distributions, see Eq.~(\ref{sdII2}), in region II}
\end{figure}

In Figs.~\ref{ortsvert1} and \ref{ortsvert2} the spatial distributions (SD) $|\psi_{1,\text{II}}(z,t)|^2$ and $|\psi_{2,\text{II}}(z,t)|^2$, respectively, are plotted as a function of the $z-$coordinate (in \si{\micro\meter}) and time $t$ (in \si{\second}). For $t=0$ the ground and the first excited state are visible. This means, that between $z=0$ and $z=h=\SI{27}{\micro\meter}$ the SD are zero. Between $z=h$ and $z=h+L=\SI{55}{\micro\meter}$ the SD have the shape of the ground state, see Fig.~\ref{ortsvert1}, or the first excited state, Figs.~\ref{ortsvert2}. For $z>\SI{55}{\micro\meter}$ the SD vanish again. While $t$ evolves, the wave function is reflected from the mirror in region II multiple times. Since the frequencies $E_n/\hbar$ vary during this process, which leads to varying superpositions of waves, more complicated SD pictures result at times after $t=0$ . 

\begin{figure}[h!]
\begin{subfigure}[t]{.49\linewidth}
\centering
\includegraphics[width=60mm]{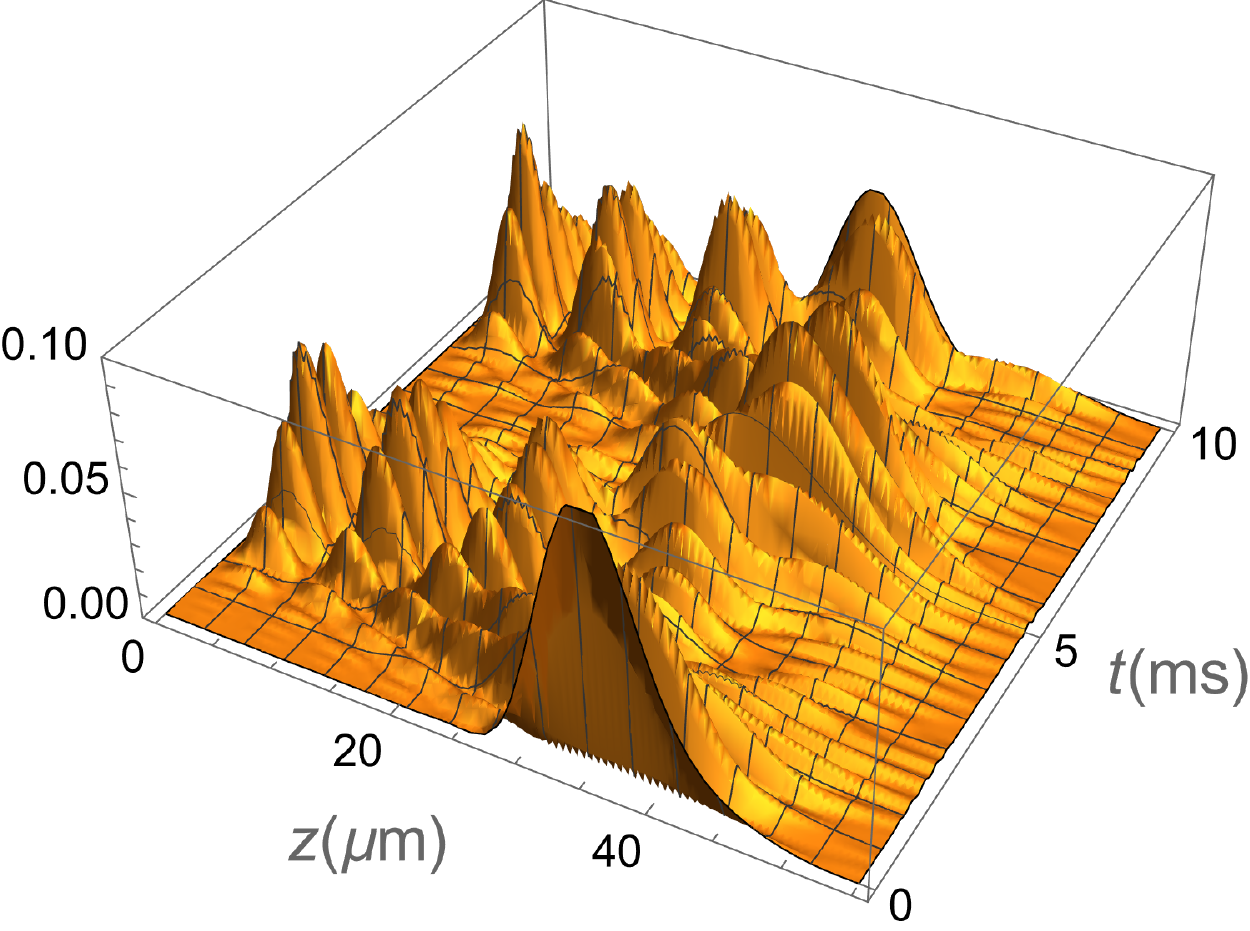}
\caption{Cosine spatial distribution $|G_{m}^c(z,t)|^2$; The ground state ($m=1$) appears at $t=0$.} \label{ortsvert1c}
\end{subfigure}%
\begin{subfigure}[t]{.49\linewidth}
\centering
\includegraphics[width=60mm]{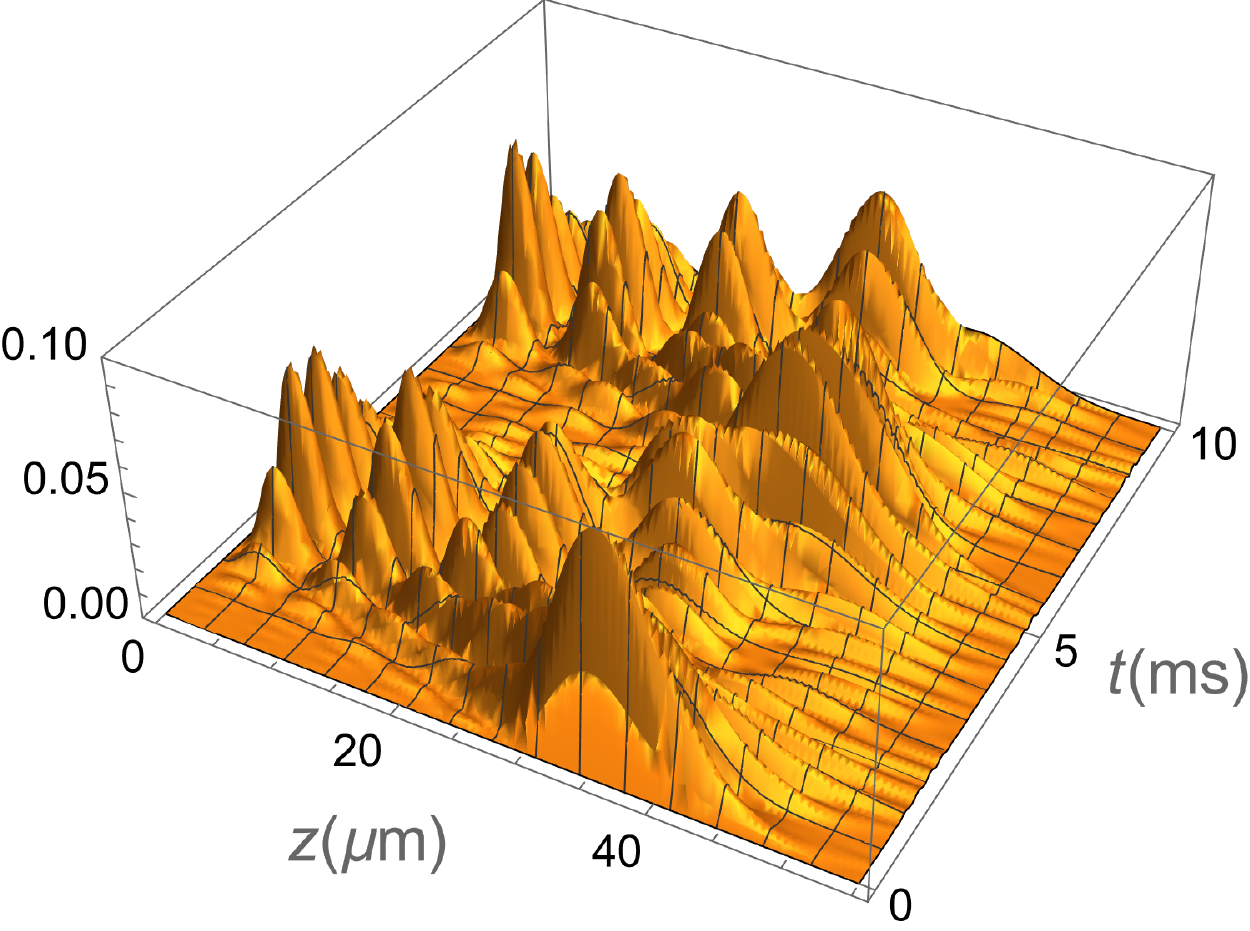}
\caption{Sine spatial distribution $|G_{m}^s(z,t)|^2$; The ground state ($m=1$) appears at $t=0$.} \label{ortsvert1s}
\end{subfigure}
\caption{Trigonometric spatial distributions, see Eq.~(\ref{sdII2}), in region II}
\end{figure}

Figs.~\ref{ortsvert1c} and  \ref{ortsvert1s} show the cosine and sine distribution functions $|G_1^c(z,t)|^2$ and $|G_1^s(z,t)|^2$ of the ground state. The sum of these functions yields Fig.~\ref{ortsvert1}. It is interesting to consider these parts of $|\psi_{1,\text{II}}(z,t)|^2$ separately since they could be important for interpreting experimental results.
 %But, as can be seen in this figure, at $z=0$ (mirror) and $t\approx0.0025$ as well as $t\approx0.0075$ the quantum wave has a maximum, which means that it is reflected at this points at the mirror. If the velocity of neutron wave amounts to \SI[per-mode=symbol]{6}{\meter\per\second} then after 0.01 seconds the wave has covered a distance of \SI{6}{\centi\meter}.

%In Fig.~\ref{ortsvert2} the SD $|\psi_{2,II}(z,t)|^2$ is plotted as a function of the $z-$coordinate and time $t$. For $t=0$ the first excited state is visible. This means that between $z=0$ and $z=h=27$ (= the step of free fall) the SD is zero. Between $z=27$ and $z=27+28=55$ the SD possesses the shape of the first excited  state which is characterized through two peaks. For $t>0$ the SD-picture therefore has been evolved a little bit more complicated compared to the previous case.
 
\subsection{Spatial distribution of mixtures}
So far, we have considered only a particular state (ground state or first excited state) which enters region II. It would be interesting to consider, for example, a mixture of these two types of quantum states arriving at the step before entering region II. This case will now be investigated. In the following, we distinguish between coherent mixtures and incoherent mixtures. We consider mixtures of the ground state with the first excited state.
\subsubsection{Coherent mixtures}
 
In the beginning, we look at the following coherent superposition state:
\begin{eqnarray}\label{mix}
\psi_{1+2,\text{II}}(z,t)=\sqrt{p_1}\,\psi_{1,\text{II}}(z,t)+\sqrt{p_2}\,\psi_{2,\text{II}}(z,t)\,\,\,,
\end{eqnarray}
where $p_1$ and $p_2$ are probabilities, and we ignore a potential phase between both terms for simplicity. Note that
\begin{eqnarray}\label{mixnorm}
\int_0^{\infty}|\psi_{1+2,\text{II}}(z,t)|^2\,\mathrm{d}z=1
\end{eqnarray}
because of the orthonormality of the eigenfunctions: $\int_0^{\infty}\psi_1\psi_2\,\mathrm{d}z=0$.

The next step is to calculate the spatial distribution of the coherent superposition $|\psi_{1+2,\text{II}}(z,t)|^2$. This can be accomplished by using Eq.~(\ref{psiII}) for $m=1$ and $m=2$:
\begin{eqnarray}\label{mix1}
\psi_{1+2,\text{II}}(z,t)&=&\sqrt{p_1}\,\bar{C}_1\sum_{n=1}^{\infty}D_{n,1}{\mathrm{Ai}}{\left(\frac{z-z_n}{z_0}\right)}\mathrm{e}^{-i\frac{E_n}{\hbar}t}+\sqrt{p_2}\,\bar{C}_2\sum_{n=1}^{\infty}D_{n,2}{\mathrm{Ai}}{\left(\frac{z-z_n}{z_0}\right)}\mathrm{e}^{-i\frac{E_n}{\hbar}t}\nonumber\\&=&\sum_{n=1}^{\infty}{\mathrm{Ai}}{\left(\frac{z-z_n}{z_0}\right)}\,(\sqrt{p_1}\,\bar{C}_1D_{n,1}+\sqrt{p_2}\,\bar{C}_2D_{n,2})\left[\,{\cos}{\left(\frac{E_n}{\hbar}t\right)}-i\,{\sin}{\left(\frac{E_n}{\hbar}t\right)}\right]\,\,\,.
\end{eqnarray}
For numerical calculations we will again cut off the infinite sums at $n=15$.
We define:
\begin{eqnarray}\label{mixSD}
|\psi_{1+2,\text{II}}(z,t)|^2&=&[H_{1+2}^c(z,t)]^2+[H_{1+2}^s(z,t)]^2\,\,\,,\nonumber
\\
H_{1+2}^c(z,t)&:=&\sum_{n=1}^{\infty}{\mathrm{Ai}}{\left(\frac{z-z_n}{z_0}\right)}\,(\sqrt{p_1}\,\bar{C}_1D_{n,1}+\sqrt{p_2}\,\bar{C}_2D_{n,2})\,{\cos}{\left(\frac{E_n}{\hbar}t\right)}\,\,\,,\nonumber
\\
H_{1+2}^s(z,t)&:=&\sum_{n=1}^{\infty}{\mathrm{Ai}}{\left(\frac{z-z_n}{z_0}\right)}\,(\sqrt{p_1}\,\bar{C}_1D_{n,1}+\sqrt{p_2}\,\bar{C}_2D_{n,2})\,{\sin}{\left(\frac{E_n}{\hbar}t\right)}\,\,\,.
\end{eqnarray}
In Figs.~\ref{coh73} and \ref{coh55} we present two examples, for which the coherent superposition of mixtures can be observed, namely for $p_1=0.7$ and $p_2=0.3$, and $p_1=0.5$ and $p_2=0.5$, respectively. Because of the superposition of waves in Eq.~(\ref{mix}), in both cases, only the ground state has been amplified at $t=0$. Altogether, we conclude that using a coherent superposition is not very useful for obtaining a damped behavior of the oscillations. We expect this to be different when using incoherent mixtures.
\begin{figure}[h!]
\centering

\end{figure}

\begin{figure}[h!]
\begin{subfigure}[t]{.49\linewidth}
\centering
\includegraphics[width=60mm]{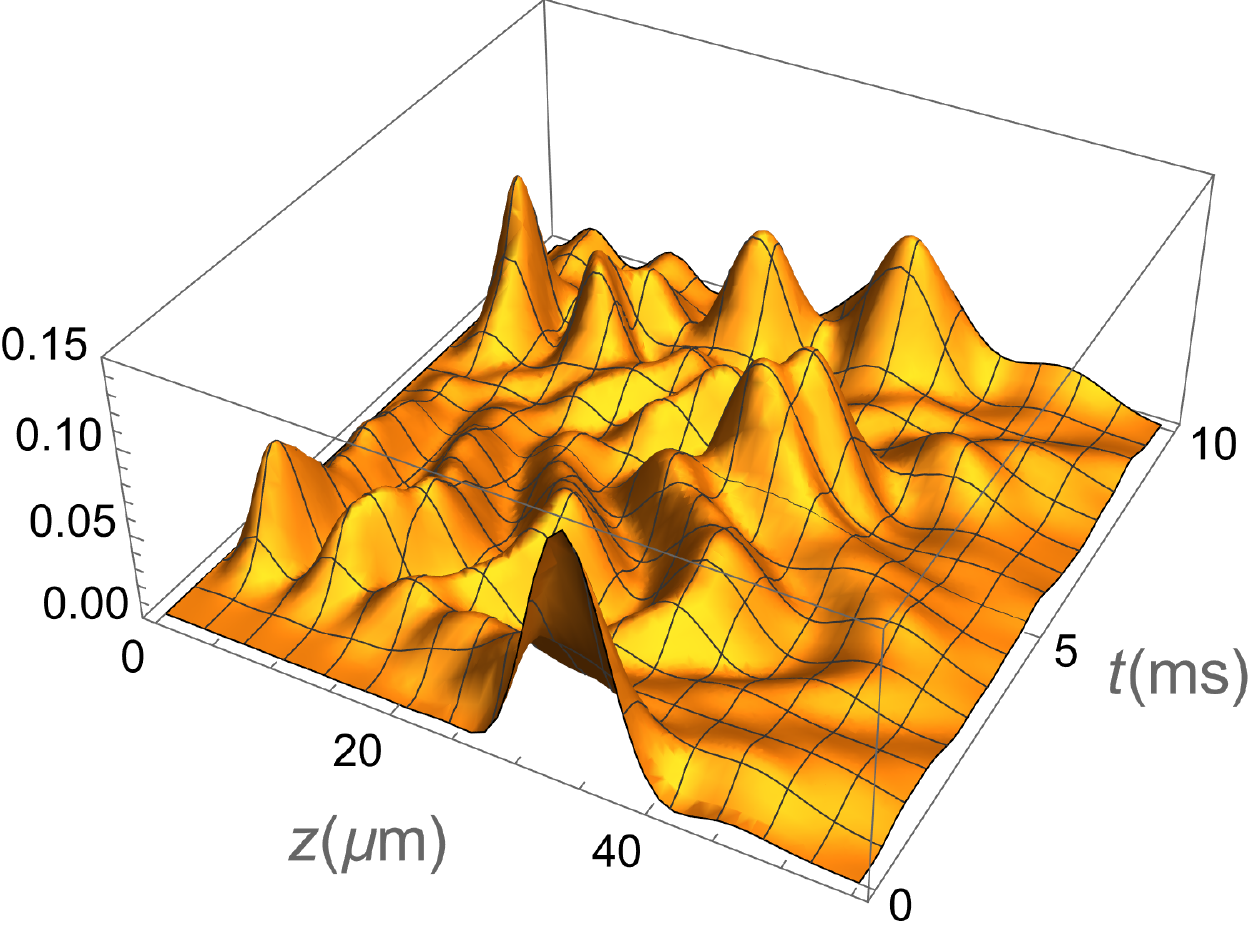}
\caption{Coherent mixture $|\psi_{1+2,\text{II}}(z,t)|^2$ with $p_1=0.7$ and $p_2=0.3$} \label{coh73}
\end{subfigure}%
\begin{subfigure}[t]{.49\linewidth}
\centering
\includegraphics[width=60mm]{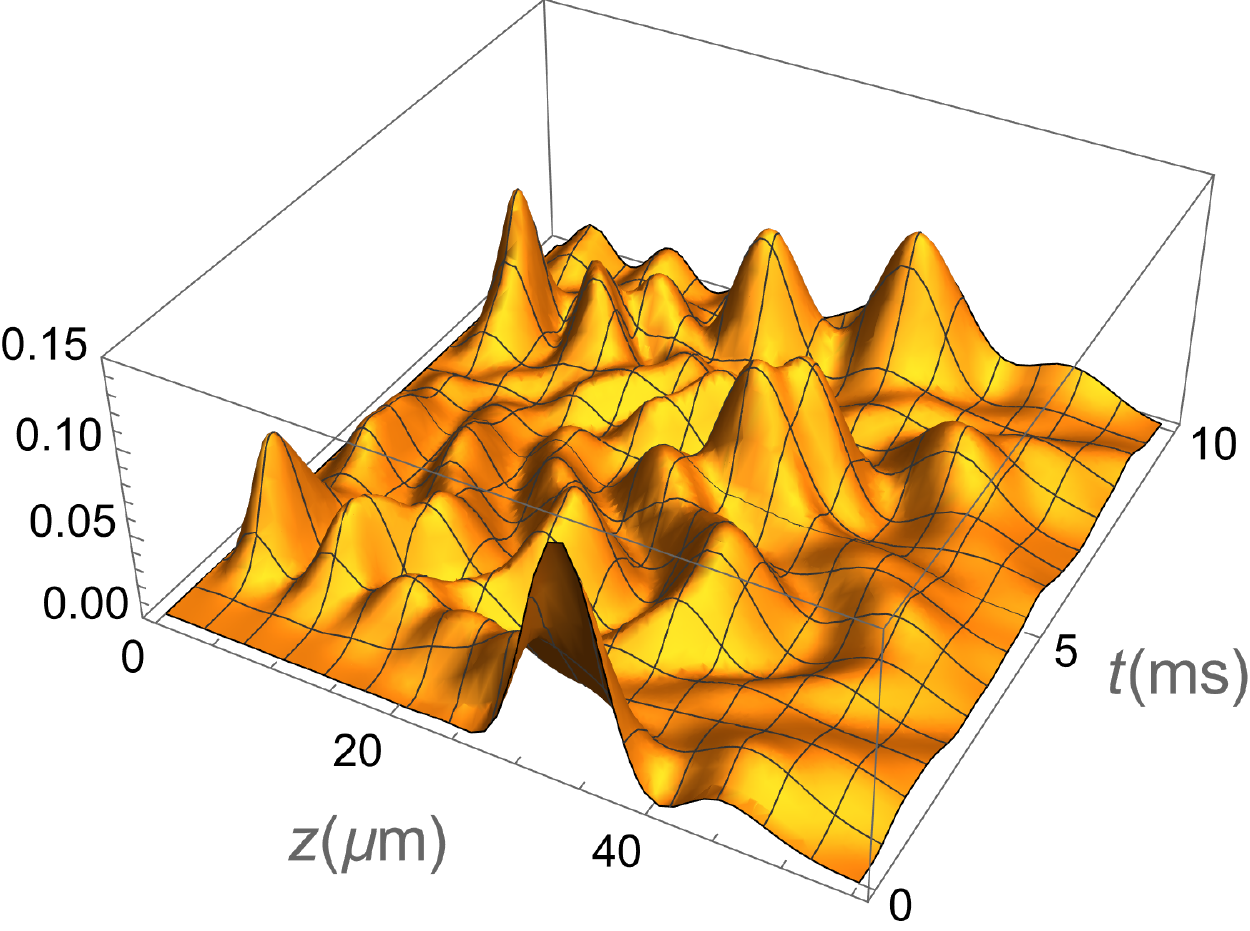}
\caption{Coherent mixture $|\psi_{1+2,\text{II}}(z,t)|^2$ with $p_1=p_2=0.5$} \label{coh55}
\end{subfigure}
\caption{Coherent mixtures, see Eq.~(\ref{mixSD}), in region II}
\end{figure}

\subsubsection{Incoherent mixtures}
\begin{figure}[h!]
\begin{subfigure}[t]{.49\linewidth}
\centering
\includegraphics[width=60mm]{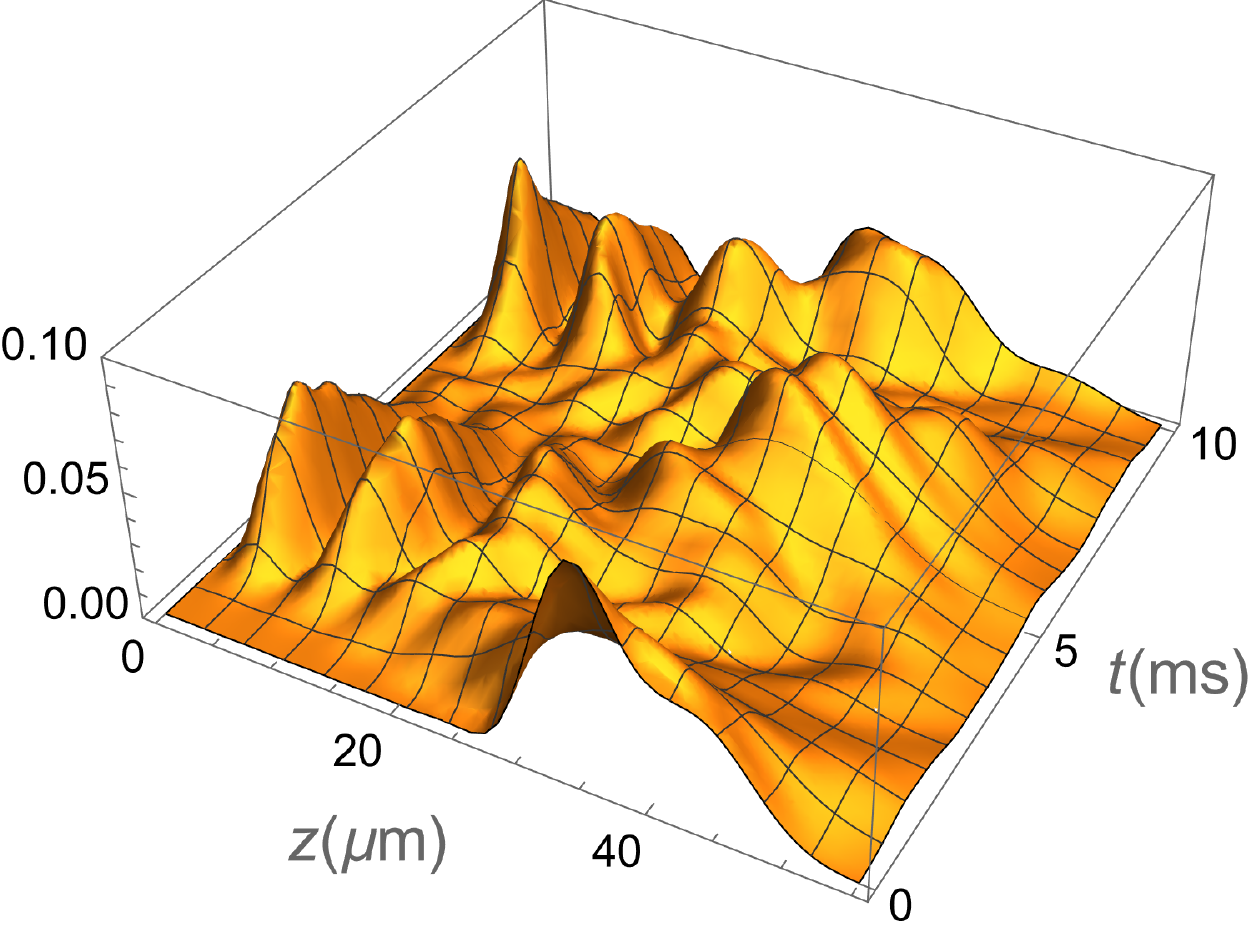}
\caption{Incoherent mixture $|\psi_{\text{mix},\text{II}}^{\text{incoh}}(z,t)|^2$ with $p_1=0.7$ and $p_2=0.3$} \label{incoh73}
\end{subfigure}%
\begin{subfigure}[t]{.49\linewidth}
\centering
\includegraphics[width=60mm]{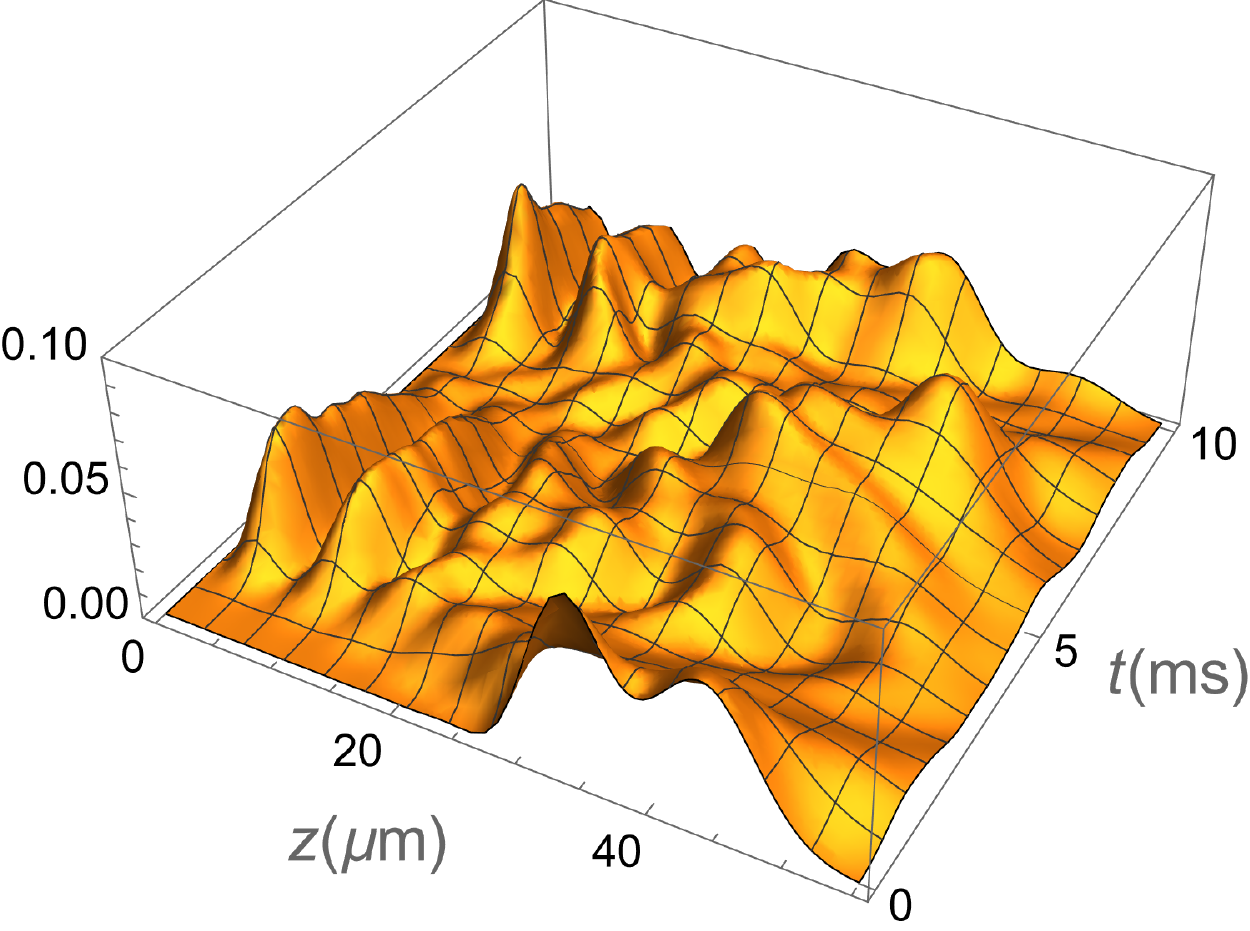}
\caption{Incoherent mixture $|\psi_{\text{mix},\text{II}}^{\text{incoh}}(z,t)|^2$ with $p_1=p_2=0.5$} \label{incoh55}
\end{subfigure}
\caption{Incoherent mixtures, see Eq.~(\ref{incohmix}), in region II}
\end{figure}

Incoherent mixtures can be described by the following formula:
\begin{eqnarray}\label{incohmix}
|\psi_{\text{mix},\text{II}}^{\text{incoh}}(z,t)|^2=p_1|\psi_{1,\text{II}}(z,t)|^2+p_2|\psi_{2,\text{II}}(z,t)|^2\,\,\,.
\end{eqnarray}
In Figs.~\ref{incoh73} and \ref{incoh55} two examples are presented, for which incoherent mixtures can be observed, namely for $p_1=0.7$ and $p_2=0.3$, and $p_1=0.5$ and $p_2=0.5$, respectively. In Fig.~\ref{incoh73}, for example, at $t=0$ the shape of the SD is qualitatively the same as in Fig.~\ref{pmix73}.
Altogether, the oscillations are less distinct than in the previous case.
 
\begin{figure}[h!]
\begin{subfigure}[t]{.49\linewidth}
\centering
\includegraphics[width=60mm]{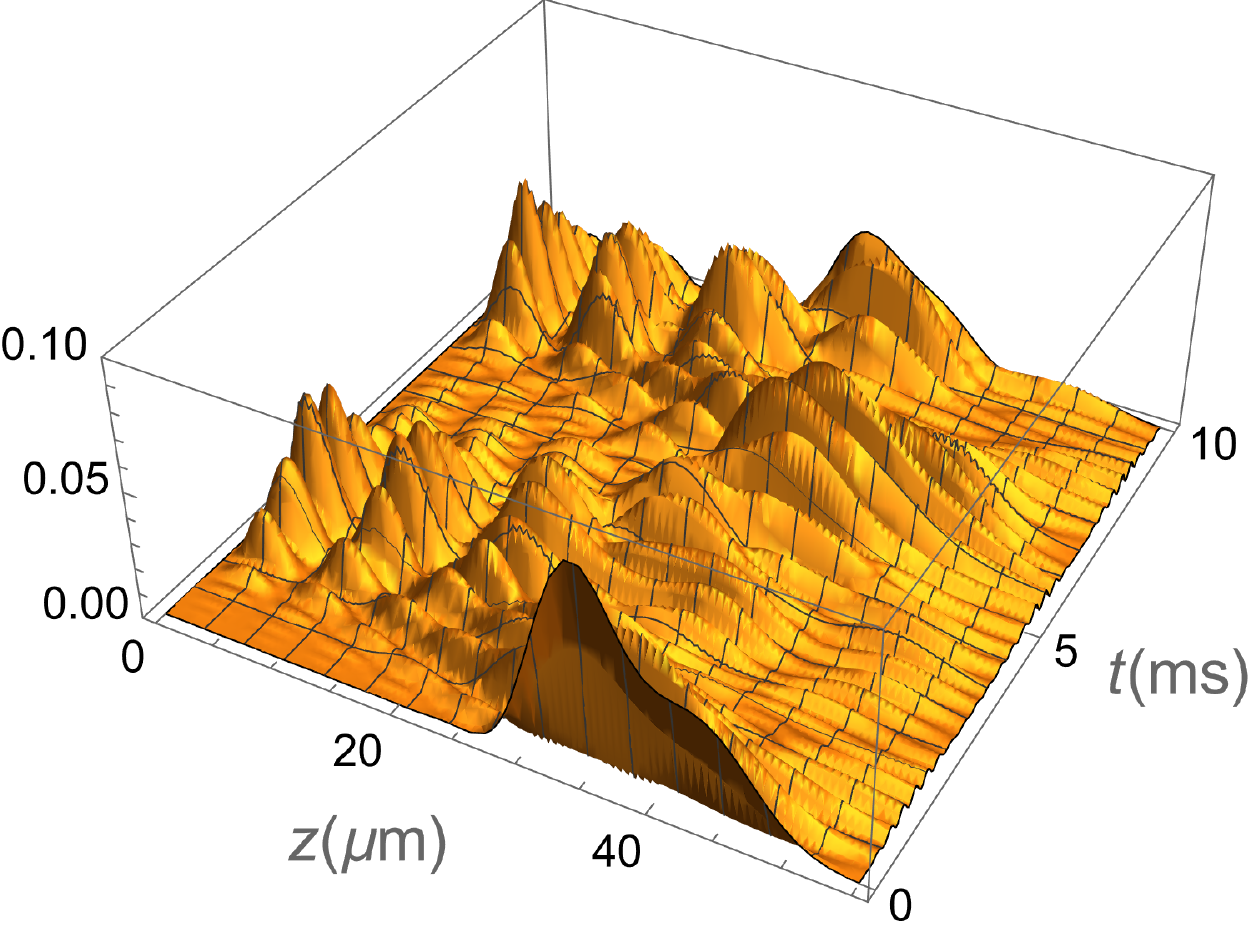}
\caption{Incoherent cosine mixture $p_1|G_1^c(z,t)|^2+p_2|G_2^c(z,t)|^2$} \label{incoh73c}
\end{subfigure}%
\begin{subfigure}[t]{.49\linewidth}
\centering
\includegraphics[width=60mm]{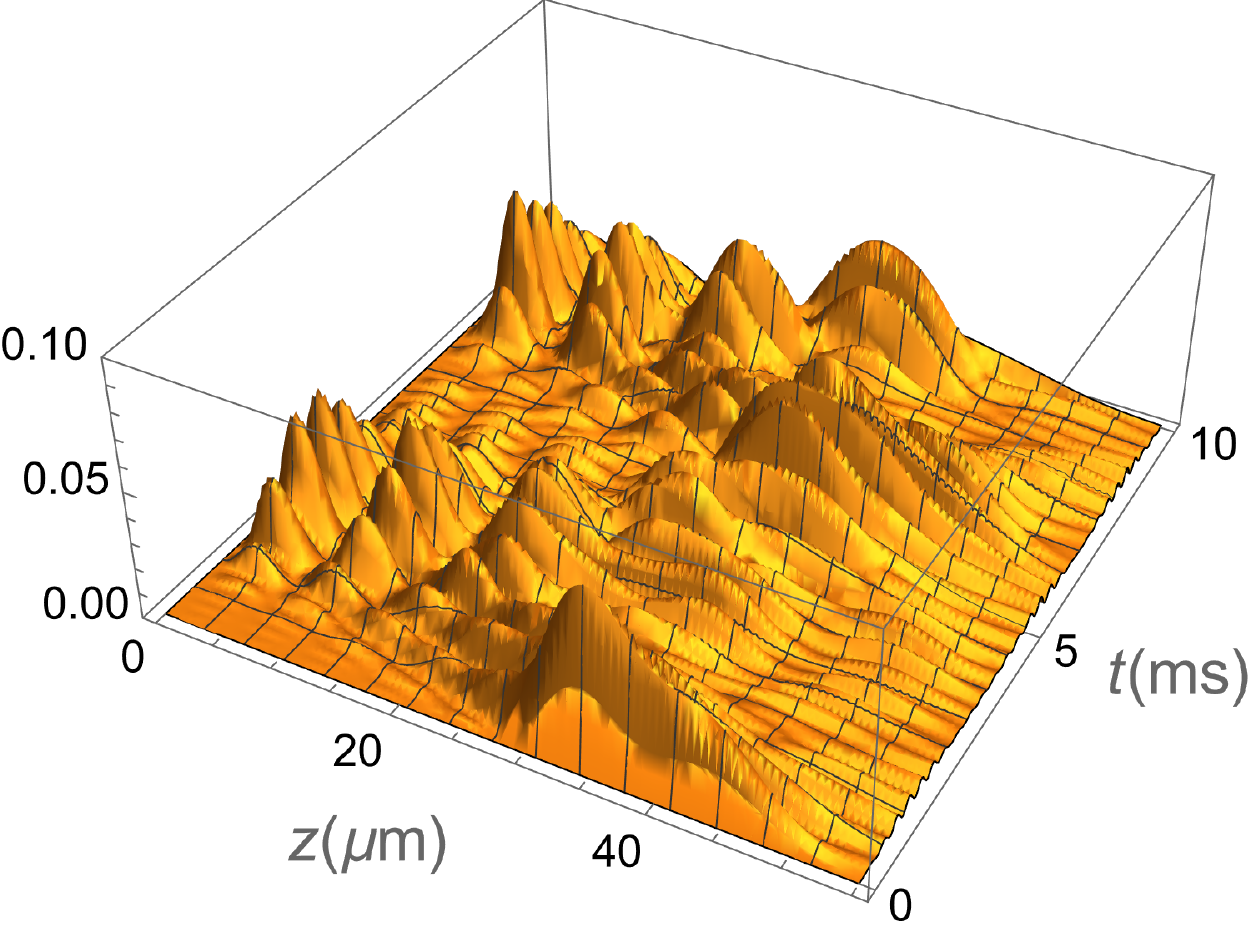}
\caption{Incoherent sine mixture $p_1|G_1^s(z,t)|^2+p_2|G_2^s(z,t)|^2$} \label{incoh73s}
\end{subfigure}
\caption{Incoherent trigonometric mixtures, see Eqs.~(\ref{sdII1}) and (\ref{incohmix}), in region II with $p_1=0.7$ and $p_2=0.3$}
\end{figure}

Figs.~\ref{incoh73c} and \ref{incoh73s} show two examples, in which $|\psi_{\text{mix},\text{II}}^{\text{incoh}}(z,t)|^2$ is separated into a cosine and a sine part. The formulas are included in the corresponding figure captions, and the sum of both of these plots gives the result in Fig.~\ref{incoh73}.

\subsection{Momentum distribution}
\begin{figure}[h!]
\begin{subfigure}[t]{.49\linewidth}
\centering
\includegraphics[width=60mm]{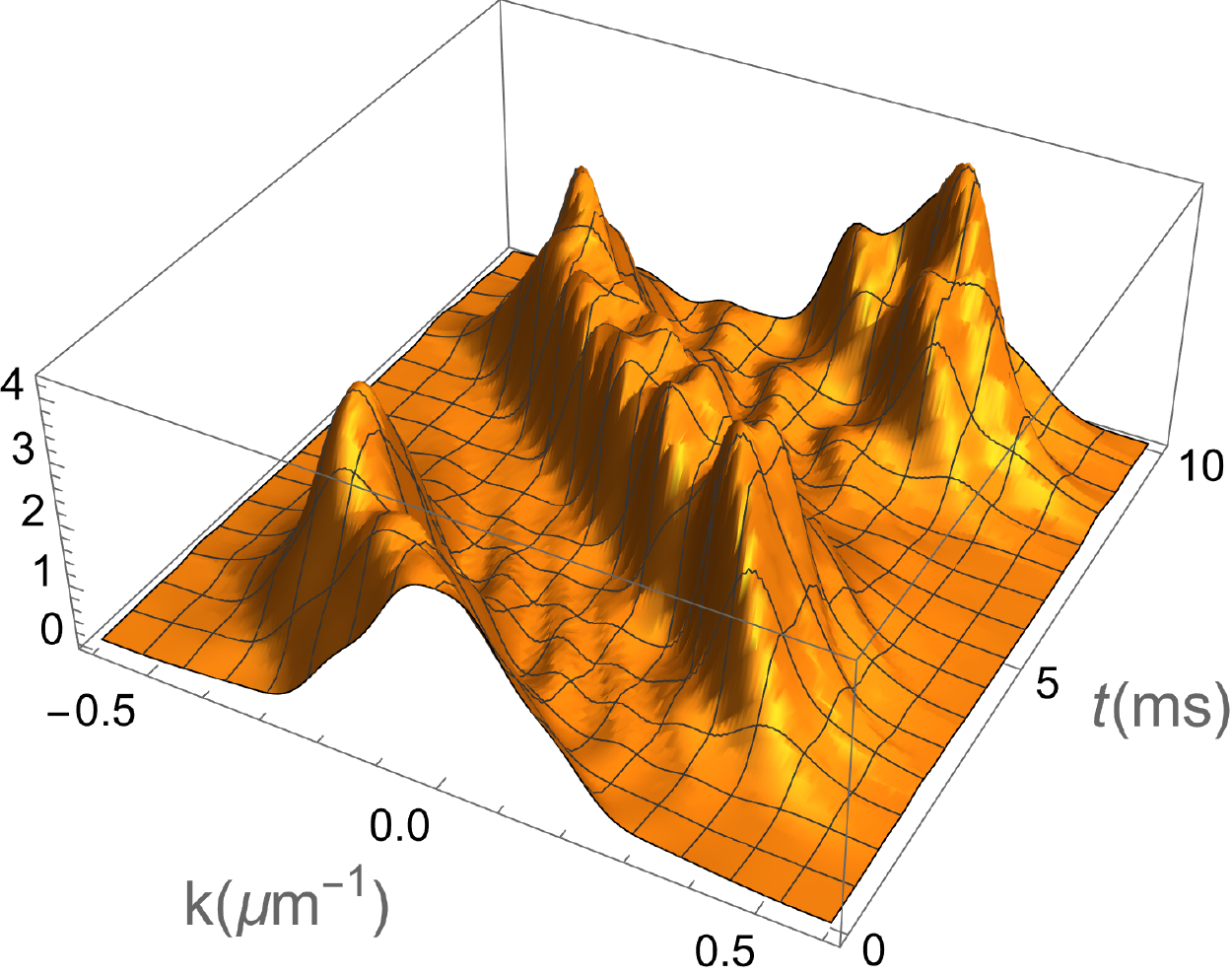}
\caption{Momentum distribution $|F_{1,\text{II}}(k,t)|^2$; Positive and negative $k-$values are included. We can nicely observe the linear increase of momentum with time and the abrupt
sign changes of momentum at the bounces.} \label{FmIIQ1bild}
\end{subfigure}%
\begin{subfigure}[t]{.49\linewidth}
\centering
\includegraphics[width=60mm]{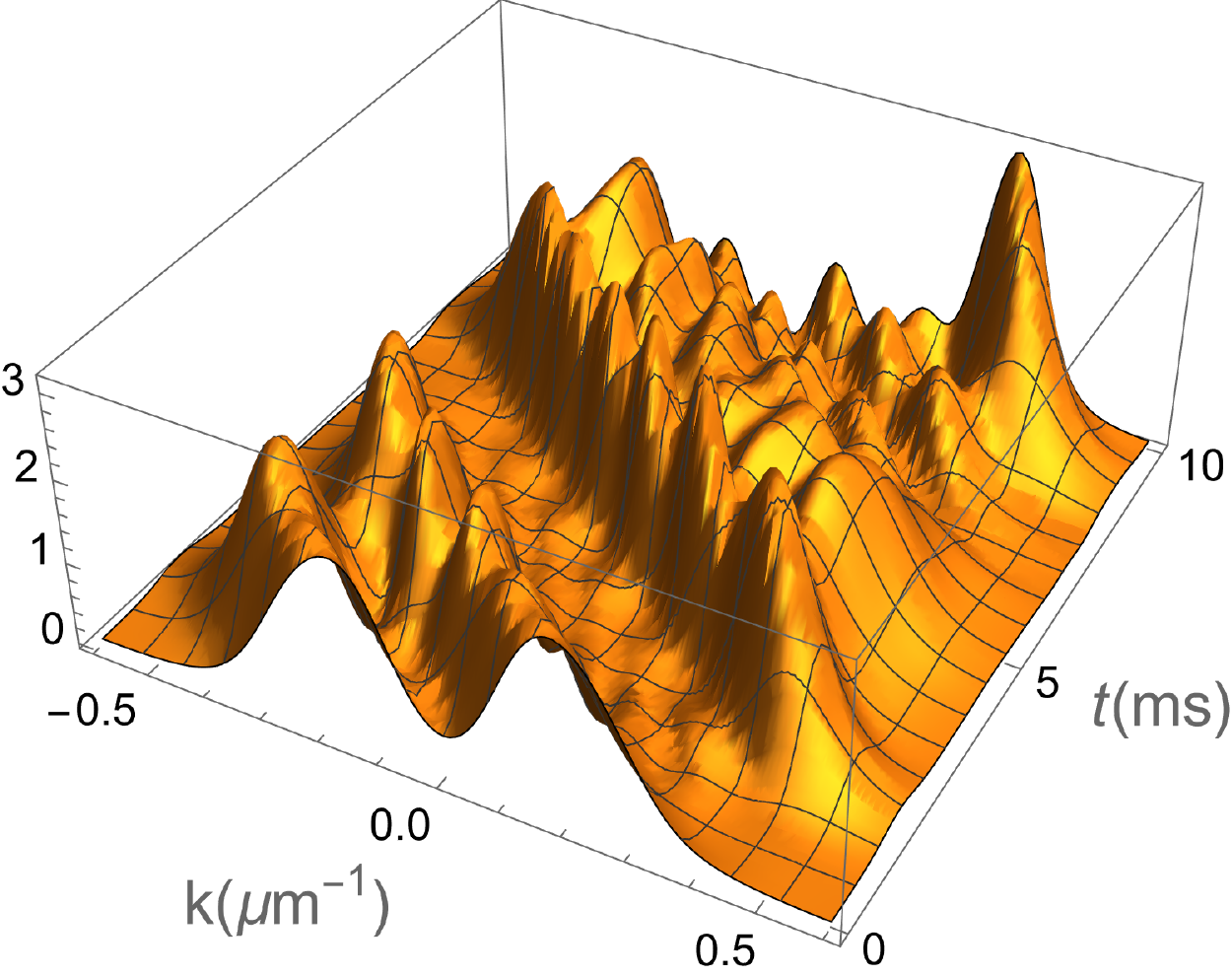}
\caption{Momentum distribution $|F_{2,\text{II}}(k,t)|^2$} \label{FmIIQ2bild}
\end{subfigure}
\caption{Momentum distributions, see Eq.~(\ref{mdIIfinal}), in region II for the ground state and the first excited state}
\end{figure}

The momentum distribution in region II is calculated through a Fourier transformation:
\begin{eqnarray}\label{mdII}
|F_{m,\text{II}}(k,t)|^2&=&\left|\frac{1}{\sqrt{2\pi}}\int_{-\infty}^{\infty}\mathrm{e}^{-ikz}\psi_{m,\text{II}}(z,t)\,\mathrm{d}z\right|^2\nonumber
\\
&=&\left|\frac{1}{\sqrt{2\pi}}\bar{C}_m\sum_{n=1}^{\infty}D_{n,m}\mathrm{e}^{-\frac{i}{\hbar}E_nt}\,[\,f_{c,\text{II}}^{\mathrm{Ai}}(k,n)-i\,f_{s,\text{II}}^{\mathrm{Ai}}(k,n)\,]\,\right|^2\,\,\,, \nonumber
\\
f_{c,\text{II}}^{\mathrm{Ai}}(k,n)&:=&\int_0^{\infty}\cos(kz)\,{\mathrm{Ai}}{\left(\frac{z-z_n}{z_0}\right)}\,\mathrm{d}z\,\,\,,\,\,\,f_{s,\text{II}}^{\mathrm{Ai}}(k,n):=\int_0^{\infty}\sin(kz)\,{\mathrm{Ai}}{\left(\frac{z-z_n}{z_0}\right)}\,\mathrm{d}z\,\,\,.\,\,\,
\end{eqnarray}
This can be expressed as
\begin{eqnarray}\label{mdIIfinal}
|F_{m,\text{II}}(k,t)|^2&=&[\,F_{m,\text{II}}^{\text{Re}}(k,t)\,]^2+[\,F_{m,\text{II}}^{\text{Im}}(k,t)\,]^2\,\,\,,\nonumber
\\
F_{m,\text{II}}^{\text{Re}}(k,t)\,&:=&\frac{\bar{C}_m}{\sqrt{2\pi}}\sum_{n=1}^{\infty}D_{n,m}\left[\,{\cos}{\left(\frac{E_n}{\hbar}t\right)}f_{c,\text{II}}^{\mathrm{Ai}}(k,n)-{\sin}{\left(\frac{E_n}{\hbar}t\right)}f_{s,\text{II}}^{\mathrm{Ai}}(k,n)\right]\,\,,\nonumber
\\
F_{m,\text{II}}^{\text{Im}}(k,t)\,&:=&-\frac{\bar{C}_m}{\sqrt{2\pi}}\sum_{n=1}^{\infty}D_{n,m}\left[\,{\cos}{\left(\frac{E_n}{\hbar}t\right)}f_{s,\text{II}}^{\mathrm{Ai}}(k,n)+{\sin}{\left(\frac{E_n}{\hbar}t\right)}f_{c,\text{II}}^{\mathrm{Ai}}(k,n)\right]\,\,.
\end{eqnarray}
In  Figs.~\ref{FmIIQ1bild} and \ref{FmIIQ2bild} we present two examples of $|F_{m,\text{II}}(k,t)|^2$ for $m=1$ and $m=2$. Fig.~\ref{FmIIQ1bild} shows the time dependence of the ground state momentum distribution in region II. At $t=0$ the function $|F_{1,\text{II}}(k,0)|^2$ is the Fourier transform of the approximated wave function of Fig.~\ref{psig15} squared and has a maximum exactly at $k=0$. 
%This can be proved by differentiating the expression
%\begin{eqnarray}\label{mdIIfinalt=0}
%|F_{1,\text{II}}(k,0)|^2=\left[\,\frac{\bar{C}_1}{\sqrt{2\pi}}\sum_{n=1}^{\infty}D_{n,1}\,f^{\mathrm{Ai}}_{c,\text{II}}(k,n)\right]^2+\left[\,\frac{\bar{C}_1}{\sqrt{2\pi}}\sum_{n=1}^{\infty}D_{n,1}\,f^{\mathrm{Ai}}_{s,\text{II}}(k,n)\right]^2\,\,\,
%\end{eqnarray}
%with respect to $k$ and setting the result equal to $0$. This function is seen in Fig.~\ref{FmIIQ1bild} (if $t=0$) upfront. 
During the evolution of time $t$ the maxima are displaced to larger values of $k$. This causes an inclined periodic $(t-k)$-pattern of the momentum distribution. For the first excited state ($m=2$) the pattern is shifted once more, as can be seen in Fig.~\ref{FmIIQ2bild}. At $t=0$ we can observe a double peak in accordance with the first excited state at the boundary between regions I and II. As expected, there is a distinct minimum at $k=0$.
 
%
%
%It is not difficult to show that the normalization condition 
%\begin{eqnarray}\label{easy}
%\int_{-\infty}^{\infty}|F_{m,II}(k,t)|^2\,\mathrm{d}k=\int_{-\infty}^{\infty}|\psi_{m,II}(z,t)|^2\,\mathrm{d}z=1\,\,
%\end{eqnarray}
%is fulfilled.
%
\subsection{Wigner function }
\begin{figure}[h!]
\begin{subfigure}[t]{.49\linewidth}
\centering
\includegraphics[width=60mm]{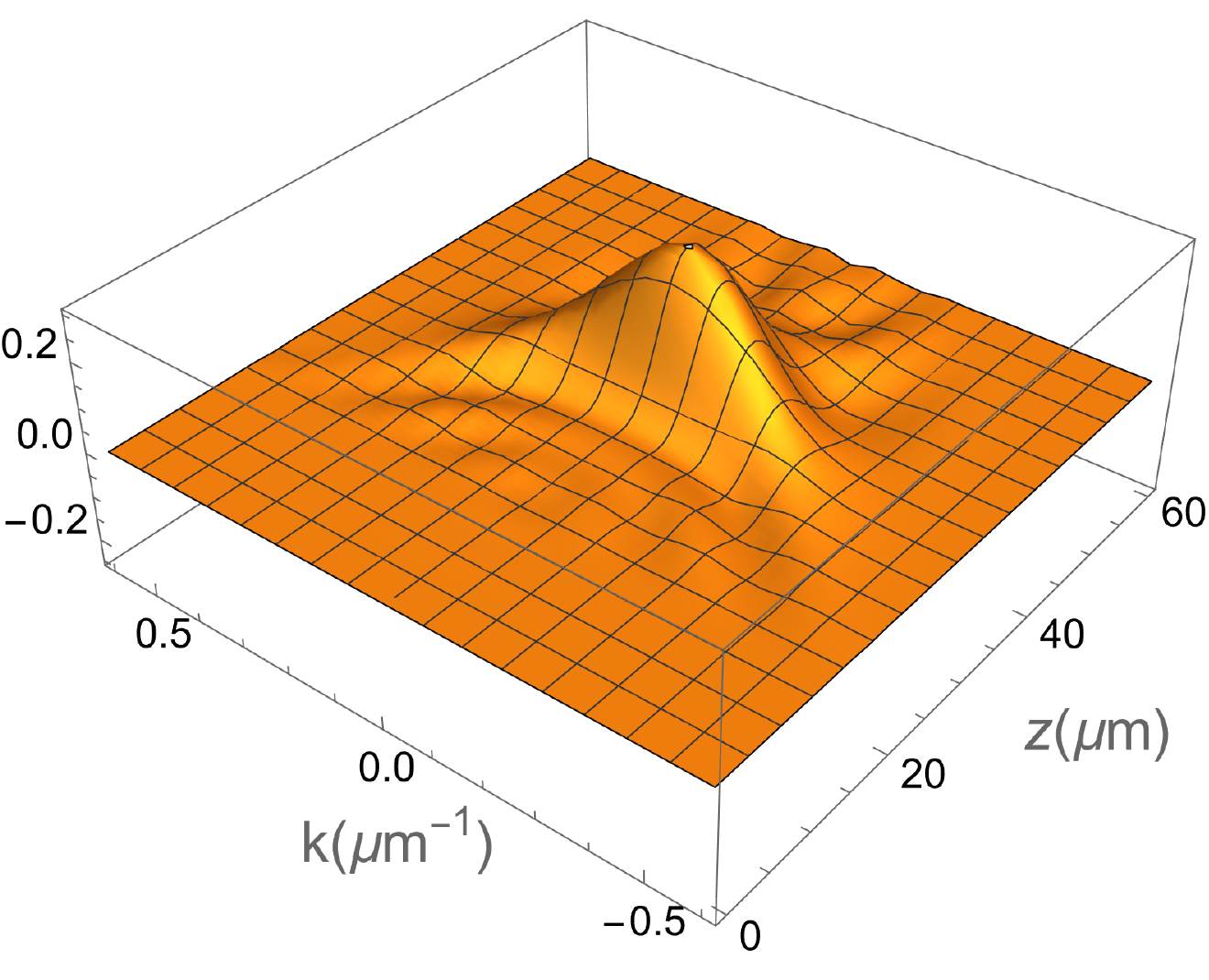}
\caption{$t=\SI{0}{\second}$} \label{WignerIIm1t0Genau}
\end{subfigure}%
\begin{subfigure}[t]{.49\linewidth}
\centering
\includegraphics[width=60mm]{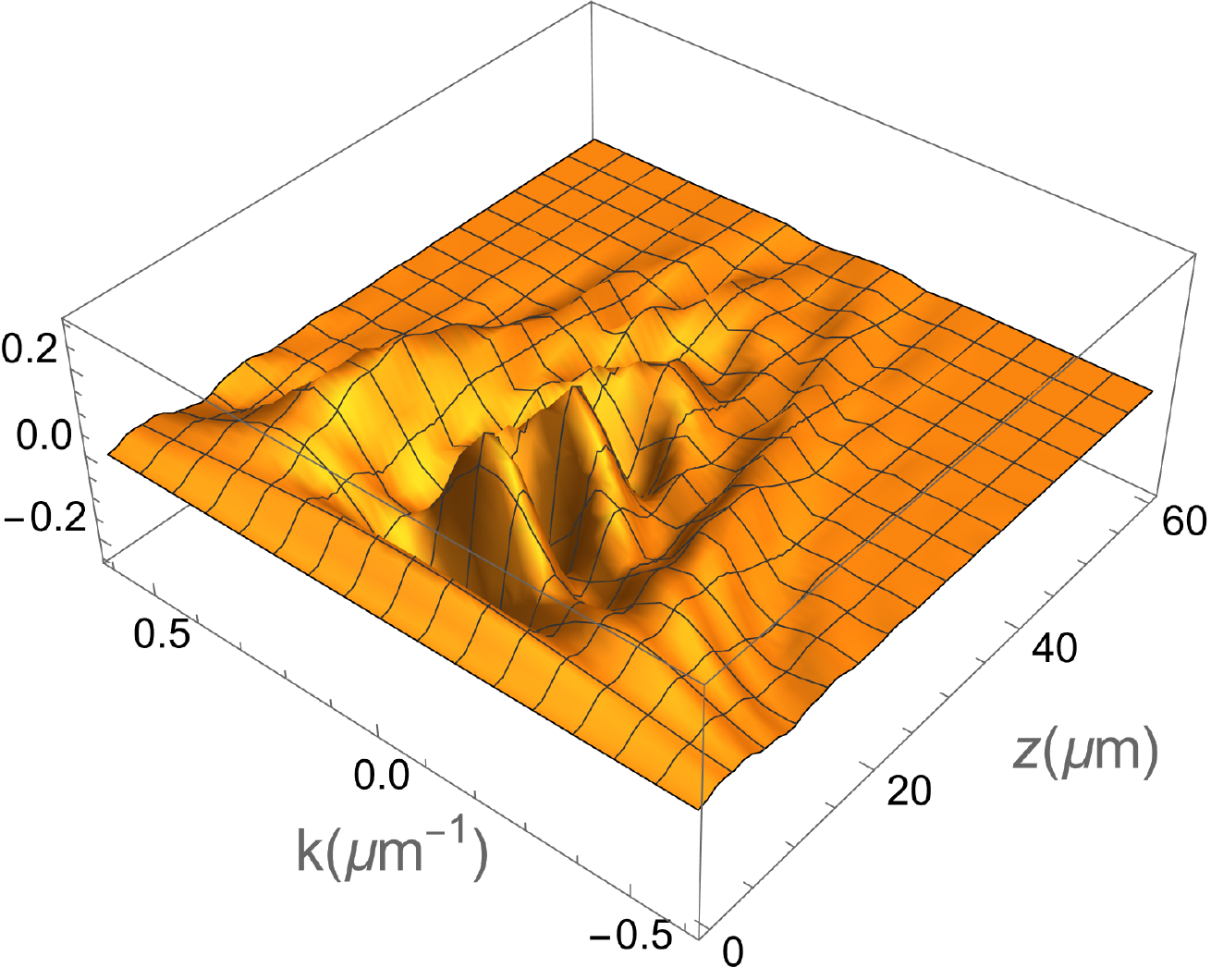}
\caption{$t=\SI{0.003}{\second}$} \label{WignerIIm1t003Genau}
\end{subfigure}
\caption{Evolution of the Wigner function $W_{1,\text{II}}(z,k,0)$ of the ground state in region II, see Eq.~(\ref{WigII})}
\end{figure}

The Wigner function $W_{m,\text{II}}(z,k,t)$ in region II is given in terms of the corresponding wave function $\psi_{m,\text{II}}(z,t)$ from Eq.~(\ref{psiII}):
\begin{eqnarray}\label{WII}
W_{m,\text{II}}(z,k,t)=\frac{1}{2\pi}\int_{-\infty}^{\infty}\mathrm{e}^{iz'k}{\psi^\ast_{m,\text{II}}}{\left(z+\frac{z'}{2},t\right)}{\psi_{m,\text{II}}}{\left(z-\frac{z'}{2},t\right)}\,\mathrm{d}z'\,\,\,,\,\,z\ge{0}\,\,\,.
\end{eqnarray}
Proceeding similarly to how we did in Eq.~(\ref{boundAB}), we find the limits of integration to fulfill $-2z\le{}z'\le{}2z$. Moreover, we can separate the wave function into two parts using Eq.~(\ref{sdII1}):
\begin{eqnarray}\label{psisep}
\psi_{m,\text{II}}(z,t)=G_m^c(z,t)+i\,G_m^s(z,t)\,\,\,.
\end{eqnarray}
After a simple calculation we obtain the following result for the Wigner function in region II:
\begin{eqnarray}\label{WigII}
W_{m,\text{II}}(z,k,t)&=&\frac{1}{\pi}\int_{0}^{2z}\mathrm{d}z'\left\{\cos(kz')\left[{G_m^c}{\left(z+\frac{z'}{2},t\right)}\,{G_m^c}{\left(z-\frac{z'}{2},t\right)}+{G_m^s}{\left(z+\frac{z'}{2},t\right)}\,{G_m^s}{\left(z-\frac{z'}{2},t\right)}\right]\right.\nonumber
\\
&\phantom{=}&+\left.\sin(kz')\left[{G_m^s}{\left(z+\frac{z'}{2},t\right)}\,{G_m^c}{\left(z-\frac{z'}{2},t\right)}-{G_m^c}{\left(z+\frac{z'}{2},t\right)}\,{G_m^s}{\left(z-\frac{z'}{2},t\right)}\right]\right\}\,\,\,.
\end{eqnarray}
Figs.~\ref{WignerIIm1t0Genau} and \ref{WignerIIm1t003Genau} show two examples for $m=1$ at time $t=0$ and $t = \SI{0.003}{\second}$, respectively. At first, we will discuss the case $t=0$. In this case, the Wigner function reads
\begin{eqnarray}\label{WigIIt=0}
W_{1,\text{II}}(z,k,0)&=&\frac{1}{\pi}\int_{0}^{2z}\mathrm{d}z'\cos(kz')\,{G_1^c}{\left(z+\frac{z'}{2},0\right)}\,{G_1^c}{\left(z-\frac{z'}{2},0\right)}\,\,\,,\nonumber
\\
G_1^c(z,0)&=&\frac{\bar{C}_1}{\sqrt{2\pi}}\sum_{n=1}^{\infty}D_{n,1}{\mathrm{Ai}}{\left(\frac{z-z_n}{z_0}\right)}\,\,\,.
\end{eqnarray}
$W_{1,\text{II}}(z,k,0)$ is depicted in Fig.~\ref{WignerIIm1t0Genau}. It is non-vanishing between $z=\SI{27}{\micro\meter}$ and $z=\SI{55}{\micro\meter}$ and should be positive almost everywhere. This Wigner function should be approximately the same as in Fig.~\ref{W1zweiSp}, where only positive $k-$values have been taken into account and the $z-$values are shifted ($0\le{z}\le{}\SI{28}{\micro\meter}$) because no step has been considered there. 
 
At $ t= \SI{0.003}{\second}$ the Wigner function takes on the shape shown in Fig.~\ref{WignerIIm1t003Genau}. It is clearly more complicated than the shape of the ground state in Fig.~\ref{WignerIIm1t0Genau}, and has multiple local extrema. 

%The wave function has arrived at the mirror, and the corresponding spatial distribution $|\psi_{1,\text{II}}(z,t)|^2$ of Fig.~\ref{ortsvert1} has a maximum shortly before $z=0$ and $t=0.003\,s$. At the point $z=0$ the quantum wave is reflected. The Wigner function looks like a turtle or, better expressed, like a trilobite.

\section{Including a Yukawa-like term in region II}
It is interesting to include a small perturbation to the potential in the Schr\"odinger equation in order to simulate a small variation of the gravitational field near the mirror. The basic idea behind this proposal is to modify gravity at small distances and determine the limits on non-Newtonian gravitation below \SI{10}{\micro\meter}.

\subsection{First order perturbation calculation and a new wave function}

When a neutron with mass $m_{\text{N}}$ approaches the mirror, the mass of this extended source might modify the gravitational acceleration of Earth $g$ due to a potentially present non-Newtonian force with range $\delta$, see Ref.~\cite{Abele:2010a}. This modification would lead to an additional, Yukawa-type interaction
\begin{eqnarray}\label{pertW}
W(z)=W_0\,{\mathrm{e}^{-\frac{\zeta}{(\delta/z_0)}}}\,\,\,.
\end{eqnarray}
The parameter $W_0$ is a positive or negative constant with the dimension of an energy and $\delta$ is called Yukawa-distance over which the corresponding force acts. $\delta$ is measured in units of $z_0$.  The stationary  Schr\"odinger equation~(\ref{zeitabh}) reads now
\begin{eqnarray}\label{pert}
\left[-\frac{\hbar^2}{2m_{\text{N}}}\frac{\mathrm d^2}{\mathrm dz^2}+m_{\text{N}}gz+W(z)\right]\Psi_n^{\text{Yu}}(z)
=\epsilon_n\Psi_n^{\text{Yu}}(z)\,\,\,.
\end{eqnarray}
$\Psi_n^{\text{Yu}}(z)$ and $\epsilon_n$ are the corresponding wave functions and energy eigenvalues. The quantity $W(z)$ has to be a small correction to $m_{\text{N}}gz$. In the basis of normalized Airy-functions given in Eq.~(\ref{EFnorm}) we obtain at first order:
\begin{eqnarray}\label{pert1eps}
\epsilon_n&=&E_n+E_n^{(1)}\,\,\,,
\end{eqnarray}
\begin{eqnarray}\label{pert1}
\Psi_n^{\text{Yu}}(z)&=&\psi_n(z)+\psi_n^{(1)}(z)\,\,\,,
\end{eqnarray}
where
\begin{eqnarray}
E_n^{(1)}&=&\left<\psi_n|W|\psi_n\right>=\int_0^{\infty}|\psi_n(z)|^2\,W(z)\,\mathrm{d}z\,\,\,,
\end{eqnarray}
\begin{eqnarray}
\psi_n^{(1)}(z)&=&\sum_{n'\ne{}n}\frac{\left<\psi_{n'}|W|\psi_n\right>}{E_n-E_{n'}}\psi_{n'}(z)
\nonumber
\\
&=&
\sum_{n'\ne{}n}\frac{J_{n',n}}{E_n-E_{n'}}\psi_{n'}(z)
\end{eqnarray}
with
\begin{eqnarray}\label{Yuk4}
J_{n',n}&:=&\int_0^\infty[\psi_{n'}(z')]^*\,W(z')\,\psi_n(z')\,\mathrm{d}z'\,\,\,.
\end{eqnarray}
Note that $\left<\psi_n|\psi_n\right>=1$, $\langle\psi_n|\psi_n^{(1)}\rangle=0$, $\left<\psi_n|\Psi_n^{\text{Yu}}\right>=1$ and $\left<\Psi_n^{\text{Yu}}|\Psi_n^{\text{Yu}}\right>\approx1 $. 

\begin{figure}[h!]
\begin{subfigure}[t]{.49\linewidth}
\centering
\includegraphics[width=80mm]{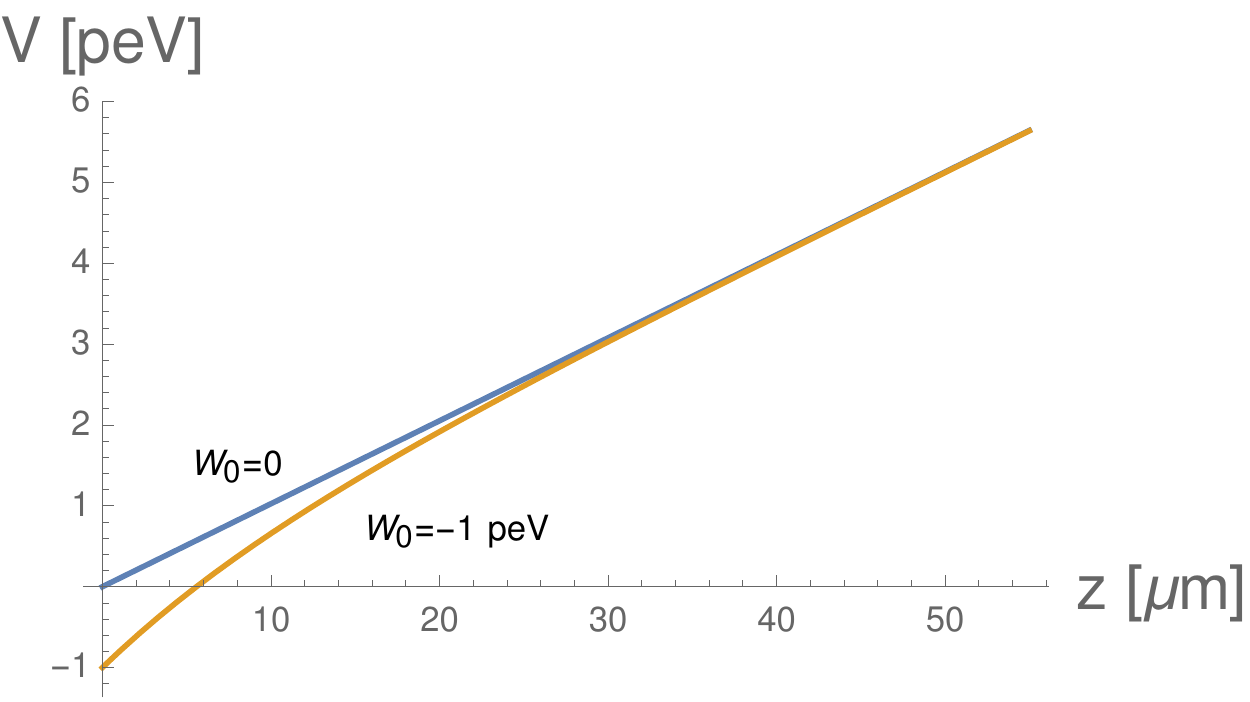}
\caption{Total potential $V(z)$} \label{potential}
\end{subfigure}%
\begin{subfigure}[t]{.49\linewidth}
\centering
\includegraphics[width=80mm]{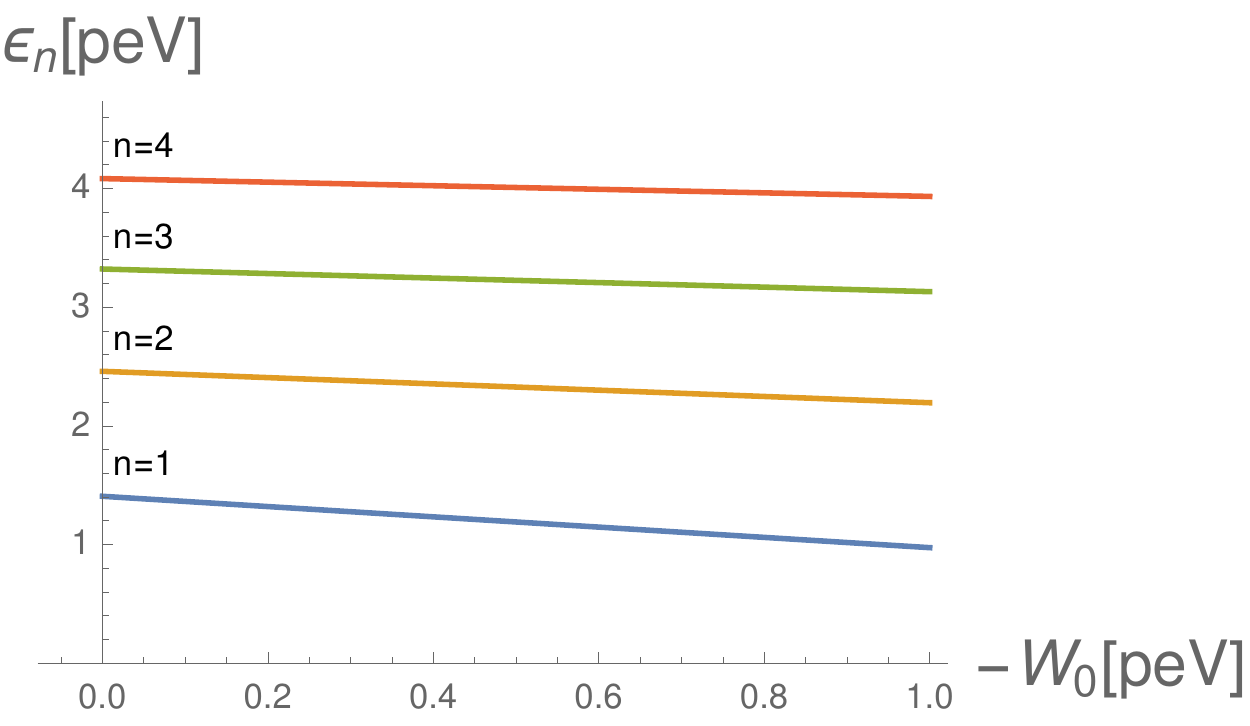}
\caption{Energy eigenvalues $\epsilon{}_n$ according to Eq.~(\ref{pert1eps})} \label{eigenwerte}
\end{subfigure}
\caption{Total potential $V(z)=m_{\text{N}}gz+W(z)$ and energy eigenvalues $\epsilon_n(W_0)$ for a Yukawa-distance $\delta=\SI{10}{\micro\meter}$ }
\end{figure}

In  Fig.~\ref{potential} the total potential $V(z)=m_{\text{N}}gz+W(z)$ is drawn for $W_0=0$ and for an additional attractive potential with the strength $W_0=-\SI{1}{\pico\electronvolt}$. Since the total force is therefore given by  $\vec{K}(z)=-\vec{\nabla} V(z)=[-mg+\frac{W_0}{\delta}\,e^{-z/\delta}]\,\boldsymbol{e}_z$, we can see that the Yukawa-potential leads to an additional force. Fig.~\ref{eigenwerte} shows $\epsilon_n$ as a function of $-W_0$. The eigenvalues $\epsilon_n$ decrease with increasing $W_0$. 

Next, we write the eigenfunctions~(\ref{pert1}) in the form
\begin{eqnarray}\label{Dss1}
\Psi_n^{\text{Yu}}(z)=\psi_n(z)+
\sum_{n'\ne{}n}\frac{J_{n',n}}{E_n-E_{n'}}\psi_{n'}(z)=:
\sum_{n'}\psi_{n'}(z)\,\,T_{n',n}\,\,\,.
\end{eqnarray}
$T_{n',n}$ is a unit matrix but with small off-diagonal terms proportional to $W_0$ that represent the small numbers $J_{n',n}$.

For simplicity, we neglect the modifications of the neutron wave function at $t=0$ by the Yukawa interaction in region I and consequently assume that the wave function in region II resembles the one in Eq.~(\ref{psiII}), but with a yet unknown form of $\psi_{n'}(z)$:
\begin{eqnarray}\label{unmod}
\psi_{m,\text{II}}^{\text{Yu}}(z)&=&\bar{C}_m\sum_{n'}\psi_{n'}(z)\,\bar{D}_{n',m}\quad\textrm{with}\quad
\bar{D}_{n',m}:=D_{n',m}\,{\sqrt{z_0}{\mathrm{Ai'}\left(-\frac{z_{n'}}{z_0}\right)}}\,\,\,.
\end{eqnarray}
From Eq.~(\ref{Dss1}) we can extract $\psi_{n'}(z)$ by means of the following approach: multiplying from the right by $T^{-1}_{n,n''}$ and summing over $n$ we obtain
\begin{eqnarray}\label{unmod1}
\sum_{n}\Psi_n^{\text{Yu}}(z)T^{-1}_{n,n''}=\sum_{n'}\psi_{n'}(z)\sum_{n}T_{n',n}T^{-1}_{n,n''}=\sum_{n'}\psi_{n'}(z)\,\delta_{n',n''}=\psi_{n''}(z)\,\,\,.
\end{eqnarray}
Inserting this relation into Eq.~(\ref{unmod}), we get the modification of Eq.~(\ref{psiII}) for Yukawa forces
\begin{eqnarray}\label{Yuka4ddd}
\psi_{m,\text{II}}^{\text{Yu}}(z)=\bar{C}_m\sum_{n,n'}\Psi_n^{\text{Yu}}(z)T^{-1}_{n,n'}\bar{D}_{n',m}\,\,\,.
%=\overline{C}_m\sum_{\text{N}}\Psi^{\text{Yu}}_n(z)\,D^{\text{Yu}}_{n,m}
\end{eqnarray}
%with the modified expansion coefficients
%\begin{eqnarray}\label{DSS5}
%D^{\text{Yu}}_{n,m}:=\sum_{n'}T^{-1}_{n,n'}\,\overline{D}_{n',m}\,\,\,.
%\end{eqnarray}
Taking into account the time dependence of the eigenfunctions from Eq.~(\ref{Dss1}) in region II
\begin{eqnarray}\label{TDII}
\Psi_n^{\text{Yu}}(z,t)=\Psi_n^{\text{Yu}}(z)\,\exp[-\frac{i}{\hbar}\epsilon_nt]\,\,\,,
\end{eqnarray}
we get the time evolution of the perturbed wave function from Eq.~(\ref{Yuka4ddd}) in region II:
\begin{eqnarray}\label{Yuka4e}
\psi_{m,\text{II}}^{\text{Yu}}(z,t)&=&\bar{C}_m\sum_{n',n''}\psi_{n'}(z)\;\Big\{\sum_{n}T_{n',n}\,\exp[-\frac{i}{\hbar}\epsilon_nt]\;T^{-1}_{n,n''}\Big\}\;\bar{D}_{n'',m}\,\,\,.
\end{eqnarray}
For vanishing Yukawa forces, $T_{n',n}=\delta_{n',n}$ and $\epsilon_n=E_n$, this equation reduces to Eq.~(\ref{psiII}).

%****************************************************************
%
\subsection{Space distribution with Yukawa correction in region II}
We separate the real and imaginary parts of the wave function in Eq.~(\ref{Yuka4e}):
\begin{eqnarray}\label{YukReImpsi}
\text{Re}_{\psi}&=&\bar{C}_m\sum_{n',n''}\psi_{n'}(z)\;\Big\{\sum_{n}T_{n',n}\,\cos\Big(\frac{\epsilon_n}{\hbar}t\Big)\;T^{-1}_{n,n''}\Big\}\;\bar{D}_{n'',m}\,\,\,,\nonumber\\
\text{Im}_{\psi}&=&\bar{C}_m\sum_{n',n''}\psi_{n'}(z)\;\Big\{\sum_{n}T_{n',n}\,\sin\Big(\frac{\epsilon_n}{\hbar}t\Big)\;T^{-1}_{n,n''}\Big\}\;\bar{D}_{n'',m}\,\,\,,
\end{eqnarray}
such that
\begin{eqnarray}\label{YukaAbso}
|\psi_{m,\text{II}}^{\text{Yu}}(z,t)|^2&=&\text{Re}_{\psi}^2+\text{Im}_{\psi}^2\,\,\,.
\end{eqnarray}
An interesting way of comparing space distributions with and without Yukawa interaction is the following quantity: 
\begin{eqnarray}\label{YukDiff}
\Delta_\text{Yu}(z,t):=|\psi_{m,\text{II}}^{\text{Yu}}(z,t)|^2-|\psi_{m,\text{II}}(z,t)|^2\,\,\,.
\end{eqnarray}
Here $|\psi_{m,\text{II}}(z,t)|^2$ is given in Eq.~(\ref{sdII2}) or, equivalently, by Eq.~(\ref{YukaAbso}) when setting $W_0=0$.

\begin{figure}[h!]
\begin{subfigure}{.49\linewidth}
\includegraphics[width=60mm]{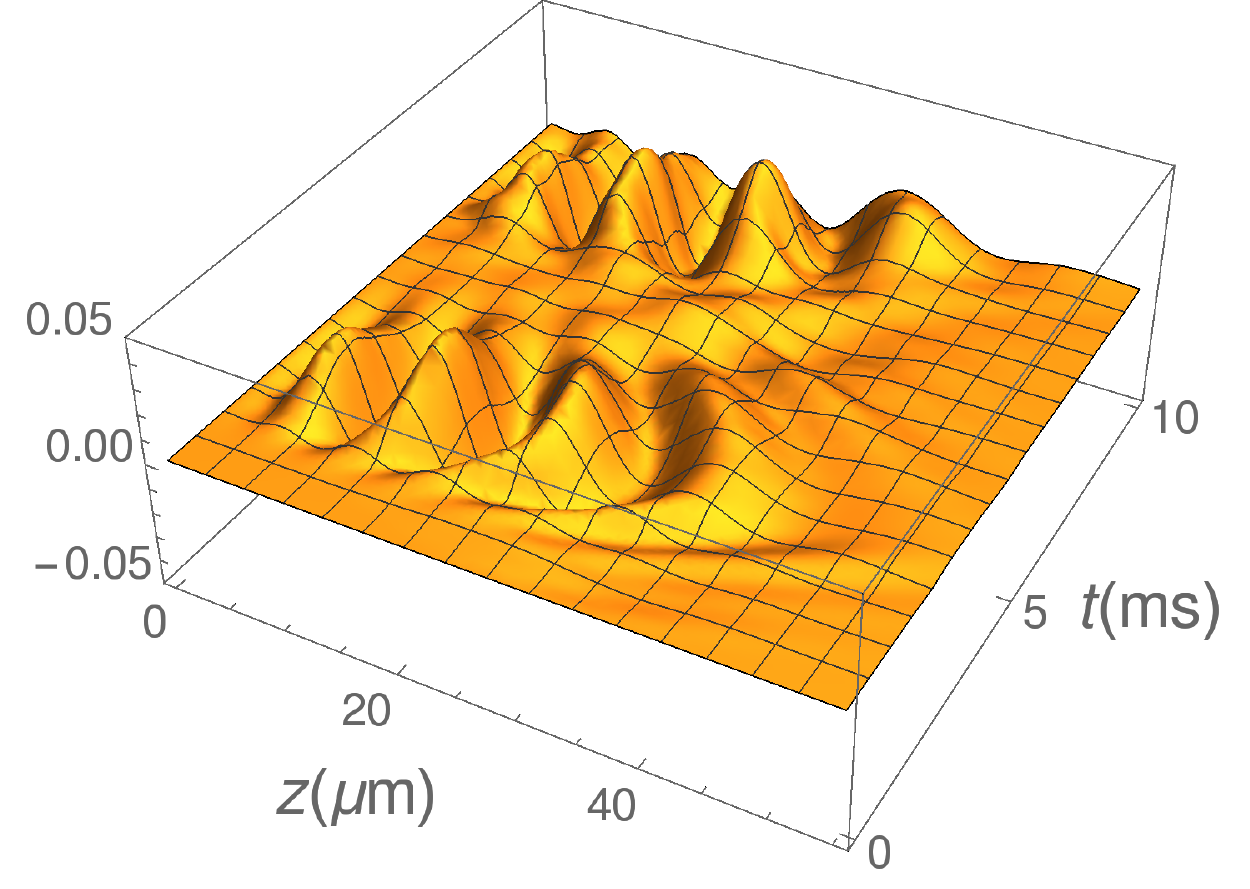}
\caption{$\Delta_\text{Yu}(z,t)$, see Eq.~(\ref{YukDiff})} \label{sdyukdiff1}
\end{subfigure}
\begin{subfigure}{.49\linewidth}
\includegraphics[width=60mm]{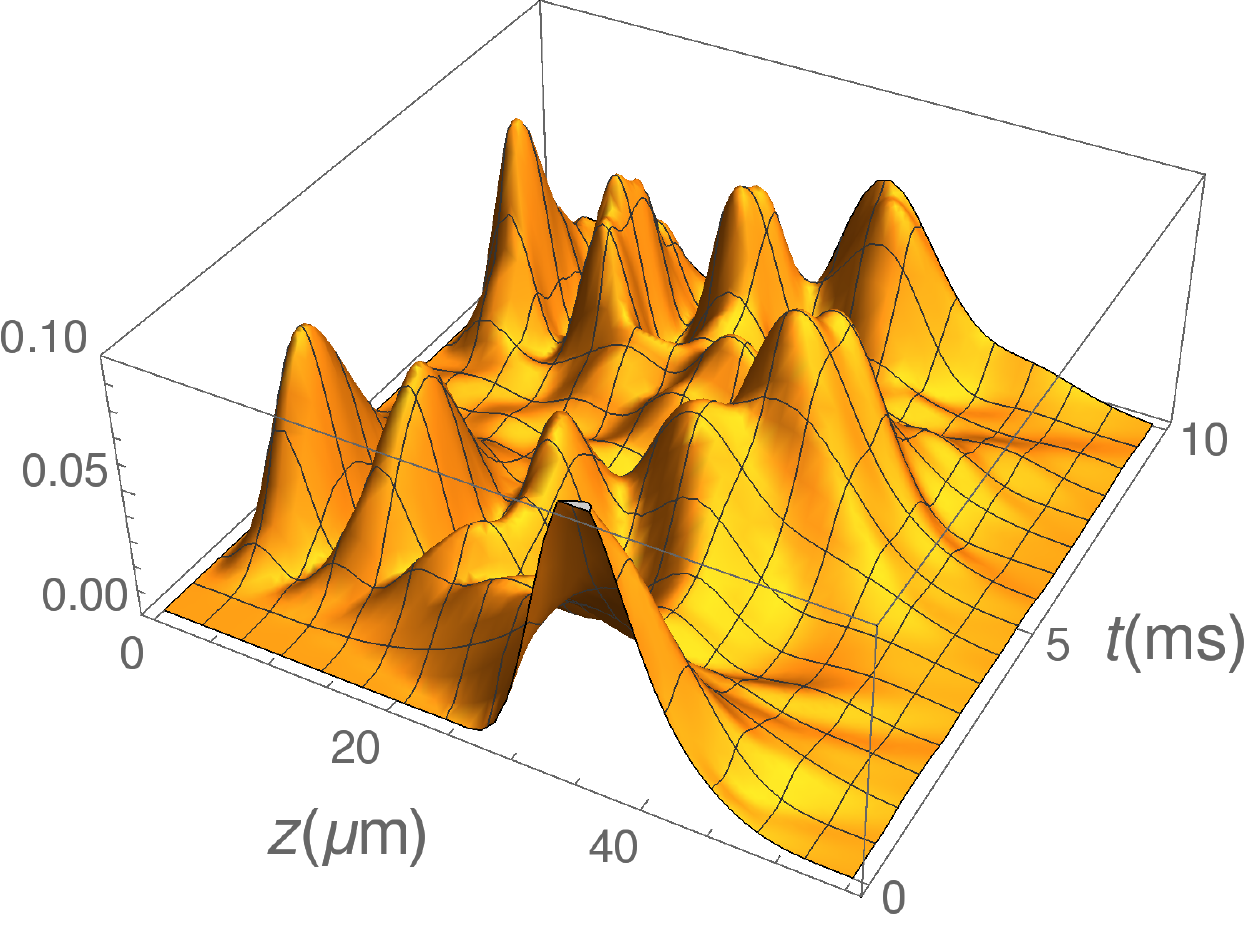}
\caption{$|\psi_{m,\text{II}}^{\text{Yu}}(z,t)|^2$, see Eq.~(\ref{YukaAbso})} \label{sdyukges1}
\end{subfigure}
\caption{Difference in spatial distributions with and without Yukawa interaction, and spatial distribution with Yukawa interaction; $m=1$,  $\delta=\SI{10}{\micro\meter}$ and $W_0=-\SI{1}{\pico\electronvolt}$}
\end{figure}

In  Fig.~\ref{sdyukdiff1} we assume $W_0=-\SI{1}{\pico\electronvolt}$ and draw the corresponding $\Delta_\text{Yu}(z,t)$. For comparison, we also depicted $|\psi_{m,\text{II}}^{\text{Yu}}(z,t)|^2$ in Fig.~\ref{sdyukges1}. For $t=0$ this function is exactly the same as in Fig.~\ref{ortsvert1}. However, for $t>0$, small differences to the function shown in Fig.~\ref{ortsvert1} can be noticed. 
\subsection{Momentum distribution of Yukawa correction in region II}
Next, we want to look at the momentum distribution and therefore consider the Fourier transform of the wave function in Eq.~(\ref{Yuka4e}), which is given by (compare with Eq.~(\ref{mdII})) 
%\begin{eqnarray}\label{FTpsi}
%FT[\psi_{n'}(z)]&=&\frac{1}{\sqrt{2\pi}\sqrt{z_0}Ai'\Big(-\frac{z_{n'}}{z_0}\Big)}\Big[f^{Ai}_{c,II}(k,n')-if^{Ai}_{s,II}(k,n')\Big]\;,
%\end{eqnarray}
\begin{eqnarray}\label{YukaF1}
%FT\Big[\psi_{m,\text{II}}^{Yu}(z,t)\Big]=
F_{m,\text{II}}^{\text{Yu}}(k,t)
&=&\frac{\bar{C}_m}{\sqrt{2\pi}\sqrt{z_0}}\sum_{n',n''}\frac{1}{\text{Ai}'\Big(-\frac{z_{n'}}{z_0}\Big)}\Big[\,f^{\text{Ai}}_{c,\text{II}}(k,n')-if^{\text{Ai}}_{s,\text{II}}(k,n')\,\Big]
\nonumber
\\
&\phantom{=}&
\phantom{\frac{\bar{C}_m}{\sqrt{2\pi}\sqrt{z_0}}\sum_{n',n''}\frac{1}{\text{Ai}'\Big(-\frac{z_{n'}}{z_0}\Big)}}
\times
\Big\{\sum_{\text{N}}T_{n',n}\Big[\,\cos\Big(\frac{\epsilon_n}{\hbar}t\Big)-i\sin\Big(\frac{\epsilon_n}{\hbar}t\,\Big)\Big]\}T^{-1}_{n,n''}\Big\}\bar{D}_{n'',m}\,\,\,.
\end{eqnarray}
The real and the imaginary part of this expression are
\begin{eqnarray}\label{YukReImF}
\text{Re}_{F}&=&\frac{\bar{C}_m}{\sqrt{2\pi}\sqrt{z_0}}\sum_{n',n''}\frac{1}{\text{Ai}'\Big(-\frac{z_{n'}}{z_0}\Big)}
\Big\{\,f^{\text{Ai}}_{c,\text{II}}(k,n')\sum_{n}T_{n',n}\,\cos\Big(\frac{\epsilon_n}{\hbar}t\Big)\;T^{-1}_{n,n''}
\nonumber
\\
&\phantom{=}&
\phantom{\frac{\bar{C}_m}{\sqrt{2\pi}\sqrt{z_0}}\sum_{n',n''}\frac{1}{\text{Ai}'\Big(-\frac{z_{n'}}{z_0}\Big)}}
-f^{\text{Ai}}_{s,\text{II}}(k,n')\sum_{n}T_{n',n}\,\sin\Big(\frac{\epsilon_n}{\hbar}t\Big)\;T^{-1}_{n,n''}\,\Big\}\bar{D}_{n'',m}\,\,\,,
\nonumber
\\
\text{Im}_{F}&=&-\frac{\bar{C}_m}{\sqrt{2\pi}\sqrt{z_0}}\sum_{n',n''}\frac{1}{\text{Ai}'\Big(-\frac{z_{n'}}{z_0}\Big)}\Big\{\,f^{\text{Ai}}_{s,\text{II}}(k,n')\sum_{n}T_{n',n}\,\cos\Big(\frac{\epsilon_n}{\hbar}t\Big)\;T^{-1}_{n,n''}
\nonumber
\\
&\phantom{=}&
\phantom{-\frac{\bar{C}_m}{\sqrt{2\pi}\sqrt{z_0}}\sum_{n',n''}\frac{1}{\text{Ai}'\Big(-\frac{z_{n'}}{z_0}\Big)}}
+f^{\text{Ai}}_{c,\text{II}}(k,n')\sum_{n}T_{n',n}\,\sin\Big(\frac{\epsilon_n}{\hbar}t\Big)\;T^{-1}_{n,n''}\,\,\Big\}\bar{D}_{n'',m}\,\,\,,
\end{eqnarray}
such that
\begin{eqnarray}\label{FYu2}
|F_{m,\text{II}}^{\text{Yu}}(k,t)|^2&=&\text{Re}_{F}^2+\text{Im}_{F}^2\,\,\,.
\end{eqnarray}
Setting $W_0=0$ in Eq.~(\ref{YukReImF}) recovers Eq.~(\ref{mdIIfinal}), which gives the expression for $|F_{m,\text{II}}(k,t)|^2$.

An interesting way of comparing the momentum distributions with and without Yukawa interaction is the following quantity: 
\begin{eqnarray}\label{YukDiffM}
\Delta_{\text{Yu}}(k,t)=|F_{m,\text{II}}^{\text{Yu}}(k,t)|^2-|F_{m,\text{II}}(k,t)|^2\,\,\,.
\end{eqnarray}

\begin{figure}[h!]
\begin{subfigure}{.49\linewidth}
\includegraphics[width=60mm]{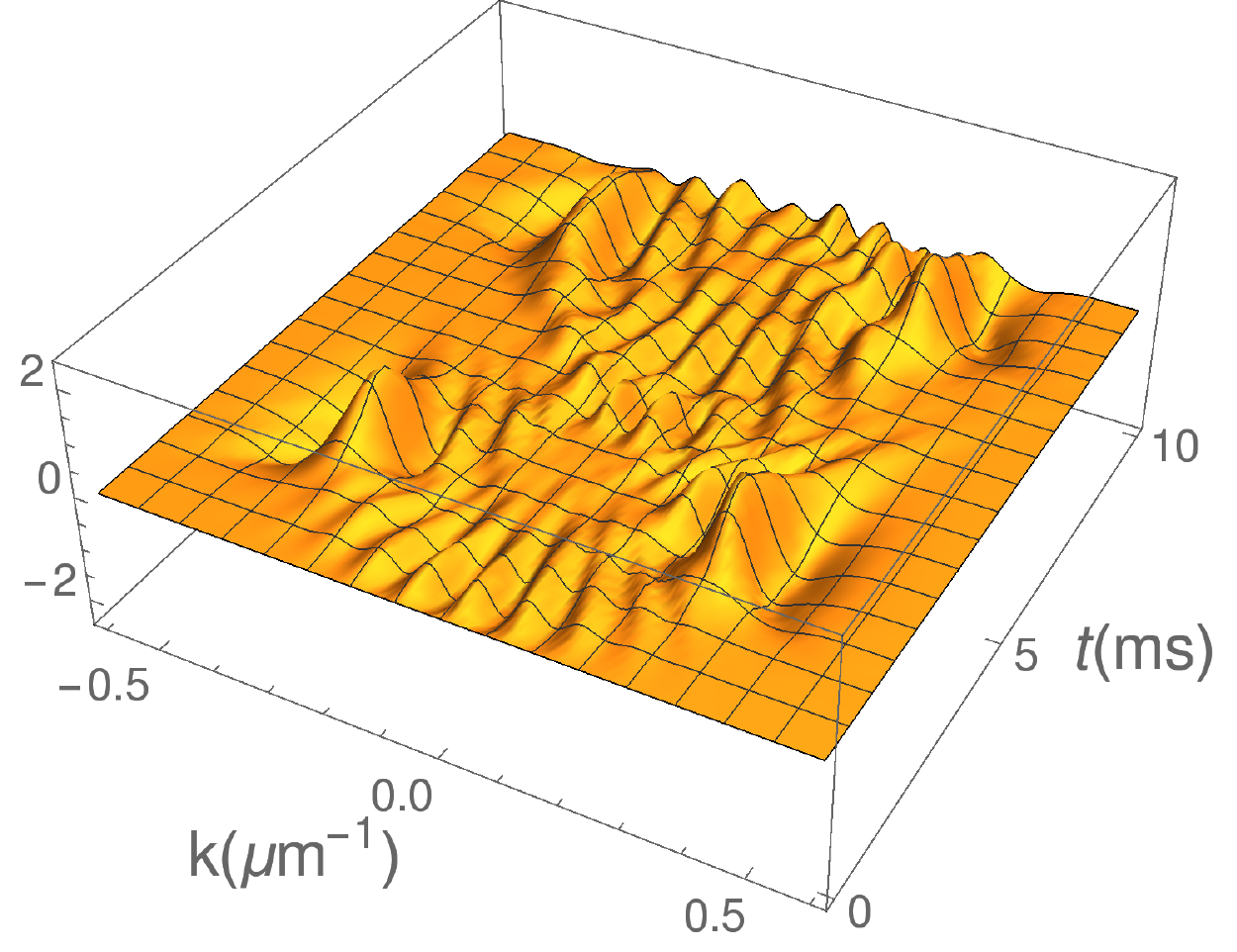}
\caption{$\Delta_\text{Yu}(k,t)$, see Eq.~(\ref{YukDiffM})} \label{impyukdiff1}
\end{subfigure}
\begin{subfigure}{.49\linewidth}
\includegraphics[width=60mm]{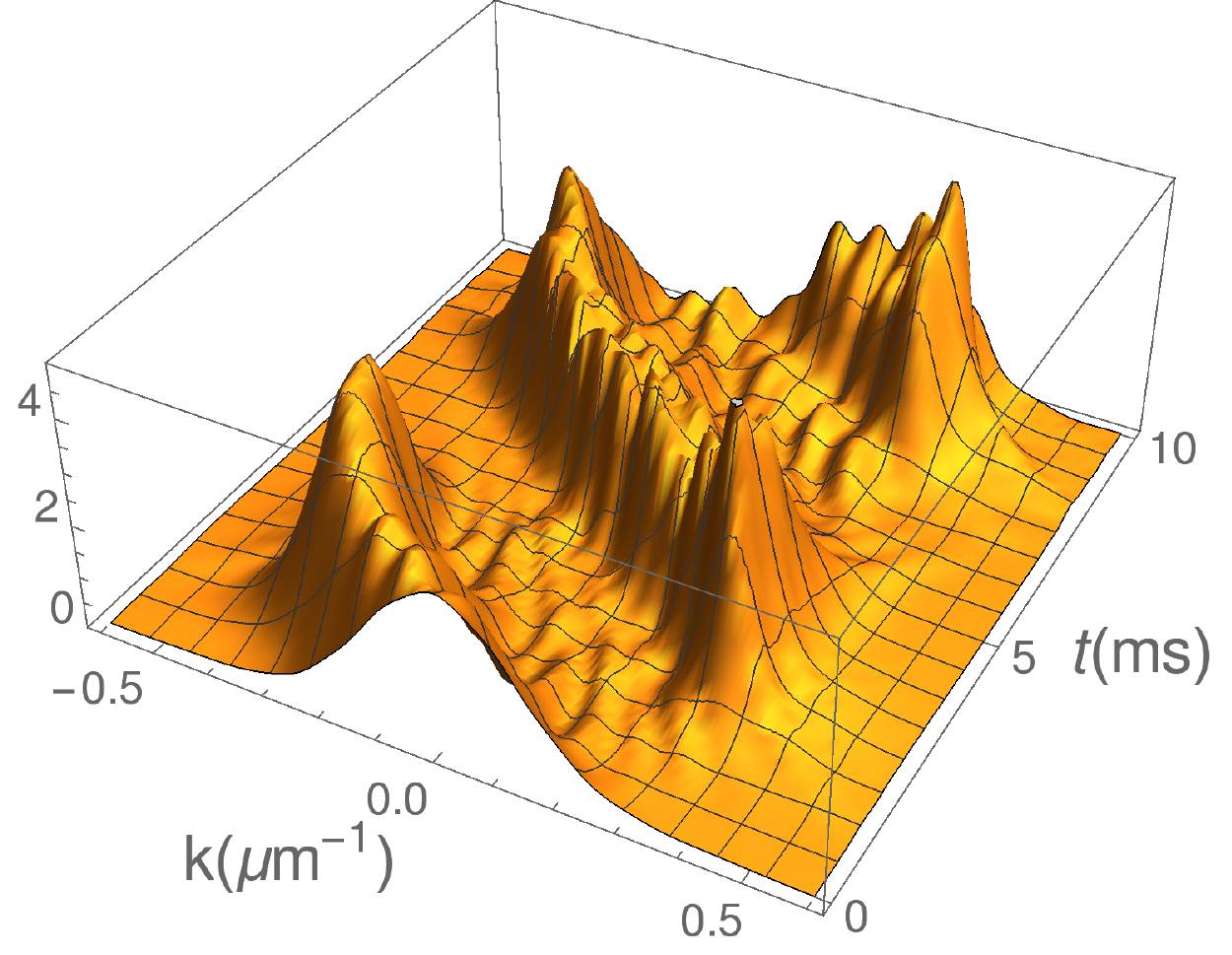}
\caption{$|F_{m,\text{II}}^{\text{Yu}}(k,t)|^2$, see Eq.~(\ref{FYu2})} \label{impyukges1}
\end{subfigure}
\caption{Difference in momentum distributions with and without Yukawa interaction, and momentum distribution with Yukawa interaction; $m=1$,  $\delta=\SI{10}{\micro\meter}$ and $W_0=-\SI{1}{\pico\electronvolt}$}
\end{figure}

In Fig.~\ref{impyukdiff1} this difference of the momentum distributions with and without Yukawa interaction is depicted. At the jump discontinuities $t\approx\SI{3}{\milli\second}$ and $t\approx\SI{9}{\milli\second}$ distinct differences are visible. At these times the wave function is reflected at the mirror where the Yukawa potential is the strongest. This can also be observed in Fig.~\ref{impyukges1}, in which $|F_{m,\text{II}}^{\text{Yu}}(k,t)|^2$ is shown. At these points in time $|F_{m,\text{II}}^{\text{Yu}}(k,t)|^2$ exhibits distinct maxima compared to the momentum distribution without Yukawa interaction depicted in Fig.~\ref{FmIIQ1bild}. %Obviously one can somewhat better observe  differences in momentum distribution compared to space distribution. 
\section{Conclusion}
%------------------------------------------------------------------------------
In this theoretical treatise the wave function of the \textit{q}\textsc{Bounce} experiment has been investigated in detail. The gravitational field of the Earth constitutes the potential used in the Schr\"odinger equation. This yields solutions for the wave function which correspond to the Airy function. Since the wave function has to vanish at the mirror surface, a ground state and excited states evolve. These states have been analyzed with respect to spatial and momentum distributions. For this purpose, the Wigner function has also been used. It was shown that the distribution spectra of the ground and excited states exhibit the anticipated properties, which are reproduced in the marginal distribution functions. Furthermore, the time dependence of a mixture of the ground state and the first excited state has been considered. The \textit{q}\textsc{Bounce}-problem in which the neutron wave is enclosed between 2 mirrors has been analyzed as well. Finally, the case where the wave function exits the double mirror system and freely falls on a subsequent mirror has been considered. The purpose of these calculations was to motivate measurements both in real space and in momentum space for comparison between experimental findings and theoretical results. Finally and in addition, we made an attempt at a first order perturbation calculation in order to describe a very small change in the potential near the mirror due to a Yukawa-like coupling. Already from this very simplified calculation we predicted differences in the spatial and momentum distributions between cases with and without a Yukawa-like interaction. However, in order to make a statement about a realistic experimental situation, the probability distribution of neutrons at the transition from region I to region II should be known in detail when also taking into account the Yukawa-like interaction in region I. Though, this is beyond the scope of this article since a much more intricate computation would be required.

\vspace{1.0cm}
%\textbf{Acknowledgement} 
%\vspace{0.1cm}
%\\
%We want to thank Mario Pitschmann for %beneficial discussions and helpful %suggestions. 
%
% Add the bibliography as a section "References" to the table of contents
\addcontentsline{toc}{section}{References}
\bibliography{Airy}{}
\bibliographystyle{unsrt}
 
\end{document}